\documentclass[twocolumn]{aastex62} 

\usepackage{bm}
\usepackage{natbib}
\usepackage{amsmath}
\usepackage{graphicx}
\usepackage{etoolbox}
\newcounter{lsaref}
\makeatletter
\apptocmd{\@lbibitem}{\stepcounter{lsaref}[\arabic{lsaref}]}{}{}
\makeatother

\setcitestyle{round}

\makeatletter
\let\frontmatter@title@above=\relax
\makeatother

\begin{document}

\title{\large OGLE-2017-BLG-1038: A Possible Brown-dwarf Binary Revealed by Spitzer Microlensing Parallax}


\author[0000-0001-5924-8885]{Amber Malpas}
\affiliation{University of Canterbury, Department of Physics and Astronomy, Private Bag 4800, Christchurch 8020, New Zealand}

\author[0000-0003-3316-4012]{Michael D. Albrow}
\affiliation{University of Canterbury, Department of Physics and Astronomy, Private Bag 4800, Christchurch 8020, New Zealand}

\author[0000-0001-9481-7123]{Jennifer C. Yee}
\affiliation{Center for Astrophysics $|$ Harvard \& Smithsonian, 60 Garden St.,Cambridge, MA 02138, USA}

\author{Andrew Gould}
\affiliation{Max-Planck-Institute for Astronomy, K\"onigstuhl 17, 69117 Heidelberg, Germany}
\affiliation{Department of Astronomy, Ohio State University, 140 W. 18th Ave., Columbus, OH 43210, USA}
\affiliation{Korea Astronomy and Space Science Institute, Daejon 34055, Korea}

\author[0000-0001-5207-5619]{Andrzej Udalski}
\affiliation{Astronomical Observatory, University of Warsaw, Al. Ujazdowskie 4, 00-478 Warszawa, Poland}

\author[0000-0002-3654-4662]{Antonio Herrera Martin}
\affiliation{University of Canterbury, Department of Physics and Astronomy, Private Bag 4800, Christchurch 8020, New Zealand}

\nocollaboration


\author{Charles A. Beichman}
\affiliation{IPAC, Mail Code 100-22, Caltech, 1200 E. California Blvd., Pasadena, CA 91125, USA}

\author{Geoffery Bryden}
\affiliation{Jet Propulsion Laboratory, California Institute of Technology, 4800 Oak Grove Drive, Pasadena, CA 91109, USA}

\author[0000-0002-7669-1069]{Sebastiano Calchi Novati}
\affiliation{IPAC, Mail Code 100-22, Caltech, 1200 E. California Blvd., Pasadena, CA 91125, USA}

\author[0000-0002-0221-6871]{Sean Carey}
\affiliation{IPAC, Mail Code 100-22, Caltech, 1200 E. California Blvd., Pasadena, CA 91125, USA}

\author[0000-0001-8877-9060]{Calen~B.~Henderson}
\affiliation{IPAC, Mail Code 100-22, Caltech, 1200 E. California Blvd., Pasadena, CA 91125, USA}

\author[0000-0003-0395-9869]{B.~Scott~Gaudi}
\affiliation{Department of Astronomy, Ohio State University, 140 W. 18th Ave., Columbus, OH  43210, USA}

\author[0000-0003-1525-5041]{Yossi Shvartzvald}
\affiliation{Department of Particle Physics and Astrophysics, Weizmann Institute of Science, Rehovot 76100, Israel}

\author[0000-0003-4027-4711]{Wei Zhu}
\affiliation{Department of Astronomy, Tsinghua University, Beijing 100084, China}

\collaboration{{\it Spitzer} team}

\author[0000-0002-7511-2950]{Sang-Mok Cha}
\affiliation{Korea Astronomy and Space Science Institute, Daejon 34055, Republic of Korea}
\affiliation{School of Space Research, Kyung Hee University, Yongin, Kyeonggi 17104, Republic of Korea} 

\author[0000-0001-6285-4528]{Sun-Ju Chung}
\affiliation{Korea Astronomy and Space Science Institute, Daejon 34055, Republic of Korea}
\affiliation{University of Science and Technology, Korea, (UST), 217 Gajeong-ro Yuseong-gu, Daejeon 34113, Republic of Korea}

\author[0000-0002-2641-9964]{Cheongho Han}
\affiliation{Department of Physics, Chungbuk National University, Cheongju 28644, Republic of Korea}

\author[0000-0002-9241-4117]{Kyu-Ha Hwang}
\affiliation{Korea Astronomy and Space Science Institute, Daejon 34055, Republic of Korea}

\author[0000-0002-0314-6000]{Youn Kil Jung}
\affiliation{Korea Astronomy and Space Science Institute, Daejon 34055, Republic of Korea}

\author{Dong-Jin Kim}
\affiliation{Korea Astronomy and Space Science Institute, Daejon 34055, Republic of Korea}

\author[0000-0001-8263-1006]{Hyoun-Woo Kim}
\affiliation{Korea Astronomy and Space Science Institute, Daejon 34055, Republic of Korea}

\author[0000-0003-0562-5643]{Seung-Lee Kim}
\affiliation{Korea Astronomy and Space Science Institute, Daejon 34055, Republic of Korea}
\affiliation{University of Science and Technology, Korea, (UST), 217 Gajeong-ro Yuseong-gu, Daejeon 34113, Republic of Korea}

\author[0000-0003-0043-3925]{Chung-Uk Lee}
\affiliation{Korea Astronomy and Space Science Institute, Daejon 34055, Republic of Korea}
\affiliation{University of Science and Technology, Korea, (UST), 217 Gajeong-ro Yuseong-gu, Daejeon 34113, Republic of Korea}

\author{Dong-Joo Lee}
\affiliation{Korea Astronomy and Space Science Institute, Daejon 34055, Republic of Korea}

\author[0000-0001-7594-8072]{Yongseok Lee}
\affiliation{Korea Astronomy and Space Science Institute, Daejon 34055, Republic of Korea}
\affiliation{School of Space Research, Kyung Hee University, Yongin, Kyeonggi 17104, Republic of Korea}

\author[0000-0002-6982-7722]{Byeong-Gon Park}
\affiliation{Korea Astronomy and Space Science Institute, Daejon 34055, Republic of Korea}
\affiliation{University of Science and Technology, Korea, (UST), 217 Gajeong-ro Yuseong-gu, Daejeon 34113, Republic of Korea}

\author[0000-0003-1435-3053]{Richard W. Pogge}
\affiliation{Department of Astronomy, Ohio State University, 140 W. 18th Ave., Columbus, OH  43210, USA}

\author[0000-0001-9823-2907]{Yoon-Hyun Ryu}
\affiliation{Korea Astronomy and Space Science Institute, Daejon 34055, Republic of Korea}

\author[0000-0002-4355-9838]{In-Gu Shin}

\affiliation{Center for Astrophysics $|$ Harvard \& Smithsonian, 60 Garden St.,Cambridge, MA 02138, USA}
\affiliation{Department of Physics, Chungbuk National University, Cheongju 28644, Republic of Korea}

\author[0000-0001-6000-3463]{Weicheng Zang}
\affiliation{Department of Astronomy, Tsinghua University, Beijing 100084, China}

\collaboration{KMTNet Collaboration}

\author[0000-0002-6212-7221]{Patryk Iwanek}
\affiliation{Astronomical Observatory, University of Warsaw, Al. Ujazdowskie 4, 00-478 Warszawa, Poland}

\author[0000-0003-4084-880X]{Szymon Koz{\l}owski}
\affiliation{Astronomical Observatory, University of Warsaw, Al. Ujazdowskie 4, 00-478 Warszawa, Poland}

\author[0000-0001-7016-1692]{Przemek Mr\'{o}z}
\affiliation{Astronomical Observatory, University of Warsaw, Al. Ujazdowskie 4, 00-478 Warszawa, Poland}
\affiliation{Division of Physics, Mathematics, and Astronomy, California
Institute of Technology, Pasadena, CA 91125, USA}

\author[0000-0002-2339-5899]{Pawe{\l} Pietrukowicz}
\affiliation{Astronomical Observatory, University of Warsaw, Al. Ujazdowskie 4, 00-478 Warszawa, Poland}

\author[0000-0002-9245-6368]{Radoslaw~Poleski}
\affiliation{Astronomical Observatory, University of Warsaw, Al. Ujazdowskie 4, 00-478 Warszawa, Poland}

\author{Krzysztof A. Rybicki}
\affiliation{Astronomical Observatory, University of Warsaw, Al. Ujazdowskie 4, 00-478 Warszawa, Poland}

\author[0000-0002-2335-1730]{Jan~Skowron}
\affiliation{Astronomical Observatory, University of Warsaw, Al. Ujazdowskie 4, 00-478 Warszawa, Poland}

\author[0000-0002-7777-0842]{Igor Soszy\'{n}ski}
\affiliation{Astronomical Observatory, University of Warsaw, Al. Ujazdowskie 4, 00-478 Warszawa, Poland}

\author[0000-0002-0548-8995]{Micha{\l}~K.~Szyma\'{n}ski}
\affiliation{Astronomical Observatory, University of Warsaw, Al. Ujazdowskie 4, 00-478 Warszawa, Poland}

\author[0000-0001-6364-408X]{Krzysztof Ulaczyk}
\affiliation{Department of Physics, University of Warwick, Gibbet Hill Road, Coventry, CV4~7AL,~UK}

\collaboration{OGLE Collaboration}

\begin{abstract}
We report the analysis of microlensing event OGLE-2017-BLG-1038, observed by the Optical Gravitational Lensing Experiment, Korean Microlensing Telescope Network, and Spitzer telescopes. The event is caused by a giant source star in the Galactic Bulge passing over a large resonant binary lens caustic. The availability of space-based data allows the full set of physical parameters to be calculated. 
However, there exists an eightfold degeneracy in the parallax measurement. The four best solutions correspond to very-low-mass binaries near ($M_1 = 170^{+40}_{-50} M_J$ and $M_2 = 110^{+20}_{-30} M_J$), or well below ($M_1 = 22.5^{+0.7}_{-0.4} M_J$ and $M_2 = 13.3^{+0.4}_{-0.3} M_J$) the boundary between stars and brown dwarfs. 
A conventional analysis, with scaled uncertainties for Spitzer data, implies a very-low-mass brown dwarf binary lens at a distance of 2 kpc. Compensating for systematic Spitzer errors using a Gaussian process model suggests that a higher mass M-dwarf binary at 6 kpc is equally likely.
A Bayesian comparison based on a galactic model favors the larger-mass solutions.
We demonstrate how this degeneracy can be resolved within the next ten years through infrared adaptive-optics imaging with a 40 m class telescope.

\end{abstract}


\section{Introduction} \label{sec:Intro}

Microlensing is a phenomenon in which the path of light emitted from a distant star (the source) is bent by a curve in space-time, caused by a massive object (the lens). If the source is approximately behind the lens, as seen by an observer, it brightens as unresolved images of the source are formed about the Einstein ring that has angular radius
\begin{equation}
    \theta_{\rm E} = \sqrt{\frac{4GM}{c^2}\left(\frac{{\rm au}}{D_{\rm L}} - \frac{{\rm au}}{D_{\rm S}}\right)}=\sqrt{\kappa M\pi_{\rm rel}},
    \label{E:thetaE}
\end{equation}
where $\pi_{\rm rel}= {\rm au}(D_{\rm L}^{-1} - D_{\rm S}^{-1})$, $M$ is the mass of the lens system, $D_{\rm L}$ and $D_{\rm S}$ are the distance to the lens and source, respectively, and $\kappa=4G/(c^2{\rm {\rm au}})\sim8.14 \, {\rm mas}/M_\odot$. 

For transient alignments, where the closest angular separation of the source and lens is on the order of $\theta_{\rm E}$ or smaller, photometric microlensing events can be observed as increasing and decreasing apparent brightness of the combination of the source star and unresolved neighbors, including the lens. Because only the source light is magnified, the luminosity of the lens system does not directly contribute to the event detection rate. As a result, microlensing is uniquely sensitive to the detection of low-mass, dim lenses such as brown dwarfs (BD; for example, \citealt{No1BD}, \citealt{SpitzerSwift1BD}, and \citealt{2020BDs}) and unbound planetary-mass objects (for example, \citealt{2020FFPMroz} and \citealt{2FFPs}) as proposed by \citet{FFPProposed}.

A limitation of the microlensing method is that, for most microlensing events, the light-curve model leaves a degeneracy between the mass and distance of the lens. This degeneracy can in principle be resolved either by measuring two other parameters (the Einstein radius $\theta_{\rm E}$, and the microlens parallax ${\bm \pi}_{\rm E}$) or by separately observing the lens and source some years after the event in high-resolution images. While $\theta_{\rm E}$ has been measured for most planetary and binary events published to date, ${\bm \pi}_{\rm E}$ has not. 

For events with an extremely dim lens, proper-motion measurement via late time imaging is not feasible at typical lens distances, given current observing capabilities. 
Breaking the mass-distance degeneracy for very faint lens systems thus requires a measurement of the microlens parallax. The spatial separation between observers required to detect parallax at a single epoch depends on characteristics of the microlensing event, such as the distance to the lens system and duration of the event. Because of the large separation between Earth and the Spitzer Space Telescope (located more than $1 \, {\rm au}$ distant from Earth), microlensing observations from Spitzer, in conjunction with those from Earth, provide a reliable means of measuring parallax. This uniquely wide separation is what motivated the Spitzer microlensing project \citep{Spitzer}.

Microlensing has been used to discover 34 BDs from beyond the local regime \citep{Chung2019}. So far, this extended population has demonstrated unusual dynamics, such as an unexpected number of counter-rotating BDs \citep{Chung2019,Shvartzvald2019,Shvartzvald2017}. It is unclear to what degree these extreme kinematics are representative of the population as a whole. 

BDs are stellar-like objects that are not massive enough to maintain a sufficient core temperature for main-sequence hydrogen fusion. Though the more massive BDs are capable of lithium fusion, and most BDs are capable of deuterium fusion, these processes do not provide sufficient heat to stop BDs from gradually cooling as they radiate the heat generated during their formation. As a result, they are very faint and become fainter as they age. Deuterium fusion occurs in objects with masses of approximately $>13\,M_J$. This is often adopted as a criterion to distinguish BDs from planets; objects below this mass are planets, be they bound to a stellar object or free floating. However this mass definition is sometimes in conflict with the formation definition: BDs form like stars and planets form in circumstellar disks.

All but five of the microlensing BDs have been detected as binary systems. The number of BDs detected in binaries makes up an artificially high proportion of the total number of detections because binary events have more easily detected finite-source effects and therefore are more likely to have their associated masses calculated. Some of these have member masses at about the deuterium fusion limit \citep{choi2013,han2017a,albrow2018}, supporting the arguments of \cite{Grether2006} and \cite{chabrier2014} for a mass overlap between the gas-giant planet and BD regimes. Deuterium fusion has become an insufficient metric for classification between BD and gas-giant planets. These populations have distinct formation histories, which, though difficult to infer, provide a more meaningful way to separate them in the mass-overlap region. 

The upper BD cutoff is defined by sustained hydrogen fusion. Studies evaluating the hydrogen burning limit are summarised in Table 5 of \cite{HydrogenBurningLimit}, from which we deduce that the BD upper limit is in the range of ($\sim70 - 95 \, M_J$). This variance has a large dependence on chemical composition (e.g., \citealt{HydrogenBurningVSMetalicity}). \cite{Forbes2019} investigate the idea of over-massive BDs. These are theoretically formed through Roche lobe overflow. The result is that, with only the mass information to draw from, this cutoff is vague. 

Little is known about the very low mass end of the stellar initial mass functions (IMF). The empirical IMFs of \cite{kroupa2001}, \cite{chabrier2005}, \cite{thies2007,thies2008}, and \cite{kroupa2013} show disparity with the theoretical IMFs deduced from analytical descriptions of pre-stellar-cloud core distributions \citep{padoan2002,hennebelle2008,hennebelle2009} at the very-low-mass end, approximately between $84 \, M_J$ and $210 \, M_J$ ($0.08 \, M_{\odot}$ and $0.2 \, M_{\odot}$). Empirical IMFs usually require assumptions about age and metalicity in order to determine the IMF from an observed luminosity function. Observationally, measuring a mass function across the entire stellar mass range is challenging because sampling the upper mass range requires massive star clusters, and sampling the lower mass range requires nearby clusters. With the closest massive clusters at distances of a few kiloparsecs, observing both ends of the mass function in one star cluster is not currently possible photometrically \citep{elmegreen2009}. \cite{wegg2017} shows one way in which microlensing surveys can be used to probe the IMF of the inner Milky Way, although this method used an existing dynamical model to infer the masses from the timescales ($t_{\rm E}$) of $\sim4000$ events and therefore is not purely empirical. The timescales considered were $2 \, {\rm days} < t_{\rm E} < 200 \, {\rm days}$, which relates to the mass via $t_{\rm E}^2 \propto M$.

Currently, photometric surveys are only capable of probing relatively bright and very local populations of BDs. For example, \cite{rosell2019} quote a distance limit in their Dark Energy Survey catalog of ``beyond $400\,{\rm pc}$''. This selection bias in observability provides a limited view of BDs, in distance, mass and age. Further detections of very-low-mass objects in binary systems, will help to clarify our understanding of the dynamical properties of BD populations and the low-mass end of the IMF, because such systems are likely to have been formed as part of the very-low-mass end of the IMF, not like planets in a circumstellar disk.

The following sections in this paper describe our analysis of microlensing event OGLE-2017-BLG-1038 and how we determined this event to be a BD binary. \S \ref{sec:Observations} describes the observations made of this event, and the data-reduction methods used. \S \ref{S:Ground} outlines our analysis of the ground-based data and resulting conclusions about source star characteristics. \S \ref{S:Satellite} details our analysis of the space-based, Spitzer data and our final modeling results.  The corresponding physical parameters for our most likely models are calculated in \S \ref{S:Physical}. In \S \ref{S:P} we compare the relative probabilities of our best model solutions and then we discuss, in \S \ref{S:Discussion}, how different assumptions of the galactic model, as well as selection effects, may influence these probabilities. 


\section{Data Collection and Reduction} \label{sec:Observations}

OGLE-2017-BLG-1038 is located at (R.A., decl.)$_{\rm J2000} = (17:58:36.55,\, -27:18:58.4)$, $(l,b) = (2.8536,-1.6382)^{\circ}$. It was first identified as a microlensing event candidate by the Optical Gravitational Lensing Experiment early warning system (OGLE; \citealt{OGLE-EWS}), on 2017 June 3, from their ongoing survey (mostly in $I$ band) using the $1.3 \, \rm{m}$ Warsaw telescope in the Las Campanas Observatory in Chile. Repeated OGLE observations of the event took place at an interval of mostly 1 day.

The Korean Microlensing Telescope Network (KMTNet; \citealt{KMTNet}) also discovered this event as KMT-2017-BLG-0363 and observed it in the \textit{V} and \textit{I} bands. OGLE-2017-BLG-1038 was observed in two overlapping KMTNet search fields (BLG03 and BLG43), from each of the three KMTNet telescopes: Cerro Tololo Inter-American Observatory (KMT-C), South African Astronomical Observatory (KMT-S), and Siding Springs Observatory (KMT-A). This resulted in a cadence of $\sim 15$ minutes between successive observations.
The KMTNet observations were also primarily made in the \textit{I} band. However, occasional \textit{V}-band observations were made to provide color information. Therefore, 12 sets of KMTNet light curves were obtained for this event. 

The end of the event was also observed by the Spitzer Space Telescope Infrared Array Camera (IRAC; \citealt{Spitzer-IRAC}) instrument at an approximately 1 day cadence. While both the KMTNet and OGLE observations were made as part of regular survey operations, the Spitzer observations were scheduled for this event specifically as part of a program to enable space-parallax measurements for microlensing events \citep{Novati2015-survey,Spitzer}. This event was selected for Spitzer observations on 2017 June 13 (HJD' = 7918.11) and met the objective criteria on 2017 June 19 (HJD' = 7923.95). Both of these selections took place before the binary nature of the event was recognized, i.e., when it was still believed to be a point lens. Members of the Spitzer Team first noticed that the event was anomalous on 2017 June 20 (HJD' 7925.04).

Kinematic measurements from the source star in this event, as well as surrounding field stars, were obtained from Gaia Early Data Release 3 \citep{GaiaEDR3,GaiaMission}. 

The ground-based data were reduced using difference imaging \citep{tomaney1996,Alard1998} procedures. The OGLE images were reduced with their custom difference image procedures (see \citealt{2000AcA....50..421W}). The KMTNet light curves were extracted from the images using pyDIA \citep{PyDIA} software, and the Spitzer light curve was extracted by the methods detailed in \citet{Novati2015-photometry}.


\section{Ground-Based Analysis} \label{S:Ground}

\begin{figure}
\centering
\includegraphics[width=0.47\textwidth]{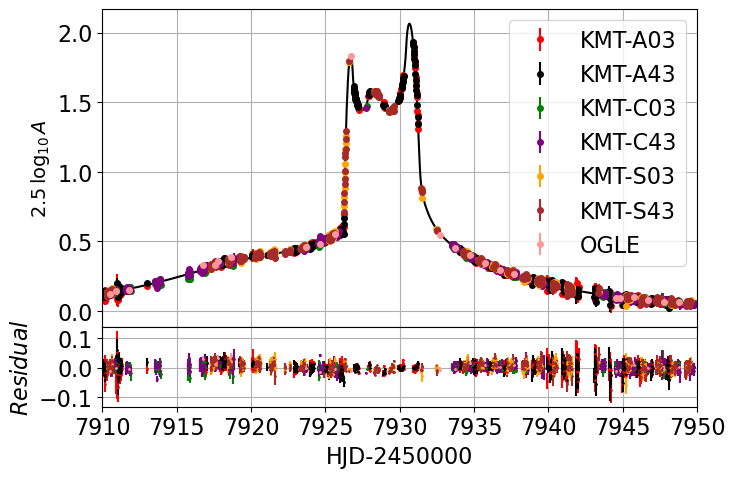}
\caption{Magnification curves resulting from the fitted static binary-lens model.} \label{Fig:StaticLC}
\end{figure}

The light curve of this event (see Figure \ref{Fig:StaticLC}) has a triple-peaked perturbation over a 5 day period (2017 June 22-27) with the three peaks showing smoothed curves, indicative of a resolved source crossing a caustic. Caustics are features of a multiple-lens system. Therefore, we began our modeling with a binary-lens model, which we ultimately found was sufficient to describe the light curves for this event. 

The binary-lens model is parameterized by ($s$, $q$, $\rho$, $u_{0}$, $\alpha$, $t_0$ , $t_{\rm E}$), where $s$ is the angular separation of the two lens masses in units of $\theta_{\rm E}$, $q$ is the mass ratio of the lens objects, $\rho$ is the source angular radius in units of $\theta_{\rm E}$, $u_0$ is the closest line-of-sight point of approach to the lens center of mass made by the source in its relative trajectory (again in units of $\theta_{\rm E}$), and $t_0$ is the time at which this happens ($\left|u_0\right| = u\left(t_0\right)$, where ${\mathbf u}(t_i)$ is the position of the source, projected onto the lens plane, at a given time, ($t$), $\alpha$ is the angle of the projected rectilinear source trajectory relative to an axis that passes through the lens masses, and $t_{\rm E}$ is the Einstein radius crossing time (the time the source takes to travel an angular distance of $\theta_{\rm E}$). For simplification, the motions in these models were considered from the reference frame of the lens system. This meant that, for modeling purposes, the relative velocities of any of the bodies involved were attributed to the “source velocity”.

Our analysis of the ground-based light curves began by performing a grid search over a fixed resolution on $s$, $q$, $u_0$, and $\alpha$, using point-source approximations away from the caustics, for their computational speed, and convolved magnification maps in high-magnification regions, where finite-source effects were significant. The other model parameters were fitted by $\chi^2$ minimization with $\rho$ values found by interpolating between grid points with discrete convolutions. These calculations used a modified version of the Microlensing Observations Rapid Search for Exoplanets code \citep{MORSE}.

\begin{figure}
\centering
\includegraphics[width=0.47\textwidth]{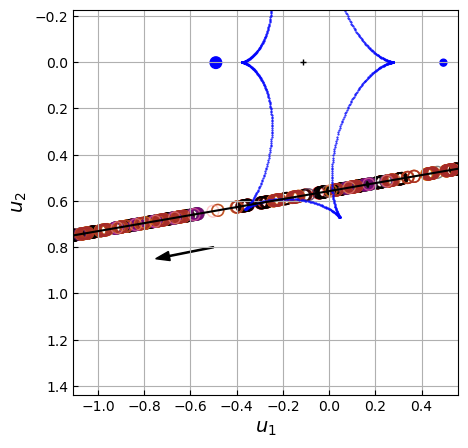}
\caption{Lens system and caustic geometry resulting from the $(s,q)=(1.0,1.0)$ grid seed, with a projected source trajectory, for the static binary-lens model, fitted to the ground-based data. Colored circles show the source position at the times of the data points, where the colors correspond to those specified in Figures~\ref{Fig:StaticLC} and~\ref{Fig:GroundCumulative}, and the circle size depicts the source size.} \label{Fig:StaticCaustic}
\end{figure}

The best 20 grid solution regions were further investigated using the Emcee sampler \citep{emcee}. For this process we used the more accurate Image Centered Inverse RAy Shooting (ICIRAS) \citep{ICIRAS}  or contour integration \citep{Bozza2010, Bozza2018}  methods to calculate the model magnification in regions close to caustics, and the hexadecapole  approximation \citep{pejcha2009,hexadecapole} otherwise. A fixed limb-darkening coefficient ($\Gamma=0.53$)\footnote{Attempts to include this limb-darkening coefficient as a free parameter in the model later in the modeling process did not result in a more likely coefficient being found.} was applied to the source in these calculations. Two of the regions converged to the same, and significantly most likely solution, while the next most likely solution had a $\Delta\chi^2$ of $\sim110\,000$, before renormalization. The geometry of this static, ground-based solution is shown in \figurename{ \ref{Fig:StaticCaustic}}, and the magnification curve, with ground-based data, is shown in \figurename{ \ref{Fig:StaticLC}}. The fitted model parameters are displayed in Table \ref{T:GroundModelComparison} as the Static model. The solution corresponds to a source passing over the edges of a large resonant caustic. We note that this solution corresponds to small negative blending for three of the data sources, though this is a normal occurrence for microlensing photometry in a very crowded bulge field \citep{Park2004}, especially for dim lenses. Table \ref{T:GroundModelComparison} shows $F_{\rm B}/F_{\rm S}$ for the OGLE source, which is within $2 \, \sigma$ of being positive. 

The source fluxes for each data set, were found from a linear fit;
\begin{equation}
    F_i=A_i\times F_{\rm S}+F_{\rm B},
\end{equation}
\noindent where $F_{\rm S}$ is the source-star flux, $F_{\rm B}$ is the blended flux\footnote{The blended flux is made up of the nonlensed contributions to the light-curve flux measurements, from light sources near the line of sight. Sometimes the largest contributor to this flux component is the lens star, though this is rarely the case.}, $A_i$ is the magnification at time $t_i$, and $F_i$ is the observed total flux at time $t_i$. 
This solution to the static model was used to renormalize the ground-based data uncertainties (see \citealt{renormerrors}), and the solution was then allowed to reconverge.


\subsection{Lens Orbital Motion or Ground-Based Parallax?}

\begin{figure} 
\centering
\includegraphics[width=0.47\textwidth]{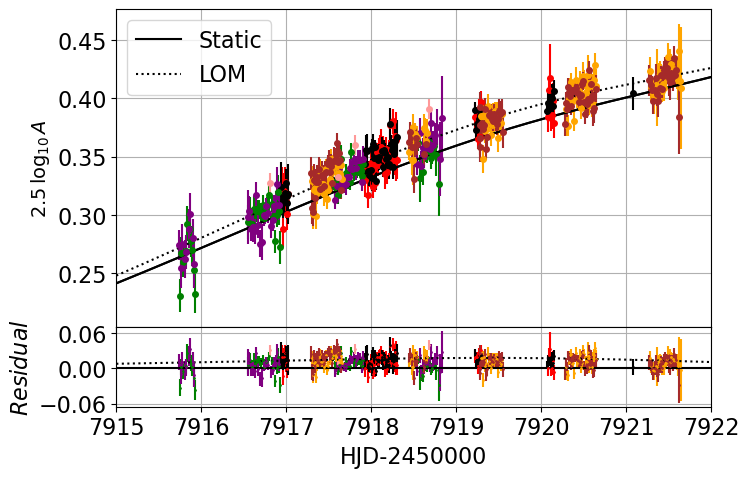}
\caption{7915-7922 HJD crop of the magnification model from the best, static fit (solid black lines) and the corresponding ground-based light-curve data with renormalized errors. The data show a clear trend above the fit line in this region. The dotted black lines show the lens-orbital-motion-inclusive magnification model used in the next step of this event analysis. Outside of this crop region the two models are visually indistinguishable. The data colors correspond to those specified in Figures~\ref{Fig:StaticLC} and~\ref{Fig:GroundCumulative}} \label{Fig:StaticProblemRegion}
\end{figure}

\begin{figure} 
\centering
\includegraphics[width=0.47\textwidth]{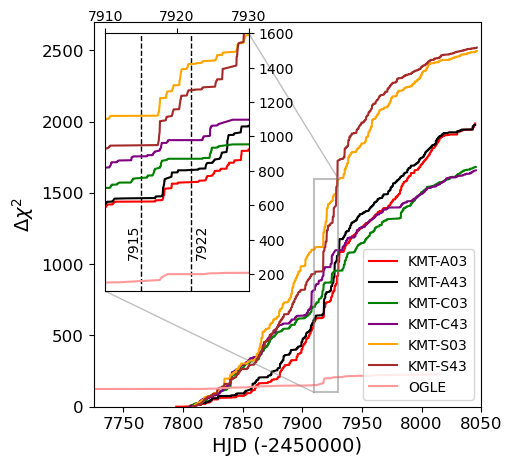}
\caption{Cumulative $\chi^2$ plot for the renormalized static model.} \label{Fig:GroundCumulative}
\end{figure}

Although the peaks of the light curve are well fitted by this static-lens, rectilinear-source model, there is a region between dates 7915-7922 where the model systematically underpredicts the data (Figure~\ref{Fig:StaticProblemRegion}). In Figure~\ref{Fig:GroundCumulative} we show the cumulative $\chi^2$ as a function of time for each individual data set. All curves show significant jumps near 7915-7922, indicating that there is a real missing feature in our static model. Higher-order effects are required for the model to provide a good description of these data.

Common high-order effects in microlensing light curves are 
orbital parallax (motion of Earth during an event) and orbital motion of the binary-lens system. A known degeneracy exists between these. Suspecting the significance of one or both of these higher-order effects, we added them to the generative model, both collectively and separately. We approximated the orbital motion of the lens objects by allowing $\alpha$ and $s$ to vary linearly with time, adding the model parameters $\dot{\alpha}$ and $\dot{s}$. Modeling the parallax effect requires the introduction of two new parameters, $(\pi_{{\rm E},N},\pi_{{\rm E},E})$, which are components of the vector ${\boldsymbol\pi_{\rm E}}$, where $\lVert{\boldsymbol\pi_{\rm E}}\rVert = \frac{\pi_{\rm rel}}{\theta_{\rm E}}$, and its direction is that of the lens-source relative proper motion. The introduction of measurable parallax breaks the reflected symmetry of the source trajectory about the lens axis; a trajectory above the lens axis is not equivalent to a trajectory below the lens axis (except in the limit that the source lies exactly on the ecliptic). We therefore modeled both positive and negative $u_0$ solutions in which parallax was considered. For those solutions with both parallax and lens orbital motion, we calculate $\beta$ (the ratio of the projected kinetic to potential energy of the lens; \citealt{An2002, Dong2009}), where values less than unity indicate a lens system consistent with a bound orbit;

\begin{equation}\beta = \frac{ 2({\rm au})^{2}}{c^{2}} \frac{\pi_{\rm E}}{\theta_{\rm E}} \frac{\left[\left(\frac{1}{s}\frac{ds}{dt}\right)^{2}+\left(\frac{d\alpha}{dt}\right)^{2}\right]s^{3}} {\left[\pi_{\rm E}+\left(\frac{\pi_{\rm S}}{\theta_{\rm E}}\right)\right]^{3}}.
\end{equation}

In our investigations of the significance of these two higher-order effects (Table~\ref{T:GroundModelComparison}), we find that, alone, lens orbital motion describes the static model discrepancies better than parallax. Including both higher-order effects yields only a minor $\chi^2$ reduction compared with the purely lens-orbital-motion model, and the lens-orbital-motion parameters change very little. (The low $\beta$ values for these models show that the implied orbits are bound.)
Conversely, the posteriors of the parallax model change drastically when lens orbital motion is added. We therefore conclude lens orbital motion is well constrained and sufficient to describe the deviation on the static model from 7915-7922. This model is illustrated by the dotted lines in Figure~\ref{Fig:StaticProblemRegion}.

\begin{table*}[t]
\caption{Comparison of the Highest Likelihood Fit Parameters for Binary-lens Models with and without the Higher-order Effects of Parallax and Lens Orbital Motion, Fit to Ground-based Data, with Renormalized Errors}
\begin{tabular}{c c c c c c c}
\hline \hline
& Static & \multicolumn{2}{c}{Parallax} & LOM & \multicolumn{2}{c}{Parallax + LOM} \\ \cline{3-4}\cline{6-7}
${\hat u}_0$ & - & + & - & - & + & - \\
\hline
$s$ & $0.9833^{+0.0006}_{-0.0005}$ 
        & $0.9757^{+0.0003}_{-0.0009}$
            & $0.9978^{+0.0011}_{-0.0005}$
                & $0.9932^{+0.0002}_{-0.0009}$
                    & $0.9916^{+0.010}_{-0.0009}$
                        & $0.9888^{+0.015}_{-0.0004}$\\
$q$ & $0.621 \pm 0.002$
        & $0.616^{+0.003}_{-0.002}$
            & $0.590^{+0.002}_{-0.003}$
                & $0.607 \pm 0.003$
                    & $0.609^{+0.004}_{-0.003}$
                        & $0.615^{+0.002}_{-0.005}$ \\
$\log_{10}\rho$ & $-1.6027^{+0.0006}_{-0.0011}$
        & $-1.6094^{+0.0011}_{-0.0008}$
            & $-1.5818^{+0.0017}_{-0.0008}$
                & $-1.5842^{+0.0004}_{-0.0012}$
                    & $-1.5859^{+0.0018}_{-0.0005}$
                        & $-1.5911^{+0.0026}_{-0.0004}$ \\
$u_0$ & $-0.5693^{+0.0008}_{-0.0006}$ 
        & $0.468^{+0.003}_{-0.009}$ 
            & $-0.430^{+0.012}_{-0.005}$
                & $-0.5551^{+0.0004}_{0.0012}$
                    & $0.552 \pm 0.005$
                        & $-0.647^{+0.029}_{-0.004}$ \\
$\alpha$ & $-2.9702^{+0.0005}_{-0.0007}$ 
        & $2.845^{+0.003}_{-0.011}$
            & $-2.853^{+0.011}_{-0.004}$
                & $-2.9710^{+0.0007}_{-0.0006}$
                    & $2.965^{+0.007}_{-0.006}$
                        & $-3.064^{+0.030}_{-0.004}$ \\
$t_0 $ & $7926.900^{+0.007}_{-0.006}$
        & $7924.47^{+0.07}_{-0.16}$
            & $7927.33^{+0.06}_{-0.04}$
                & $7927.009^{+0.003}_{-0.011}$
                    & $7926.83^{+0.10}_{-0.15}$
                        & $7927.37^{+0.03}_{-0.22}$ \\
$ t_{\rm E}$ & $11.852^{+0.015}_{-0.018}$
        & $13.57^{+0.10}_{-0.07}$
            & $10.55^{+0.02}_{-0.08}$
                & $11.855^{+0.023}_{-0.007}$
                    & $12.02^{+0.15}_{-0.10}$
                        & $12.33^{+0.11}_{-0.16}$ \\
\hline
$\pi_{{\rm E}N}$ & 0
        & $-11.5^{+0.3}_{-1.0}$ 
            & $12.6^{+1.2}_{-0.5}$ 
                & 0 
                    & $-0.6^{+0.6}_{-0.5}$ 
                        & $-9.6^{+3.1}_{-0.4}$ \\
$\pi_{{\rm E}E}$ & 0
        & $10.7 \pm 0.5$ 
            & $-10.0^{+0.2}_{-0.8}$ 
                & 0 
                    & $1.1^{+1.1}_{-0.7}$ 
                        & $1.9^{+0.8}_{-0.7}$ \\
$\dot{s}$ & 0 & 0 & 0 
                & $0.30 \pm 0.04$ 
                    & $0.31^{+0.08}_{-0.04} $ 
                        & $0.24^{+0.07}_{-0.04}$ \\
$\dot{\alpha}$ & 0 & 0 & 0 
                & $-1.27 \pm 0.04$ 
                    & $1.20^{+0.06}_{-0.05}$ 
                        & $-1.57^{+0.10}_{-0.06}$ \\
\hline
$\beta$ & & & & 
                    & 0.13 
                        & 0.01 \\
$\chi^2_{min}$ & 12592.83 
        & 11946.44 
            & 12047.65 
                & 11468.38 
                    & 11466.26 
                        & 11441.77 \\
$\Delta\chi^2_{min}$ & 0 
        & -646.40 
            & -545.19 
                & -1124.46 
                    & -1126.57 
                        & -1151.07 \\
$N$ & \multicolumn{6}{c}{$12607$} \\
\hline
$I_{\rm S,OGLE}$ & 16.4 & & & & & \\ 
$F_{\rm B,OGLE}/F_{\rm S,OGLE}$ & -0.0075 $\pm$ 0.0041 & & & & & \\
\hline%
\end{tabular}
\tablecomments{Those solutions indicated to by ``LOM'' refer to the models in which lens orbital motion was included. The source magnitude uses a zero point of $I_{ZP}=28$. N is the total number of light-curve data points. Solutions with $\beta < 1$ are consistent with a bound orbit, but can only be calculated for models including both lens orbital motion and parallax.}
\label{T:GroundModelComparison} \end{table*}

\subsection{Source Color} \label{sS:Color}
\begin{figure}[ht!]
\centering
\includegraphics[width=0.5\textwidth]{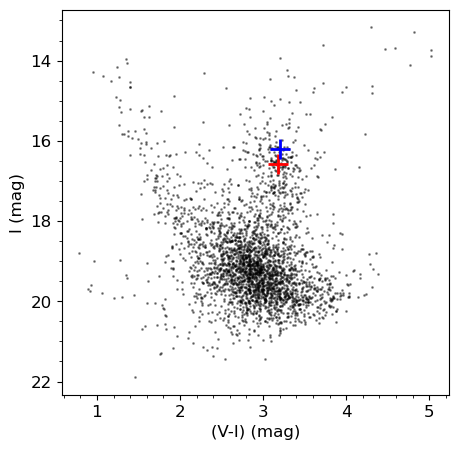}
\caption{Color-magnitude diagram from the KMTC-03 field with the fitted centroid of the red clump and the source position indicated by the red ``+'' and blue ``+'', respectively. \label{Fig:CMD}}
\end{figure}

Color-magnitude Diagrams (CMDs) were created for each KMTNet observation site and field with $I$ and $V$ data (KMTC-03; Figure \ref{Fig:CMD}, KMTC-43, KMTS-03, KMTS-43, KMTA-03, and KMTA-43). We use the normal KMT practice of adopting magnitude zero points of $I_{ZP} = 28$ and $V_{ZP} = 28.65$. The source-star fluxes, obtained from fitting the magnification model to each light curve, were used to find the source star's position on the corresponding CMDs. The source fluxes for the highest likelihood solution (ground based) are given in Table \ref{T:lmpFluxes}.

\begin{table} 
\hfill
\caption{Source fluxes, for Each Observation Source and Band}
\setlength\tabcolsep{0pt}
\begin{tabular*}{\linewidth}{@{\extracolsep{\fill}} lccr }
\hline
\hline
Source & $F_{{\rm S},I}$ & $F_{{\rm S},V}$ & $F_{{\rm S},L}$ \\
\hline
KMTC-03 & 52595.17 & 5000.67 & \\
KMTC-43 & 34761.72 & 5074.91 & \\
KMTS-03 & 50862.79 & 4125.91 & \\
KMTS-43 & 52803.33 & 4511.05 & \\
KMTA-03 & 40084.08 & 4813.62 & \\
KMTA-43 & 38048.48 & 4800.27 & \\
OGLE & 43537.75 & & \\
\textit{Spitzer} & & & 56.09 \\ 
\hline
\end{tabular*}
\tablecomments{These values were calculated using an orbiting, binary-lens model, for each of the ground-based sources. The Spitzer source flux is an estimate based on comparative CMDs between the Spitzer field and the KMTC-03 field.}
\label{T:lmpFluxes}
\end{table}

The red clump in each CMD was centroid fitted, and acted as a calibration for obtaining the intrinsic colors and magnitudes of the field.  The galactic bulge red clump can be used to calibrate the CMD because its intrinsic color and magnitude are known to high precision. The intrinsic color of the red clump is $(V-I)_{{\rm RC},0} = 1.06$ \citep{Bensby2011}. The intrinsic I-magnitude of the red-clump was found by interpolating the extinction correction table from \citep{Nataf2013a} for the target's galactic longitude $(l=2.85^{\circ}, \, b=-1.64^{\circ})$; $I_{{\rm RC},0}=14.35 \pm 0.04$. Assuming that the source is obscured by the same amount of dust as the average red clump star in this field, $(V-I)_{{\rm RC},0}$ and $I_{{\rm RC},0}$ provide an absolute color and magnitude calibration to the CMDs.

Using the mean calibrated color and magnitude, the intrinsic magnitude and color of the source was found to be $(I_0,(V-I)_0) = (14.01\pm0.05,1.11\pm0.04)$, averaged over all six CMDs. These values are very similar for each of the possible solutions for the final model. 

This source color information was also used to infer the Spitzer source flux and a color-color relation between KMTC-03 and Spitzer using the method of \citet{Novati2015-photometry}.  The expected Spitzer Source flux is $F_{{\rm S},L}=56.1\pm1.7$ and the optical-infrared source color is $(I-L)_{\rm S} = -4.43 \pm 0.03$, with an $L$-magnitude zero point of 25.

\section{Inclusion of Satellite Data} \label{S:Satellite}

Having a Spitzer light curve for this event meant that, despite there being very inconclusive orbital parallax signals in the ground-based data, parallax could still be measured \citep{Refsdal1966}. In this section we describe our analysis of the space-based Spitzer data using typical error renormalization methods, discuss concerns over systematics errors in the data, and present an alternate approach to coping with such systematics.

\subsection{Satellite Parallax Degeneracies}\label{satellitedegen}

\begin{figure}
\centering
\includegraphics[width=0.5\textwidth]{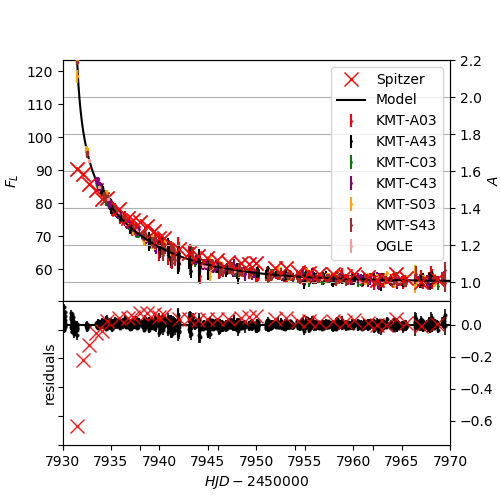}
\caption{Raw Spitzer light curve and model light curve resulting from the static binary-lens model (no parallax), fitted to the ground-based data, and transforming to the Spitzer flux system assuming $F_{\rm L} \equiv F_{{\rm S},L} A$, where $F_{{\rm S},L} = 56.1$ and $(I - L)_{\rm S} = -7.4$ . The ground-based observations have also been scaled to the Spitzer flux system. The residuals between the model and the data are depicted in black for ground-based data and red for \textit{Spitzer} data. These show a dramatic difference for the $t<7935$ data. \label{Fig:SpitzerLC}}
\end{figure}

\figurename{ \ref{Fig:SpitzerLC}} shows the raw Spitzer data and a corresponding magnification curve from estimating $F_{\rm S}=56.1$ (as is suggested by the color comparisons of \S \ref{sS:Color}), $F_{\rm B}=0$, and adopting the ground-based model. In this figure, we can see a clear, decreasing signal that has $\Delta F>30$ Spitzer flux units. The Spitzer data are inconsistent with very small parallax, as the shape of the magnification curve is not well represented by the static ground-based model, and no alternative values of $F_{\rm S}$ and $F_{\rm B}$ could bring them into agreement. At the time of the first Spitzer observation, the ground-based light
curve is still exiting the cusp while the Spitzer data are
clearly not. This is strong evidence for a parallax
effect. At the same time, the required magnification change as seen from Spitzer ($\Delta A \sim 1.6$) indicates that the parallax cannot be too large. 

When viewed from Spitzer, the angular source trajectory across the lens plane is offset by a vector 
$(\Delta \beta, \Delta \tau)/ \theta_{\rm E}$, in
directions (perpendicular, parallel) to ${\bm D}_\perp$, the separation between Spitzer and Earth projected onto the lens plane. This vector is related to the parallax measurement, but can be more useful in understanding the parallax likelihood space in comparison with the caustic diagram representation of the event. The two parameters $(\Delta \beta, \Delta \tau)$ can be mapped onto $\pi_{{\rm E},E}$ and $\pi_{{\rm E},N}$, 
via ${\boldsymbol\pi_{\rm E}} = \frac{{\rm au}}{D_{\perp}} \left(\Delta\tau,\Delta\beta\right).$
The parallel offset is simply

\begin{equation}
\Delta \tau = \frac{t_{0,{\rm Spitzer}} - t_{0,{\rm Earth}}}{t_{\rm E}}.
\end{equation}

In the case of a single lens, the perpendicular offset suffers from a four-fold satellite parallax degeneracy, 

\begin{equation}
\Delta \beta = \pm u_{0, {\rm Spitzer}} - \pm u_{0, {\rm Earth}},
\label{eqn:satellite_degeneracy}
\end{equation}

due to the exact circular symmetry of the magnification field about the lens \citep{Refsdal1966}, as illustrated in \citet{Gould1994}. (The sign convention we adopt here is that a positive value of $u_{0}$ indicates that, during its projected trajectory, the source approaches the lens center of mass on its right hand side.) In general, this fourfold degeneracy usually reduces to twofold with the addition of a second lens body, as the resulting caustic features break the symmetry of the magnification field.  However, for binary-lens events in which the trajectory runs approximately parallel to the lens axis (such as the current case), trajectories reflected about the lens axis result in similar magnification curves, in which case the four-fold degeneracy is retained \citep{2015ApJ...805....8Z}.

A grid-search approach was used to determine the most likely parallax-solution regions. With the inclusion of space-based data, the two parallax parameters ($\pi_{{\rm E},N}$ and $\pi_{{\rm E},E}$) were added to the model.

When performing the parallax grid search, the ground-based model parameters (including lens orbital motion) were fixed, and a maximum-likelihood search was performed for the Spitzer light curve over a large range of discrete $\pi_{{\rm E},N}$ and $\pi_{{\rm E},E}$ values. 

\begin{figure*}
\plottwo{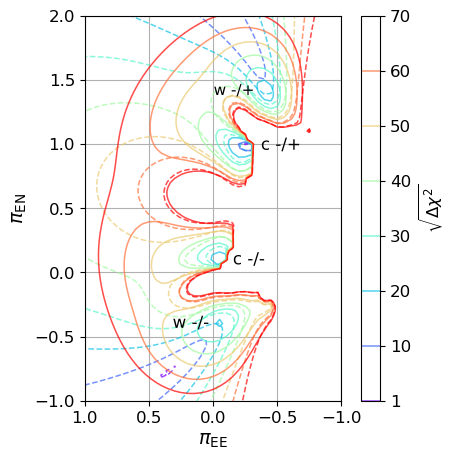}{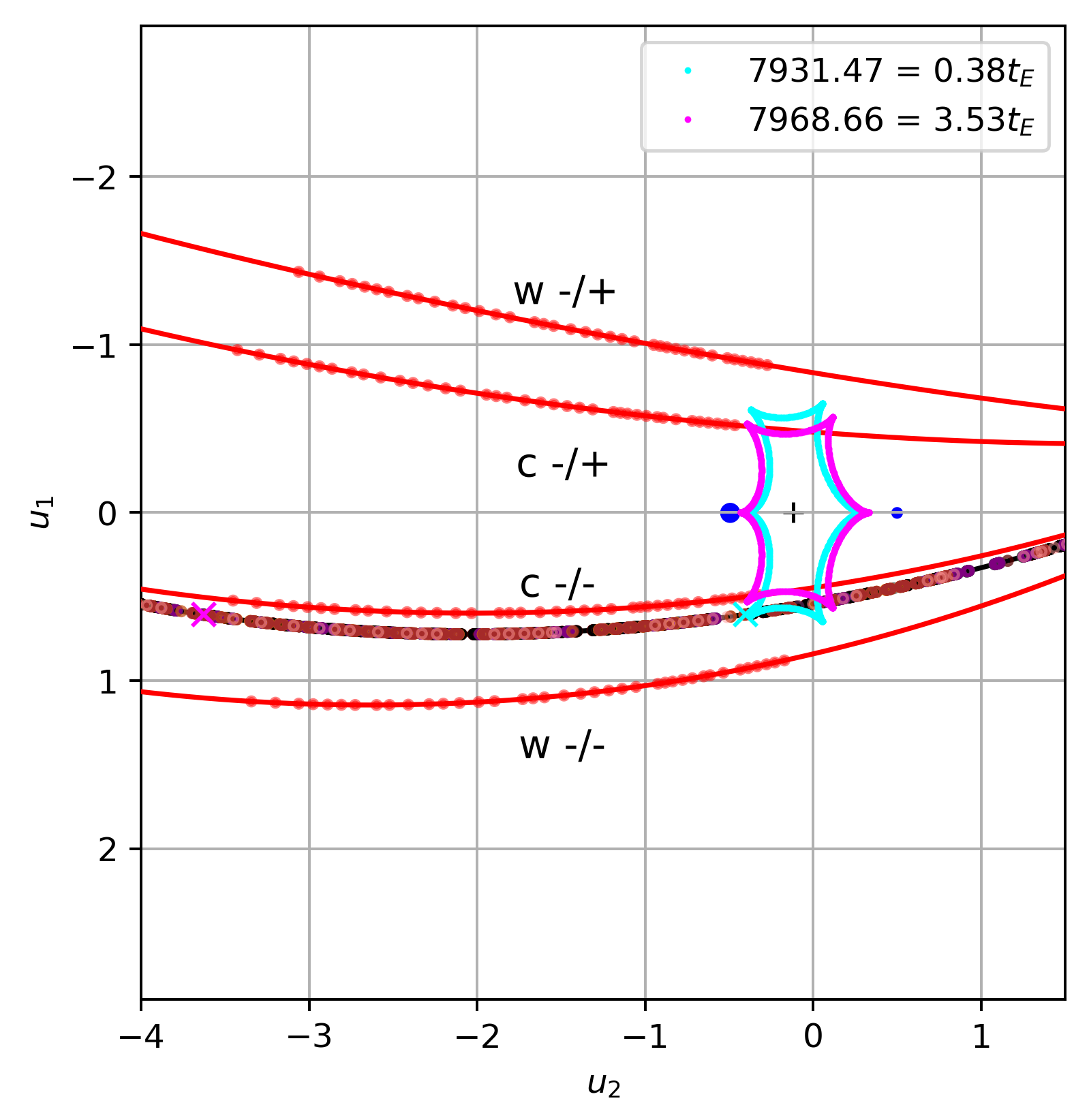}
\caption{\emph{Left}: contour maps demonstrating the results of the parallax grid searches over discretely varied $\pi_{{\rm E},E}$ and $\pi_{{\rm E},N}$ for the $-u_{0,Earth}$ configuration, including only the Spitzer $\chi^2$ components. The dashed contours show the $\chi^2$ landscape without a color constraint and the solid lines with the constraint. Note that the $\pi_{{\rm E},N}$-axis  of this figure is reversed from the usual orientation so that the two figures approximately align. \emph{Right}: caustic diagram with projected ground-based and Spitzer-based trajectories (black and red, respectively).  The four Spitzer trajectories are the result of minimization from the local $\chi^{2}$ minima from the left figure, with all modeling parameters free to evolve.  The data points are represented by colored circles on the trajectories, where the colors correspond to the observation site and field, as specified in Figures~\ref{Fig:StaticLC} and~\ref{Fig:GroundCumulative}. The caustics change with the orbiting of the lens bodies and are depicted here at the instances of the first and last Spitzer data points, specifically for the c -/+ solution (all four $u_{0,Earth}<0$ solutions looks very diagrammatically similar to the one shown). These epochs are represented on grounds-based trajectory (also from the c -/+ solution) with colours matching their corresponding caustics.}  
\label{Fig:RSGrid}
\end{figure*}

This grid search indicated that there were four solution regions for the given ground-based model, with the two outer regions having much higher likelihoods (i.e, lower $\chi^2$) than the two inner regions (\figurename{ \ref{Fig:RSGrid}a}). These four solutions regions represent the $\pm u_{0,{\rm Spitzer}}$ degenerate trajectories relating to two distinct solution families. We refer to these families as close (c) and wide (w). The four solutions regions result from only $-u_{0,Earth}$ and indicate that, including the $+u_{0,Earth}$ trajectory, we have an eightfold degeneracy for this particular geometry.

Because the Spitzer data only cover the falling part of the light curve and cover no caustic feature, the light curve alone does not contribute very strong constraints on the parallax measurement. We have thus implemented in the modeling an additional $\chi^2$ penalty term that weighted the fit toward a source-flux ratio (between KMT-C03 and Spitzer $L$) matching that inferred by the calculated $(I-L)_0$ source color, found in \S\ref{sS:Color}. This color-constraint \citep{shin2017ogle} term was of the form

\begin{equation}
\chi^2_{constraint} = \frac{\left(2.5\log_{10}\left(\frac{\left(\frac{F_{I}}{F_L}\right)_{model}}{\left(\frac{F_{I}}{F_L}\right)_{constraint}}\right)\right)^2} {\sigma^2_{constraint}}.
\end{equation}

The constraint changed the likelihood space of the parallax model. The four solutions-regions from the unconstrained grid, remained as features in the constrained grid.  However, the close set of solutions have more comparable likelihoods to the wide set than in the unconstrained grid. 

\begin{figure}
\plotone{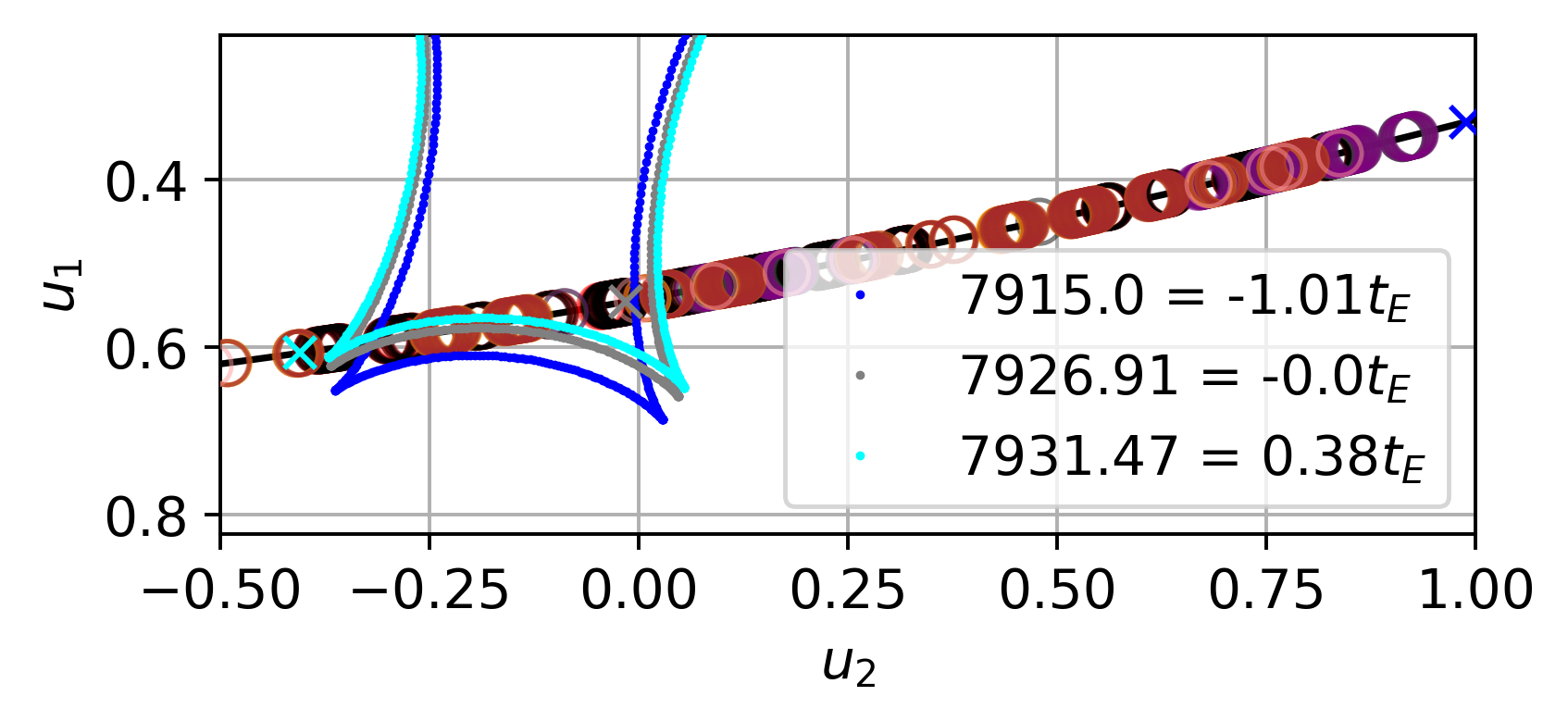}
\caption{Caustic diagram with projected ground-based trajectory for the c -/+ solution. The ground-based data points are represented by colored circles on the trajectories, where the colors correspond to the observation site and field, as specified in Figures~\ref{Fig:StaticLC} and~\ref{Fig:GroundCumulative}. The caustics are depicted here at three instances corresponding to the start of the ``problem region'' (7915 = $t_0-1.01 t_{\rm E}$), $t_0$ (7926.91), and the time of the first Spitzer data point ($7931.47 = t_0+0.38 t_{\rm E}$)). These epochs are represented on the grounds based trajectory with colors matching their corresponding caustics.}
\label{Fig:c-/+Caustic}
\end{figure}

When comparing \figurename{ \ref{Fig:RSGrid}}a and \figurename{ \ref{Fig:RSGrid}}b, the reason for the four lobes in the likelihood space becomes apparent. For this event, $\Delta \beta$ approximately aligns with $\pi_{{\rm E},N}$ and $\Delta \tau$ with $\pi_{{\rm E},E}$. 
Simplistically, changing $\Delta\tau$ moves the Spitzer-data nodes backward or forward in time along the Spitzer trajectory, whereas $\Delta\beta$ shifts the ``parallel'' space-based trajectory closer to or farther away from the ground-based trajectory. 
The lobes and connective contours in \figurename{ \ref{Fig:RSGrid}}a result from solutions for which the Spitzer data hug the leftmost cusps of the caustic of \figurename{ \ref{Fig:RSGrid}}b.

Figure \ref{Fig:c-/+Caustic} shows a more restricted view of the ground-based trajectory for this set of solutions, and the caustics at key epochs in the light curve, which change over the course of the event due to the orbital motion of the lenses. 

Within each wide or close set, the pairs are the previously predicted $\pm u_{0,{\rm Spitzer}}$ degenerate solutions.  
A further four degenerate solutions are obtained by reflecting all trajectories in Figure~\ref{Fig:RSGrid}b (ground and Spitzer) about the lens axis.

The eight degenerate solution regions were further investigated using emcee with both ground-based and Spitzer data, renormalized errors, and both parallax and lens orbital motion included in the model.  
All model parameters were left free to evolve for all instances.  Model parameters for the resulting solutions are given in Table~\ref{T:fullparams}. They are all somewhat similar in likelihood with an overall range in $\Delta\chi^2 \leq 87$. The best solution found was the c -/+ geometry. All close solutions were favored over the wide by a margin of $\chi^2_{w}-\chi^2_{c}\geq 8.96$. The nonfavored close solutions have a range $12.48< \Delta\chi^2 <28.06$. 

\begin{deluxetable*}{lcccccccc}[b] 
\rotate
\tablecaption{Final Model Parameters and Physical Parameters for the Eight Degenerate Solutions Utilizing Fixed Spitzer Error-bar Scaling ($S_{{\rm Spitzer}}=3.93$)}
\fontsize{6}{7.25}
\setlength\tabcolsep{2.25pt}
\tablehead{& c -/- & c -/+ & c +/- & c +/+ & w -/- & w -/+ & w +/- & w +/+}
\startdata
$s$
    & $0.9923^{+0.0007}_{-0.0004}$
        & $0.9934^{+0.0001}_{-0.0011}$ 
            & $0.9920\pm0.0006$
                & $0.9924\pm0.0006$
    &  $0.9929^{+0.0005}_{-0.0007}$
        & $0.9933^{+0.0005}_{-0.0006}$
            & $0.9919^{+0.0007}_{-0.0005}$
                & $0.9930^{+0.0006}_{-0.0005}$ \\
$q$ 
    & $0.609\pm0.003$ 
        & $0.609^{+0.004}_{-0.002}$
            & $0.607\pm0.003$ 
                & $0.608\pm0.003$ 
    & $0.607\pm0.003$                
        & $0.607^{+0.002}_{-0.003}$ 
            & $0.604^{+0.003}_{-0.002}$
                & $0.607^{+0.004}_{-0.002}$ \\
$\log_{10}\rho$ 
    & $-1.5852^{+0.0015}_{-0.0004}$
        & $-1.5835^{+0.0006}_{-0.0013}$ 
            & $-1.5858\pm 0.0010$ 
                & $-1.5851^{+0.0013}_{-0.0007}$
    & $-1.55844^{+0.0006}_{-0.0011}$
        & $-1.5835^{+0.0008}_{-0.0011}$
            & $-1.5860^{+0.0009}_{-0.0007}$
                & $-1.5843^{+0.0012}_{-0.0008}$ \\
$u_0$ 
    & $-0.5552^{+0.0011}_{-0.0006}$
        & $-0.5455^{+0.0001}_{-0.0015}$ 
            & $0.5472\pm0.0008$ 
                & $0.5552\pm0.0008$
    & $-0.5580^{+0.0006}_{-0.0010}$ 
        & $-0.5431^{+0.0008}_{-0.0007}$ 
            & $0.5439^{+0.0009}_{-0.0007}$
                & $0.5577\pm0.0008$ \\
$\alpha$ 
    & $-2.9698^{+0.0006}_{-0.0008}$
        & $-2.9612^{+0.0006}_{-0.0009}$ 
            & $2.9597^{+0.0004}_{-0.0010}$ 
                & $2.9695^{+0.0009}_{-0.0005}$ 
    & $-2.9733^{+0.0006}_{-0.0008}$
        & $-2.9589^{+0.0011}_{-0.0003}$
            & $2.9557\pm0.0007$
                & $2.9738^{+0.0005}_{-0.0008}$ \\
$t_0$ 
    & $7926.991^{+0.010}_{-0.004}$
        & $7926.973^{+0.003}_{-0.012}$ 
            & $7926.930^{+0.006}_{-0.009}$ 
                & $7926.990^{+0.008}_{-0.007}$
    & $7927.034\pm0.007$
        & $7926.982^{+0.004}_{-0.011}$
            & $7926.925^{+0.009}_{-0.006}$
                & $7927.041^{+0.007}_{-0.008}$ \\
$t_{\rm E}$ 
    & $11.883^{+0.015}_{-0.021}$ 
        & $11.825^{+0.029}_{-0.006}$ 
            & $11.868^{+0.021}_{-0.014}$ 
                & $11.881^{+0.018}_{-0.017}$ 
    & $11.845^{+0.024}_{-0.012}$
        & $11.800^{+0.013}_{-0.022}$ 
            & $11.840^{+0.017}_{-0.018}$
                & $11.847\pm0.018$ \\
$\pi_{{\rm E},N}$ 
    & $0.121^{+0.009}_{-0.017}$ 
        & $1.015^{+0.007}_{-0.015}$ 
            & $-1.028\pm0.010$ 
                & $-0.107^{+0.007}_{-0.016}$ 
    & $-0.277^{+0.007}_{-0.010}$
        & $1.295^{+0.013}_{-0.003}$ 
            & $-1.357^{+0.008}_{-0.007}$
                & $0.256^{+0.005}_{-0.011}$ \\
$\pi_{{\rm E},E}$ 
    & $0.020^{+0.013}_{-0.024}$
        & $-0.230^{+0.027}_{-0.009}$ 
            & $-0.073^{+0.017}_{-0.021}$ 
                & $0.020^{+0.026}_{-0.009}$ 
    & $-0.136^{+0.018}_{-0.023}$
        & $-0.457^{+0.019}_{-0.013}$ 
            & $-0.293^{+0.017}_{-0.021}$
                & $-0.160^{+0.022}_{-0.014}$ \\
$\dot{s}$ 
    & $0.31^{+0.05}_{-0.03}$
        & $0.32^{+0.05}_{-0.03}$ 
            & $0.25^{+0.04}_{-0.03}$ 
                & $0.31^{+0.04}_{-0.03}$
    & $0.29^{+0.04}_{-0.03}$
        & $0.33^{+0.01}_{-0.07}$
            & $0.22^{+0.03}_{-0.05}$
                & $0.31\pm0.04$ \\
$\dot{\alpha}$ 
    & $-1.25^{+0.02}_{-0.05}$ 
        & $-1.25^{+0.05}_{-0.02}$ 
            & $1.21^{+0.05}_{-0.02}$ 
                & $1.27^{+0.03}_{-0.04}$ 
    & $-1.25^{+0.02}_{-0.05}$
        & $-1.23^{+0.03}_{-0.04}$
            & $1.22^{+0.04}_{-0.03}$
                & $1.28\pm0.04$ \\
\hline
$\chi^2_{min}$
    & 11555.55
        & 11529.44
            & 11541.92
                & 11537.49
    & 11566.46
        & 11601.31
            & 11615.94
                & 11570.26 \\
$\Delta\chi^2_{min}$
    & 26.12
        & 0
            & 12.48
                & 28.06
    & 37.02
        & 71.87
            & 86.50
                & 40.83 \\
\hline
$\beta$
    & 0.40
        & 0.19
            & 0.17
                & 0.39
    & 0.43
        & 0.13
            & 0.12
                & 0.45\\
$s\,({\rm au})$ 
    & $1.77^{+0.13}_{-0.12}$
        & $0.67 \pm 0.05$ 
            & $0.68 \pm 0.05$ 
                & $1.81^{+0.12}_{-0.14}$
    & $1.33 \pm 0.10$
        & $0.55 \pm 0.05$
            & $0.54 \pm 0.04$
                & $1.34 \pm 0.10$ \\
$M_{tot}\,(M_{\odot})$ 
    & $0.29^{+0.05}_{-0.03}$ 
        & $0.034 \pm 0.002$ 
            & $0.035 \pm 0.002$ 
                & $0.33^{+0.03}_{-0.05}$
    & $0.116^{+0.009}_{-0.010}$
        & $0.0259 \pm 0.0017$ 
            & $0.0258 \pm 0.0017$
                & $0.118^{+0.010}_{-0.008}$ \\
$m_1\,(M_{J})$ 
    & $188^{+29}_{-14}$ 
        & $22.30^{+0.39}_{-0.14}$ 
            & $22.6 \pm 0.2$
                & $213^{+9}_{-32}$
    & $76^{+3}_{-4}$
        & $16.90^{+0.07}_{-0.13}$
            & $16.85^{+0.11}_{-0.10}$
                & $77^{+4}_{-2}$ \\
$m_2\,(M_{J})$ 
    & $115^{+18}_{-9}$
        & $13.53^{+0.25}_{-0.05}$ 
            & $13.73 \pm 0.14$ 
                & $130^{+5}_{-20}$ 
    & $46 \pm 2$
        & $10.26^{+0.03}_{-0.09}$
            & $10.18^{+0.07}_{-0.05}$
                & $47^{+3}_{-1}$ \\
$D_{\rm L}\,(\rm{kpc})$ 
    & $6.12^{+0.21}_{-0.14}$
        & $2.33 \pm 0.11$ 
            & $2.34 \pm 0.11$
                & $6.28^{+0.10}_{-0.23}$
    & $4.61^{+0.14}_{-0.16}$
        & $1.90^{+0.09}_{-0.10}0$
            & $1.88 \pm 0.09$
                & $4.65^{+0.16}_{-0.13}$ \\
$\mu_{rel, hel} (N,E)\,({\rm mas \, yr^{-1}})$
    & $(8.8,1.7)$ 
        & $(8.73,-0.11)$ 
            & $(-8.96,1.22)$
                & $(-8.8,1.9)$
    & $(-8.0,-3.4)$
        & $(8.46,-0.53)$
            & $(-8.81,0.60)$
                & $(7.6,-4.2)$ \\
$\delta\mu_{rel, hel} (N,E)\,({\rm mas \, yr^{-1}})$ 
    & $(^{+0.1}_{-0.2},^{+0.9}_{-1.8})$
        & $(^{+0.05}_{-0.07},^{+0.25}_{-0.15})$ 
            & $(^{+0.03}_{-0.02},^{+0.19}_{-0.22})$
                & $(^{+0.5}_{-0.1},^{+1.8}_{-0.8})$ 
    & $(^{+0.3}_{-0.2},^{+0.4}_{-0.5})$
        & $(^{+0.05}_{-0.02},^{+0.21}_{-0.18})$
            & $(\pm0.03,^{+0.20}_{-0.21})$
                & $(^{+0.3}_{-0.2},^{+0.5}_{-0.3})$ \\
$\mu_{{\rm L},hel} (N,E)\,({\rm mas \, yr^{-1}})$
    & $(3.1,-5.9)$
        & $(3.1,-7.7)$ 
            & $(-14.6,-6.4)$
                & $(-14.5,-5.8)$
    & $(-13.7,-11.0)$
        & $(2.8,-8.2)$ 
            & $(-14.5,-7.0)$
                & $(1.9,-11.8)$ \\
$\delta\mu_{{\rm L},hel} (N,E)\,({\rm mas \, yr^{-1}})$ 
    & $(^{+0.2}_{-0.3},^{+1.0}_{-1.8})$
        & $\pm (0.2,0.5)$ 
            & $\pm (0.2,0.5)$
                & $(^{+0.6}_{-0.2},^{+1.8}_{-0.9})$
    & $(\pm0.3,^{+0.5}_{-0.6})$
        & $\pm(0.2,0.5)$
            & $\pm(0.2,0.5)$
                & $(\pm0.3,^{+0.6}_{-0.5})$ \\
$v_{{\rm L},hel} (l,b)\,({\rm km\,s^{-1}})$ 
    & $(-6,195)$
        & $(-13,91)$ 
            & $(-176,-19)$
                & $(-458,-66)$
    & $(-380,59)$
        & $(-15,76)$
            & $(-143,-10)$
                & $(-93,247)$ \\
$\delta v_{{\rm L},hel} (l,b)\,({\rm km\,s^{-1}})$ 
    & $(^{+13}_{-25},^{+50}_{-28})$
        & $(^{+5}_{-4},\pm 5)$ 
            & $\pm (9,4)$ 
                & $(^{+52}_{-17},^{+20}_{-37})$
    & $(^{+15}_{-14},\pm 11)$
        & $\pm(4,5)$
            & $\pm(8,3)$
                & $(^{+12}_{-11},\pm 9)$ \\
\hline  
$-2\Delta\ln z_{galactic}$ 
    & $ 3.73 $ 
        &  $ 12.42 $ 
            & $ 25.81 $ 
                & $ 0 $ 
    & $ 10.64 $ 
        & $ 14.63 $ 
            & $ 31.45 $ 
                & $ 1.86 $ 
                \\
$-2 \Delta \ln z$ 
    & $ 13.48 $ 
        &  $ 0 $ 
            & $ 25.81 $ 
                & $ 15.53 $
    & $ 35.35 $ 
        & $ 74.13 $ 
            & $ 105.43 $ 
                & $ 30.24 $ 
                \\
\enddata
\begin{minipage}[c]{650pt}
\hspace{-74pt}
\parbox{650pt}{
\tablecomments{These models included both ground-based and space-based data, with parallax and energy-constrained lens orbital motion included in the model. The parameter values quoted in this table are those corresponding to the minimum $\chi^2$ samples from the posterior. The uncertainties recover the 16th and 84th percentiles and are asymmetric because the posteriors are not entirely Gaussian and parameter values corresponding to the minimum $\chi^2$ solution differ from the mode of the same parameter's samples. The physical parameters have additional uncertainty components, added in quadrature to the percentile uncertainties, based on the fixed uncertainties of values used in the calculations of these parameters. We use $-2\Delta\ln z$ as an effective $\chi^2$ value. $-2\Delta\ln z$ incorporates both the fit likelihood and the detection probability, based on our galactic model.}
}
\end{minipage}
\label{T:fullparams}
\end{deluxetable*}

\subsection{Spitzer Systematic Errors}
Before we can have faith in these Spitzer parallax measurements, we must first address concerns of systematics in the Spitzer light curve. 

\cite{ob190960Yee}, \cite{kb180029Gould}, \cite{ob170406Hirao}, and \cite{ob180799Zang} include detailed investigations into Spitzer systematics. These investigations point to poorly determined positions of nearby blend stars in combination with the seasonal rotation of the Spitzer camera. This has resulted in variable blended levels ($F_{\rm B}$) seen over timescales on the order of tens of days. These works conclude that Spitzer systematics are at the level of $\sim1$ Spitzer flux unit where, for a typical event, $F_{\rm B}\approx3$. Concerns have been raised for previous events (\citealt{Zhu2017, Koshimoto2019}) where the flux levels were $F_{\rm S} < 5$ and thus $F_{{\rm S}}\sim F_{\rm B}$, in which case systematics on the order of 1 could be considered fractionally significant. 

We now consider whether systematics in the Spitzer data are significant for this event. The Spitzer magnification curve has a bump between $t=7936$ and $t=7941$ (corresponding to $\Delta F\simeq 5$  Spitzer flux units; see Figure~\ref{Fig:SpitzerLC}) that is not produced by any of our best generative model solutions that incorporate satellite parallax (Section~\ref{satellitedegen}). This implies a systematic error and demonstrates the scale to which we can expect them in this specific Spitzer data set; a few flux units over timescales of around 5 days. This is a higher $\Delta F$ perturbation than is expected for Spitzer systematic on a smooth curve (typically $\Delta F\simeq 1$ Spitzer flux units)

The parallax terms in the model are sensitive to small contiguous perturbations in the data, especially for those data after $t=7955$, where flux changes of a few units change the shape of the slope enough to result in different parallax measurements, which affect the resulting physical solutions.

For this event, we have a Spitzer source flux much larger than the expected blend flux, a light curve with clearly and significantly decreasing flux, and baseline observations. Therefore, we would not ordinarily expect systematics to play a major role in this case. However, this event is somewhat sensitive to systematics in the baseline and shows evidence of similar systematics elsewhere in the light curve. We are therefore cautious of the effects systematic error in the Spitzer data may have on our conclusions. 

\subsection{Modeling Spitzer Errors}\label{SSGPS}
In an attempt to properly consider the apparent systematic errors in the Spitzer data, we have included in our model an error-bar renormalization parameter and two Gaussian process (GP) parameters.

Gaussian processes were first introduced in microlensing event analysis by \cite{Li2019}. In this paper they used a GP to model source variability, rather than systematics, as well as a traditional inflated-error-bar scaling method. The GP method achieved better results in their case, as evidenced by the residuals in their Figure 1. However, they adopt their inflated-error-bar scaling model due to multiple practical and theoretical concerns. The practical issues they raise are how to cope with different blending effects between observation sources and how to perform error re-scaling. The blending issue is not relevant in our case because we only apply a GP model on the Spitzer data set. The theoretical issues they raise are in regards to choice of GP kernel and the possibility of degeneracies between the microlensing and GP parameters, for which they saw no evidence in their posterior distributions. We also saw no evidence of degeneracies between microlensing and GP parameters in our posterior distributions. In regards to the choice of GP kernel we tested both the exponential (described below) and Matern 3/2 kernels and found no significant difference between the results. We did not test the kernel used in \cite{Li2019} as it is meant for modeling quasi-periodic variations.

The degenerate solutions of Section \ref{satellitedegen} have reduced $\chi^2$ values that imply that the Spitzer
flux uncertainties have been underestimated by factors of between 2 and 5 times before renormalization. Because these factors change for each solution we include a multiplicative Spitzer error renormalization as a free parameter and consequently the likelihood
must change to include the penalty
\[ \ln P_{{\rm S}} = -N \ln{S}, \]
where $S$ is the Spitzer error scaling factor and $N$ refers here to the number of Spitzer data points.

Simultaneously we included an exponential GP model to fit the systematic features in the Spitzer light curve using 
the Celerite package \citep{celerite}. This replaces the vector
of data variances with a data covariance matrix,
\[ K_{nm} = \sigma^2_n\delta_{nm} + k(t_n, t_m). \]
We use a GP kernel
\[ k ( \tau_{nm} ) = a \exp ( {-c\tau_{nm}} ), \]
where $\tau_{nm}=|t_n-t_m|$, and $a$ and $c$ are the GP model parameters. 

The GP likelihood is then
\[ \ln{P_{GP}} = -\frac{1}{2}{\mathbf r}^T {\mathbf K}^{-1}{\mathbf r}-\frac{1}{2}\ln{\det{\mathbf K}}-\frac{N}{2}\ln{2\pi},\]
where $\mathbf r$ is the vector of (data - model) residuals.

\begin{figure*}
\plottwo{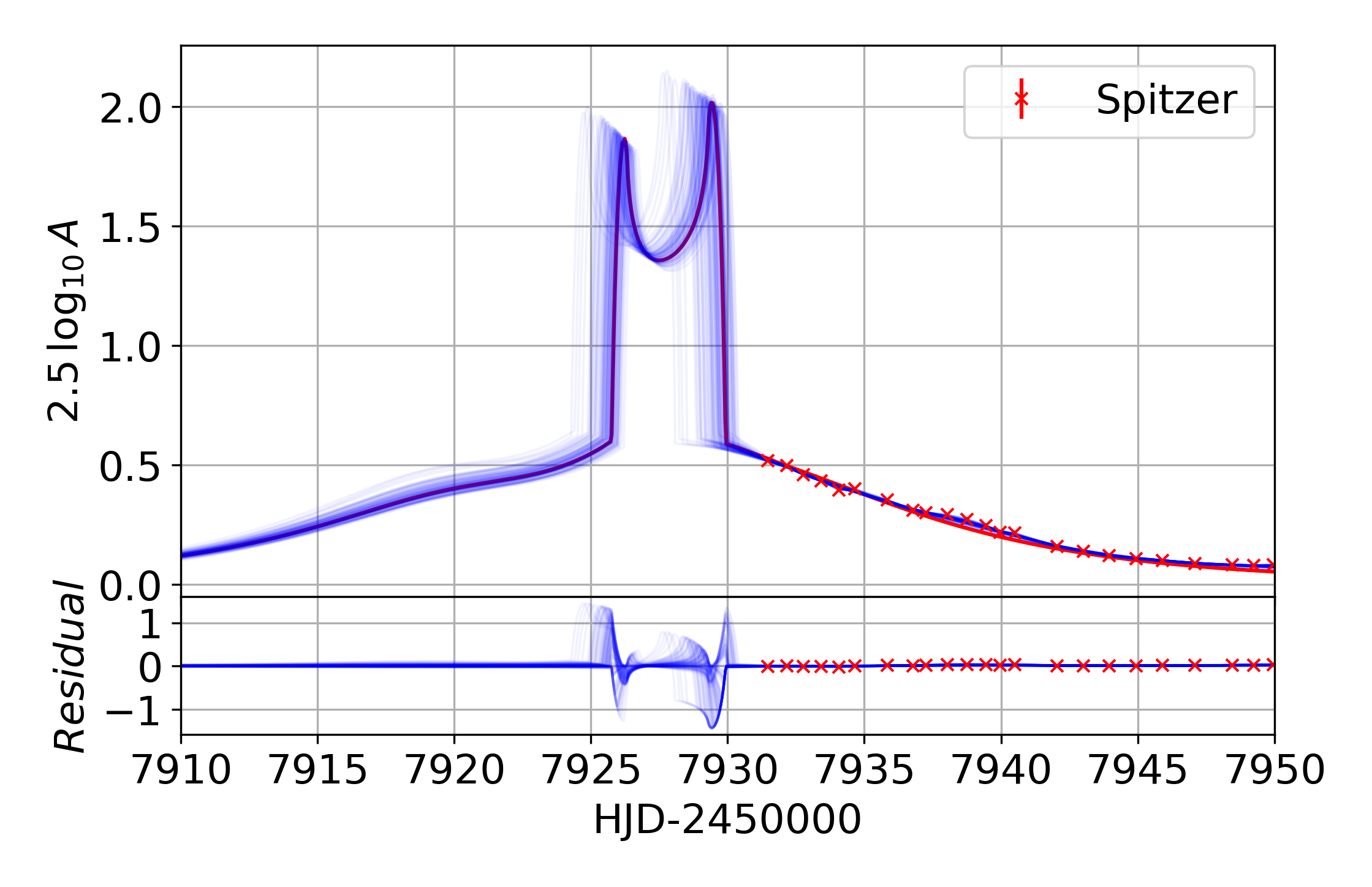}{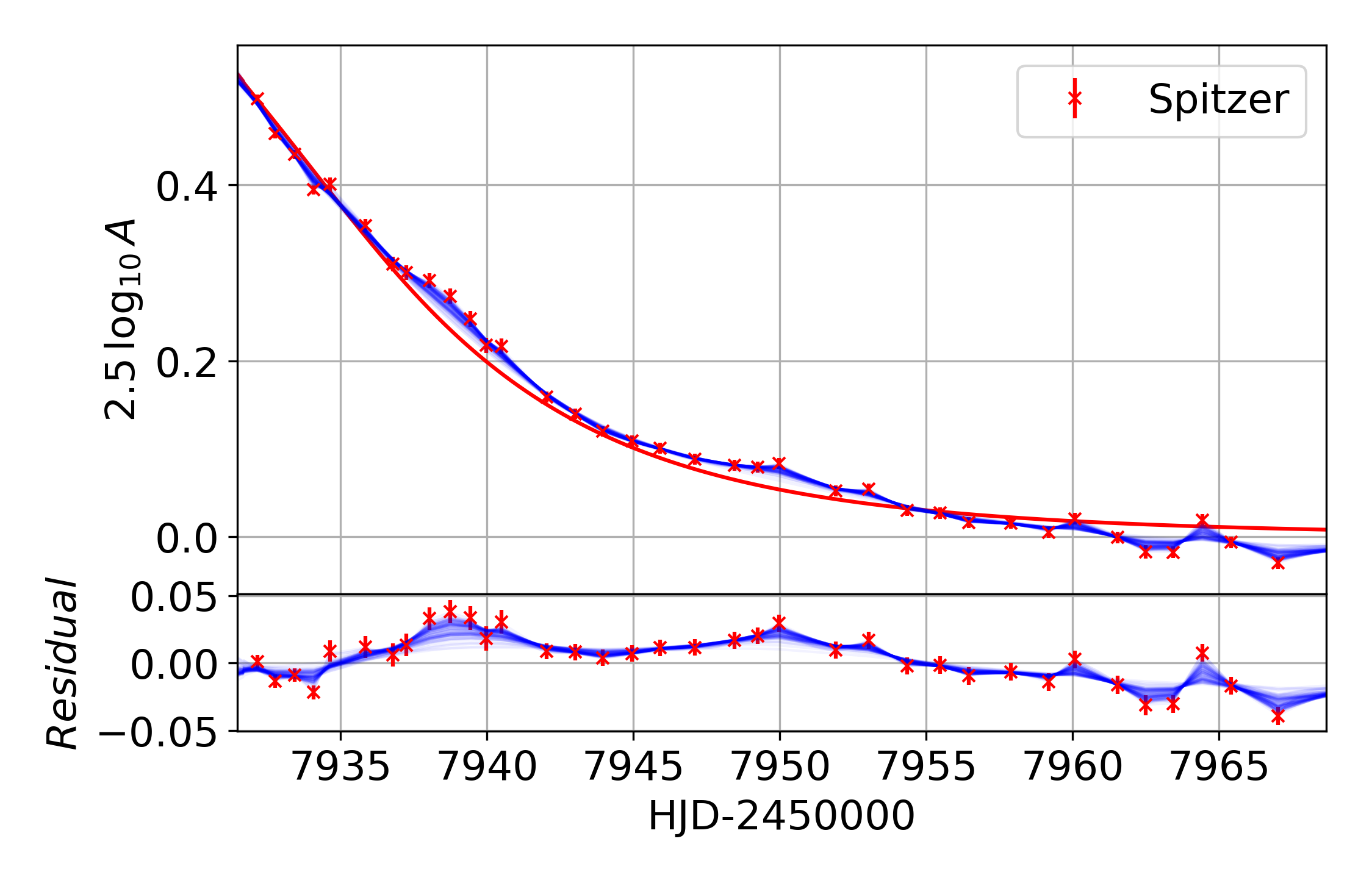}
\caption{Spitzer magnification curve for the c +/- solution. The blue lines show the fitted magnification curve for 100 samples from the posterior with the GP effects shown. The red line is the magnification curve matching the parameters in Table \ref{T:fullparamsGP}. The error-bar scaling in this figure corresponds to the red line ($S_{{\rm Spitzer}} = 2.49$), and the size of the error bars is not necessarily the scaling used for each of the blue samples. Left: shows the magnification curve over the same time period as Figure \ref{Fig:StaticLC}. Right: covers only the Spitzer data set.}
\label{Fig:GPLCs}
\end{figure*}

The results of this modeling are displayed in Table \ref{T:fullparamsGP}. We find that inclusion of these three new model parameters has little effect on the microlensing parameters of all eight solutions, although the spread of likelihood values between solutions does change. With the GP parameters included in the model, our best solution is no longer the c -/+ but the c +/-, although, by a very small margin. The light curve for this model is shown in Figure \ref{Fig:GPLCs}. The full family of close solutions are all similarly likely, $-2\Delta\ln\mathcal{L}<2.3$, where we consider $-2\Delta\ln\mathcal{L}$ as an effective $\Delta\chi^2$.

\begin{deluxetable*}{lcccccccc}[b]
\rotate
\tablecaption{Final Model Parameters, Physical Parameters and Likelihood Components for the Eight Degenerate Solutions with Three Additional Parameters, Compared with the Models of Table \ref{T:fullparams}, for the Spitzer Error-bar Scaling ($S_{{\rm Spitzer}}$) and Gaussian Process Modeling ($a$, $c$). 
\label{T:fullparamsGP}}  
\fontsize{6}{7.25}
\setlength\tabcolsep{2.9pt}
\tablehead{& c -/- & c -/+ & c +/- & c +/+ & w -/- & w -/+ & w +/- & w +/+}
\startdata
$s$ 
    & $0.9927^{+0.0006}_{-0.0005}$ 
        & $0.9933^{+0.0004}_{-0.0008}$
            & $0.9919^{+0.0007}_{-0.0004}$
                & $0.9926 \pm 0.0006$
    & $0.9932^{+0.0002}_{-0.0009}$
        & $0.9934^{+0.0005}_{-0.0007}$
            & $0.9920^{+0.0007}_{-0.0005}$
                & $0.9933^{+0.0004}_{-0.0007}$ \\
$q$ 
    & $0.6061^{+0.0049}_{-0.0010}$
        & $0.6078^{+0.0027}_{-0.0033}$ 
            & $0.6074^{+0.0015}_{-0.0042}$ 
                & $0.6094^{+0.0013}_{-0.0045}$ 
    & $0.6077^{+0.00020}_{-0.0036}$ 
        & $0.6078^{+0.0016}_{-0.0041}$
            & $0.6035^{+0.0045}_{-0.0013}$
                & $0.6065^{+0.0038}_{-0.0022}$ \\
$\log_{10}\rho$ 
    & $-1.5840^{+0.0005}_{-0.0013}$ 
        & $-1.5837^{+0.0009}_{-0.0010}$ 
            & $-1.5858^{+0.0011}_{-0.0009}$
                & $-1.5846^{+0.0001}_{-0.0009}$
    & $-1.5842^{+0.0006}_{-0.0013}$
        & $-1.5831^{+0.0005}_{-0.0014}$
            & $-1.5868^{+0.0019}_{-0.0002}$
                & $-1.5843^{+0.0013}_{-0.0007}$ \\
$u_0$ 
    & $-0.5546^{+0.0005}_{-0.00013}$ 
        & $-0.5457^{+0.0009}_{-0.0010}$ 
            & $0.5471^{+0.0011}_{-0.0009}$
                & $0.5551^{+0.0001}_{-0.0009}$
    & $-0.5582^{+0.0006}_{-0.0013}$ 
        & $-0.5431^{+0.0005}_{-0.0014}$
            & $0.5442^{+0.0019}_{-0.0002}$
                & $0.5575^{+0.0010}_{-0.0007}$ \\
$\alpha$ 
    & $-2.9703^{+0.0013}_{-0.0002}$ 
        & $-2.9612^{+0.0006}_{-0.0008}$ 
            & $2.9591^{+0.0011}_{-0.0004}$ 
                & $2.9695^{+0.0008}_{-0.0007}$ 
    & $-2.9738^{+0.0008}_{-0.0006}$
        & $-2.9590^{+0.0010}_{-0.0004}$
            & $2.9562^{+0.0006}_{-0.0008}$
                & $2.9737^{+0.0006}_{-0.0008}$ \\
$t_0$ 
    & $7926.992^{+0.006}_{-0.010}$ 
        & $7926.969^{+0.006}_{-0.010}$ 
            & $7926.924 \pm 0.008$ 
                & $7926.987^{+0.007}_{-0.010}$
    & $7927.067^{+0.001}_{-0.016}$
        & $7926.973^{+0.005}_{-0.011}$
            & $7926.911^{+0.015}_{-0.004}$
                & $7927.068^{+0.005}_{-0.023}$ \\
$t_{\rm E}$ 
    & $11.881^{+0.015}_{-0.020}$ 
        & $11.834^{+0.020}_{-0.017}$ 
            & $11.876^{+0.014}_{-0.022}$ 
                & $11.884^{+0.016}_{-0.020}$ 
    & $11.811^{+0.029}_{-0.006}$
        & $11.811^{+0.016}_{-0.021}$
            & $11.853^{+0.018}_{-0.022}$
                & $11.811^{+0.037}_{-0.009}$ \\
$\pi_{{\rm E},N}$ 
    & $0.117^{+0.045}_{-0.020}$ 
        & $1.020^{+0.013}_{-0.023}$ 
            & $-1.037^{+0.030}_{-0.013}$
                & $-0.112^{+0.022}_{-0.037}$ 
    & $-0.331^{+0.021}_{-0.006}$ 
        & $1.291^{+0.004}_{-0.010}$
            & $-1.3208^{+0.0005}_{-0.0175}$
                & $0.274^{+0.002}_{-0.031}$ \\
$\pi_{{\rm E},E}$ 
    & $0.061^{+0.027}_{-0.016}$ 
        & $-0.193^{+0.032}_{-0.025}$ 
            & $-0.050^{+0.052}_{-0.007}$ 
                & $0.075^{+0.034}_{-0.015}$ 
    & $-0.373^{+0.066}_{-0.008}$ 
        & $-0.350^{+0.024}_{-0.029}$
            & $-0.135^{+0.005}_{-0.065}$
                & $-0.393^{+0.175}_{-0.021}$ \\
$\dot{s}$ 
    & $0.30^{+0.05}_{-0.03}$ 
        & $0.33^{+0.03}_{-0.05}$ 
            & $0.25^{+0.03}_{-0.05}$
                & $0.32^{+0.03}_{-0.05}$
    & $0.29^{+0.03}_{-0.05}$
        & $0.348^{+0.004}_{-0.074}$
            & $0.21^{+0.05}_{-0.03}$
                & $0.29^{+0.05}_{-0.03}$ \\
$\dot{\alpha}$ 
    & $-1.27^{+0.04}_{-0.03}$
        & $-1.26^{+0.06}_{-0.02}$ 
            & $1.24^{+0.03}_{-0.04}$ 
                & $1.27^{+0.02}_{-0.05}$ 
    & $-1.27^{+0.06}_{-0.02}$
        & $-1.24^{+0.05}_{-0.02}$
            & $1.20^{+0.06}_{-0.02}$
                & $1.28^{+0.04}_{-0.03}$ \\
$S_{{\rm Spitzer}}$ 
    & $2.2^{+0.9}_{-0.6}$
        & $1.9^{+0.8}_{-0.7}$
            & $2.0^{+1.0}_{-0.6}$
                & $2.3^{+0.6}_{-0.8}$
    & $1.9^{+1.0}_{-0.6}$
        & $1.9^{+1.3}_{-0.8}$
            & $1.3^{+1.3}_{-0.9}$
                & $1.4^{+1.3}_{-0.4}$ \\
$a$ 
    & $1.6^{+10.4}_{-0.6}$
        & $0.8^{+2.2}_{-0.4}$ 
            & $1.1^{+4.8}_{-0.4}$ 
                & $2.0^{+10.7}_{+0.9}$
    & $4.0^{+12.9}_{-2.3}$
        & $3.4^{+18.5}_{-1.2}$
            & $4.2^{+23.1}_{-1.5}$
                & $3.0^{+15.4}_{-0.9}$ \\
$c$ 
    & $0.12^{+0.11}_{-0.10}$
        & $0.27^{+0.12}_{-0.24}$
            & $0.13^{+0.14}_{-0.11}$
                & $0.11 \pm 0.10$
    & $0.08^{+0.13}_{-0.07}$
        & $0.14 \pm 0.12$
            & $0.13 \pm 0.11$
                & $0.14^{+0.08}_{-0.12}$ \\
\hline
$-2 \ln {\mathcal L}$
    & 11559.391
        & 11559.801
            & 11557.491
                & 11558.619
    & 11564.373
        & 11588.644
            & 11587.596
                & 11572.086 \\
$-2\Delta \ln {\mathcal L}$
    & 1.90
        & 2.31
            & 0
                & 1.13
    & 6.88
        & 31.15
            & 30.11
                & 14.60 \\
\hline
$\beta$
    & 0.42
        & 0.19
            & 0.18
                & 0.43
    & 0.36
        & 0.14
            & 0.12
                & 0.37 \\
$s\,({\rm au})$ 
    & $1.74^{+0.14}_{-0.18}$ 
        & $0.67^{+0.06}_{-0.05}$ 
            & $0.67^{+0.06}_{-0.05}$
                & $1.73^{+0.14}_{-0.18}$
    & $1.06^{+0.11}_{-0.08}$
        & $0.56 \pm 0.05$
            & $0.56 \pm 0.05$
                & $1.08^{+0.23}_{-0.08}$ \\
$M_{tot}\,(M_{\odot})$ 
    & $0.27^{+0.05}_{-0.08}$ 
        & $0.034 \pm 0.002$ 
            & $0.035 \pm 0.002$ 
                & $0.27^{+0.06}_{-0.07}$
    & $0.072^{+0.011}_{-0.005}$ 
        & $0.0266 \pm 0.0018$
            & $0.0270^{+0.0018}_{-0.0019}$
                & $0.074^{+0.035}_{-0.006}$ \\
$m_1\,(M_{J})$ 
    & $180^{+30}_{-50}$ 
        & $22.4^{+0.6}_{-0.3}$
            & $22.5^{+0.7}_{-0.3}$ 
                & $170^{+40}_{-50}$
    & $46.6^{+6.6}_{-0.7}$
        & $17.38^{+0.19}_{-0.09}$
            & $17.65^{+0.08}_{-0.41}$
                & $48.6^{+22.7}_{-1.7}$ \\
$m_2\,(M_{J})$ 
    & $110^{+20}_{-30}$ 
        & $13.6^{+0.4}_{-0.2}$ 
            & $13.7^{+0.4}_{-0.2}$ 
                & $110^{+20}_{-30}$ 
    & $28.3^{+4.0}_{-0.5}$
        & $10.53^{+0.10}_{-0.08}$
            & $10.65^{+0.02}_{-0.22}$
                & $29.5^{+13.9}_{-1.0}$ \\
$D_{\rm L}\,(\rm{kpc})$ 
    & $6.04^{+0.25}_{-0.50}$ 
        & $2.33^{+0.12}_{-0.11}$
            & $2.33^{+0.12}_{-0.11}$
                & $6.00^{+0.27}_{-0.47}$
    & $3.67^{+0.29}_{-0.13}$
        & $1.94 \pm 0.10$
            & $1.94 \pm 0.10$
                & $3.75^{+0.76}_{-0.15}$ \\
$\mu_{rel, hel} (N,E)\,({\rm mas \, yr^{-1}})$
    & $(7.9,4.4)$ 
        & $(8.79,0.20)$ 
            & $(-8.97,1.44)$
                & $(-7.4,5.2)$
    & $(-5.97,-5.82)$
        & $(8.64,0.05)$
            & $(-8.97,1.48)$
                & $(5.14,-6.51)$ \\
$\delta\mu_{rel, hel} (N,E)\,({\rm mas \, yr^{-1}})$ 
    & $(\pm 0.4,^{+0.8}_{-0.9})$
        & $(^{+0.05}_{-0.04},^{+0.25}_{-0.23})$ 
            & $(\pm 0.02,^{+0.42}_{-0.13})$
                & $(^{+0.7}_{-0.4},^{+1.0}_{-0.6})$ 
    & $(^{+0.04}_{-0.45},^{+0.34}_{-0.07})$
        & $(\pm 0.05,^{+0.21}_{-0.24})$
            & $(^{+0.07}_{-0.01},^{+0.01}_{-0.41})$
                & $(^{+1.53}_{-0.22},^{+1.12}_{-0.13})$ \\
$\mu_{{\rm L},hel} (N,E)\,({\rm mas \, yr^{-1}})$
    & $(2.2,-3.3)$
        & $(3.1,-7.4)$ 
            & $(-14.6,-6.2)$
                & $(-13.1,-2.4)$
    & $(-11.6,-13.5)$
        & $(3.0,-7.6)$ 
            & $(-14.6,-6.2)$
                & $(-0.5,-14.1)$ \\
$\delta\mu_{{\rm L},hel} (N,E)\,({\rm mas \, yr^{-1}})$ 
    & $(\pm 0.5,^{+0.9}_{-1.0})$
        & $\pm (0.2,0.5)$ 
            & $(\pm 0.2,^{+0.6}_{-0.4})$
                & $(^{+0.8}_{-0.4},^{+1.0}_{-0.7})$
    & $(^{+0.2}_{-0.5},^{+0.5}_{-0.4})$
        & $\pm(0.2,0.5)$
            & $(\pm 0.2,^{+0.5}_{-0.6})$
                & $(^{+1.5}_{-0.3},^{+1.2}_{-0.4})$ \\
$v_{{\rm L},hel} (l,b)\,({\rm km\,s^{-1}})$ 
    & $(9,113)$
        & $(-11,88)$ 
            & $(-174,-21)$
                & $(-358,-126)$
    & $(-292,102)$
        & $(-11,74)$
            & $(-145,-18)$
                & $(-133,213)$ \\
$\delta v_{{\rm L},hel} (l,b)\,({\rm km\,s^{-1}})$ 
    & $(\pm 10,^{+29}_{-31})$
        & $\pm (5, 5)$ 
            & $(\pm 9,^{+3}_{-5})$ 
                & $(^{+49}_{-26},\pm 14)$
    & $(^{+12}_{-27},^{+6}_{-7})$
        & $\pm(4,5)$
            & $\pm(8,^{+5}_{-3})$
                & $(^{+16}_{-9},^{+21}_{-10})$ \\
\hline  
$-2\Delta\ln z_{galactic}$ 
    & $ 1.64 $ 
        &  $ 10.41 $ 
            & $ 25.46 $ 
                & $ 2.73 $ 
    & $ 13.69 $
        & $ 12.37 $ 
            & $ 31.65 $ 
                & $ 0 $ 
                \\
$-2\Delta \ln z$ 
    & $ 0.01 $ 
        &  $ 8.92 $
            & $ 21.83 $
                & $ 0 $
    & $ 16.97 $
        & $ 39.77 $
            & $ 57.95 $ 
                & $ 11.25 $ \\
\enddata
\begin{minipage}[c]{650pt}
\hspace{-74pt}
\parbox{650pt}{
\tablecomments{These models included both ground-based and space-based data, with parallax and energy-constrained lens orbital motion included in the model. The parameter values quoted in this table are those corresponding to the minimum $\ln\mathcal{L}$ samples from the posterior. The uncertainties recover the 16th and 84th percentiles and are asymmetric because the posteriors are not entirely Gaussian and parameter values corresponding to the minimum $\chi^2$ solution differ from the mode of the same parameter's samples. We use $-2\Delta\ln\mathcal{L}$ and $-2\Delta\ln z$ as effective $\chi^2$ values. $-2\Delta\ln\mathcal{L}$ quantifies the light-curve-fit likelihood and $-2\Delta\ln z$ incorporates both the fit likelihood and the detection probability, based on our galactic model.}
}
\end{minipage}
\end{deluxetable*}

\subsection{Model Comparison}

When we compare the likelihoods using the standard analysis approach (Table \ref{T:fullparams}) and GP analysis (Table \ref{T:fullparamsGP}), both favour the close solutions. For the close solutions, the range of $\Delta\chi^2<28$ using the standard approach and $\Delta\chi^2_{\rm eff}<3$ using GP.\footnote{Here we refer to effective $\Delta\chi^2$ values, which are calculated via $\Delta \chi^2_{eff} = -2 \Delta \ln {\mathcal L}$. This is in order to provide an equivalent scale to the regular $\Delta\chi^2$ values we use to appraise solutions in the standard approach. For the standard approach $\Delta \chi^2 = -2 \Delta \ln {\mathcal L}$ because the extra likelihood components are the same for all solutions. We use these two parameters to compare spreads between methods, as they are on the same scale, but we do not directly compare solutions between methods, as these values are not equivalent.} The physical properties $M_{tot}$ and $D_{\rm L}$ of all four close solutions are in agreement between the standard and GP approaches to within $1.2 \, \sigma$. Therefore, the physical interpretation is the same in both cases.

This is not true of all the degenerate wide-family solutions. While the large-parallax solutions remain most disfavoured between approaches, with matching $M_{tot}$ and $D_{\rm L}$ values (masses at or below the deuterium fusion limit, 2 kpc away), the small-parallax solutions tell a different story. Using the standard approach, all wide solutions are disfavoured by a $\Delta\chi^2>37$. However, using the GP approach the small-parallax solutions have $\Delta\chi^2_{\rm eff}$ values of 7 and 15, within the $\Delta\chi^2$ range of close solutions using the standard approach. The physical interpretation is also different for these two solutions. The physical properties $M_{tot}$ and $D_{\rm L}$ differ between approaches by $< 4 \sigma$. 

Our interpretation is that the physical solutions are not equally sensitive to systematic errors in the Spitzer data. The posteriors of $\pi_{{\rm E},E}$ are wider using the GP approach than the standard approach, especially the wide solutions for which the extrapolated trajectories do not cross caustics. It appears that the parallax measurement (particularly $\pi_{{\rm E},E}$) is proportionally more affected for smaller parallax solutions, making them more sensitive to systematic errors, but that the affect this has on the close solutions is limited by the nearness to a caustic crossing, which has a dominating effect on the likelihood space. Whether these conclusion are true in general is an interesting thought for future work. 

While inflating error bars may be the correct approach for accommodating noise in data that is approximately Gaussian, it is appropriate to use a correlated noise approach where there are obvious systematic trends. The apparent perturbations in our Spitzer data are not represented by any of our best model solutions and therefore show that the errors in this data set are clearly correlated on time scales of a few days. However, the importance of using a correlated noise approach varies for our different solutions families and we believe that the importance of such modelling in other Spitzer events would also be dependent on many event-specific-properties.  

Whether or not we consider the expense of a GP approach necessary, in our case, depends on the $\Delta\chi^2_{\rm eff}$ ranges we are prepared to accept. If we accept solutions at the $\Delta\chi^2_{\rm eff}\lesssim90$ level, all eight degenerate solutions are valid, whether or not a GP is included. However, at the $\Delta\chi^2_{\rm eff} \lesssim50$ level, we would reject the w -/+ and w +/- solutions using the standard approach. Using the GP approach, we would accept all of these solutions, with w -/- and w +/+ converging into significantly different physical lens compositions.


\section{Physical Parameters} \label{S:Physical}

\subsection{Angular Einstein Radius}

There exist empirical relations for determining the angular size of a star from its intrinsic color and magnitude. According to \citet{Kervella2008}, the most appropriate of these relations for non-M-type giants are those found in \citet{Nordgren2002} and \citet{vanBelle1999}.\footnote{Depending on the selection of surface brightness relation, the implied $\theta_{\rm E}$ differs by around $8\%$, in this case.} We use the \citet{Nordgren2002} surface brightness relation, specifically for non-variable giant stars (their Equation 12),
\begin{equation}
    \log_{10}(2 \theta_*) = 0.5522 + 0.246\left(V-K\right)_0 - V_0/5.
\end{equation}

Using the empirical color-color relations of \citet{Bessel1988} for giant stars we find the $(V-K)_0$ equivalent of the intrinsic source color, $(V-I)_0$, that was calculated from the CMDs, $(V-K)_{{\rm S},O} = 2.57 \pm 0.09$.
The solutions for the models including higher-order effects have effectively identical $\theta_* = 7.6 \pm 0.5 \, \mu {\rm as}$. 

$\theta_{\rm E}$ was calculated using the fitted $\rho$ value for each solution, where $\theta_{\rm E} = \theta_*/\rho$. The light-curve data provided good coverage of the caustic crossing and therefore $\rho$ was well constrained, and almost identical, in our models. The calculated value of $\theta_{\rm E}$ for all solutions is
\[ \theta_{\rm E}=0.29\pm0.02\,{\rm mas}. \] 
Knowing $\theta_{\rm E}$ gives an angular scale to the geometric models.


\subsection{Mass, Distance and Separation}

The intrinsic I-band magnitude of the source star was previously calculated by comparing its fitted $I$-band magnitude to the mean red-clump magnitude on a CMD. By assuming the intrinsic red-clump magnitude and that the source star is at the distance of the average red-clump star in the CMD field, we find $D_{\rm S}=7.85\pm0.06\,\rm{kpc}$.

With values for $D_{\rm S}$, $\theta_{\rm E}$ and $\pi_{\rm E}$, the degeneracy in Equation \ref{E:thetaE} is broken, and the mass and distances can be calculated for each solution. Given the fitted parameters $\pi_{{\rm E},E}$ and $\pi_{{\rm E},N}$, $\theta_{E}$, and $D_{\rm S}$, the distance to the lens was found, using
\[ \frac{1}{D_{\rm L}[\rm{kpc}]}=\pi_{\rm rel}[{\rm mas}]+\frac{1}{D_{{\rm S}}[{\rm kpc}]},\]
where $\pi_{\rm rel}=\theta_{\rm E} \pi_{\rm E}$.
Knowing the distance to the lens system and $\theta_{\rm E}$ in angular units, the lens geometry can be calculated in absolute terms. The masses for each of the lens components, their projected separations, and the distances to the lens system are given in Tables \ref{T:fullparams} and \ref{T:fullparamsGP}. All large-parallax solutions, and both small-parallax wide solutions, are consistent with BD binary lenses of varying masses. However, the small-parallax close solutions are consistent with M-dwarf binaries, where the mass of the smaller of the binary objects ($m_2 = 110^{+20}_{-30} \, M_J$) is very near the BD upper cut-off ($\sim70 - 95 \, M_J$) and therefore may or may not be large enough for hydrogen fusion, depending mostly on its chemical composition.  


\subsection{Proper Motion and Velocity}

The relative lens-source heliocentric proper motion was determined via
\begin{equation}
    {\bm \mu}_{rel,hel} = \frac{\theta_{\rm E}}{t_{\rm E}} \hat{\bm \pi}_{\rm E} + \frac{\pi_{\rm rel}}{{\rm au}}{\bm v}_{\oplus,\perp},
\end{equation}
for each solution, where ${\bm v}_{\oplus,\perp}$ is the projected velocity of Earth at $t_0$, parallel to the lens plane, ${\bm v}_{\oplus,\perp}(N,E)=(-0.104,29.296)\,{\rm km\,s}^{-1}$, and 
$\hat{\bm \pi}_{\rm E} = {\bm \pi_{\rm E}} / \left| \pi_{\rm E} \right| $ is a unit vector in the microlensing parallax direction.

The $\mu_{rel,hel}$ values for each solution are shown in Tables \ref{T:fullparams} and \ref{T:fullparamsGP}.
All of the degenerate solutions have high relative proper motions, $\mu_{rel} > 8\,{\rm mas}\,{\rm yr}^{-1}$. A proper motion of $\mu_{rel}\lesssim10\,{\rm mas}\,{\rm yr}^{-1}$ does not innately give any information on the location of the lens i.e. disk vs. bulge. 
However, if one adds knowledge of the source proper motion, $\mu_{\rm S}$, then $\mu_{rel} = 8\,{\rm mas}\,{\rm yr}^{-1}$ may give such information. For example, if $\mu_{\rm S}$ were at the center of the bulge distribution then a bulge lens would be very unlikely because a proper motion of $8\,{\rm mas}\,{\rm yr}^{-1}$ from the centroid is extreme compared with the bulge dispersion of $\sigma(l,b)=(3.0,2.5)\,{\rm mas}\,{\rm yr}^{-1}$. Therefore this hypothetical case would favor a disk lens.

\begin{figure}[ht!]
\centering
\includegraphics[width=0.5\textwidth]{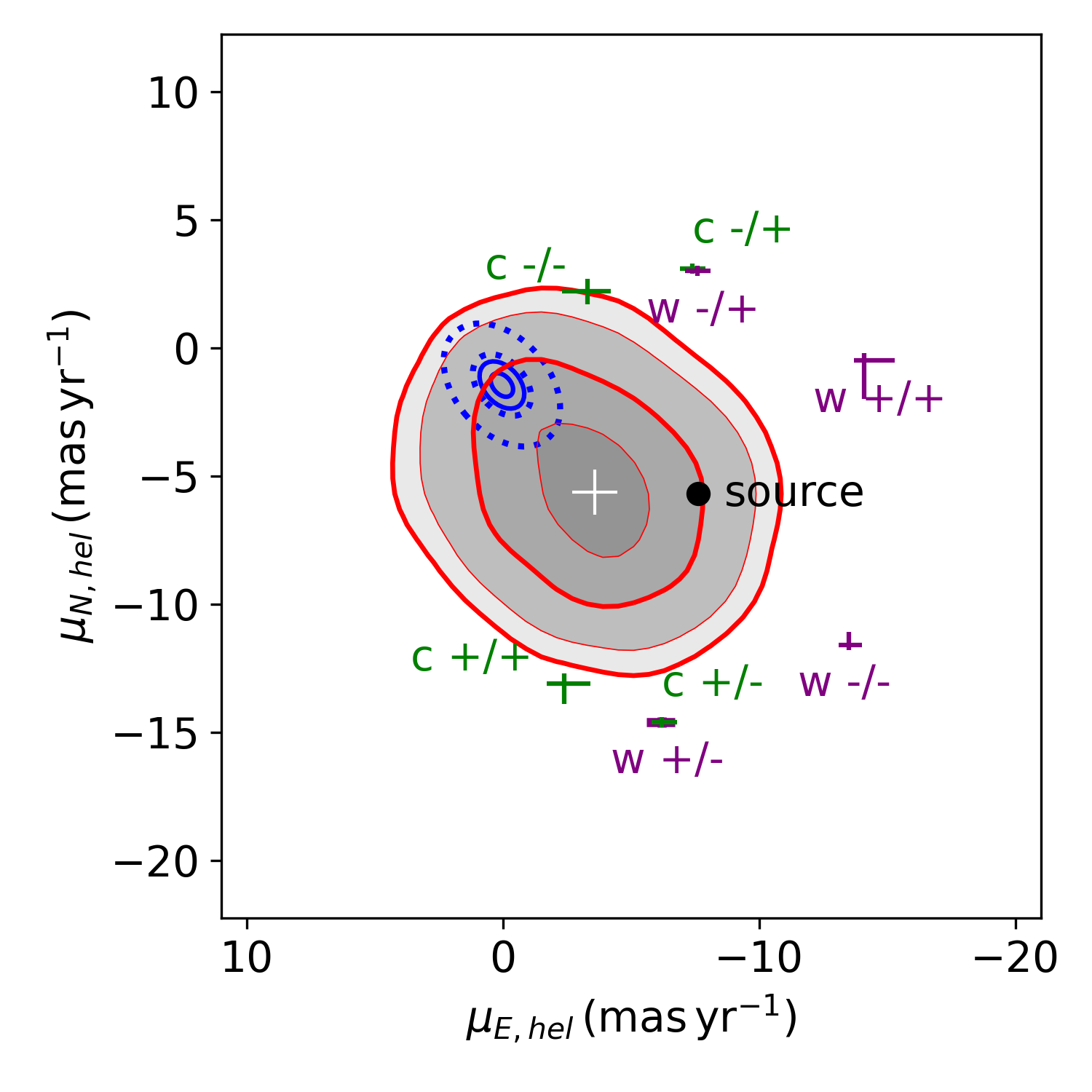}
\caption{Proper motions of the lens solutions and source star. For context, we also include contour representations of the disk (blue) and bulge (red) distributions. The bulge contours are from histograms of the red-clump stars from Gaia EDR3, selecting stars within a $0.2^o$ radius cone centred on the lens. The distribution of red-clump stars is the results of a Gaussian fit to the red clump on the field's CMD. The innermost thicker line of the red-clump distribution contains approximately 68\% of the population samples. The outermost thicker line contains approximately  95\% of the population samples. The blue contours depict the theoretical distribution of the disk stars used in our galactic model. The solid ellipses correspond to the 1 and 2 $\sigma$ proper-motion dispersions of disk stars at $D=6$ kpc. The dotted ellipses show the same for disk stars at $D=2.3$ kpc. 
}
\label{F:pm}
\end{figure}

The source star for this event was observed by Gaia (EDR3 4063557344313009920), and hence its heliocentric proper motion is precisely measured as $\mu_{{\rm S},hel}(N,E)=(-5.7,-7.7)\pm(0.2,0.3)\,{\rm mas\,yr^{-1,\,}}$\footnote{Here we have doubled the published errors, as recommended by \cite{Ryb2021}.} relative to quasars in the distant universe. The source is $\sim1\sigma$ due west of the centroid (see Figure \ref{F:pm}). This means that a bulge lens is more easily accommodated, provided that direction of $\mu_{rel}$ is roughly east. Similarly, the $\mu_{rel}$ direction most consistent with a disk lens is northeast, although this direction is also very plausible for a bulge lens.

The heliocentric lens proper motion is calculated via
\begin{equation}
    \mu_{{\rm L},hel} = \mu_{{\rm S},hel} + \mu_{rel,hel}.
\end{equation}
The unexpected outcome of our $\mu_{\rm L}$ calculations is that none of the eight degenerate solutions align well with the disk or bulge dispersions, as shown in Figure \ref{F:pm}. However, this demonstrates a misleading aspect of proper motion comparisons in that closer objects have higher proper- motions given the same tangential velocity.

The lens proper motion relates to the heliocentric lens velocity via
\begin{equation}
    v_{{\rm L},hel} = 4.74\times D_{\rm L}\mu_{{\rm L},hel},
\end{equation}
where distance is expressed in kiloparsecs, $\mu_{{\rm L},hel}$ is in miliarcseconds per year, and 4.74 is a conversion factor so that $v_{{\rm L},hel}$ is in kilometers per second. These physical parameters for each solution can also be found in Tables~\ref{T:fullparams} and \ref{T:fullparamsGP}.

From Figure \ref{F:pm} we can see that the source is a fairly kinematically typical bulge star, lying on the $1\sigma$ contour of the Gaia field bulge dispersion.

\begin{figure*}[ht!]
\plottwo{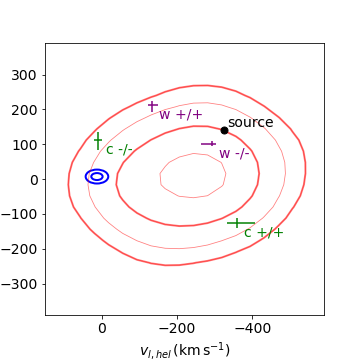}{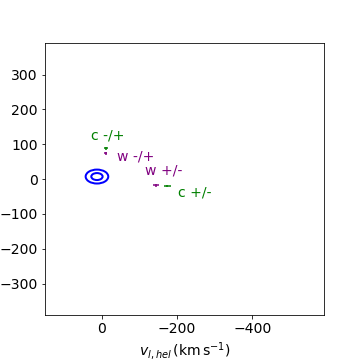}
\caption{Heliocentric velocity of the most likely lens star solutions. As previously, we also include contour representations of the bulge (red) and disk (blue) distributions. The bulge contours are from histograms of the red-clump stars from Gaia EDR3. The red-clump distribution was selected as in Figure \ref{F:pm}. Due to the unreliable nature of Gaia distances obtained from parallax, especially at large distances, $D_{\rm RC}=D_{\rm S}=7.85\,\rm{kpc}$ was used to estimate the red-clump velocities from the Gaia proper motion measurements. The outermost thicker line contains approximately  95\% of the population samples. The blue contours depict the theoretical distribution of disk stars, used in our galactic model. The solid ellipses correspond to the 1 and 2 $\sigma$ velocity dispersions. Left: The small-parallax solutions and both distributions. Right: Only the disk distribution, with the large-parallax solutions, as these solutions are at distances not compatible with a lens belonging to the bulge.}
\label{F:V} 
\end{figure*}

Comparisons of the lens velocities, from each of the eight degenerate solutions, with disk and bulge dispersions from Gaia EDR3 are shown in \figurename{ \ref{F:V}}. These empirical dispersions are used for demonstrative purposes only. All eight lens solutions have unusual velocities when compared to typical disk stars, with the w +/+ and both +/- lens solutions rotating about the galactic center more slowly than typical disk stars, the w -/- and c +/+ counterrotating, and the -/+ and c -/- solutions seemingly moving through the disk, with large $b$ velocities. The solutions are all less exceptional when compared with bulge kinematics, although only the small-parallax solutions have distances that allow for the lens to be a bulge member according to current galactic density models (e.g., \citealt{Han2003}). The velocities of the w -/-, c +/+, w +/-, and c +/- solutions also appear consistent with the retrograde microlensing group.

\section{Solution Probabilities} \label{S:P}

The somewhat uncommon physical parameters compel us to look at our solution probabilities more cautiously and holistically than a purely likelihood-based comparison. One problem with the likelihood calculation is that, formally, it relies on the assumption that our data are Gaussian distributed, with accurate uncertainties. Practically, this is never true for microlensing photometry. However, for this analysis, we apply Bayes theorem as though they were Gaussian. 

The probability of a system having the solution-specific proper motion or velocity, mass, and distance is also an important factor. We therefore calculate the probability factor $\ln{z}$ that determines the relative detection probability of each solution given a galactic model, with a bias to incorporate their relative light-curve-fit likelihoods.

We compute the galactic probability (Equation 15 of \cite{AndyRich2020}) using a modified version of the Galactic Bayesian code described in \citet{Herrera2020}. This model is based on the stellar Milky Way density model of \citet{Han2003} and mass functions of \citet{Chabrier2003} using the prescription of \citet{Dominik2006}. 

There is a common wisdom in microlensing analysis that small-parallax events are more probable than their large-parallax degenerate counterparts. This is known as the Rich argument, as detailed in \citet{Novati2015-survey}. For single-lens events and binary-lens events for which the lens axis and source trajectory are approximately parallel (as in this case), if the true parallax solution is the smaller parallax solution it will always generate a large-parallax degenerate counterpart. The reverse, however, is not always true. The ratio of these probabilities (Rich factor) is implicitly accounted for in our galactic models \citep{AndyRich2020}.

Our calculated $-2\Delta\ln{z}$ values for each degenerate solution, and both error approaches, are displayed in Tables \ref{T:fullparams} and \ref{T:fullparamsGP}. \cite{BayesFactors} interpret $-2\Delta\ln{z} <  (2.3, 4.6, 9.2)$ as (``substantial",``strong", ``decisive") evidence favoring one solution over another. By their metric, we would decisively consider the c -/+ the best solution given a standard Spitzer error approach. Using this approach, the probability of the c -/+ solution, compared with the next most probable solution (c -/-), is $-2\Delta\ln{z}$=13.48. However, using the GP error approach with the \cite{BayesFactors} interpretation, the evidence supporting the equally favoured, small-parallax, close solutions (c -/-, c +/+; $-2\Delta\ln{z}$=0.01) over the c -/+ is ``strong" ($-2\Delta\ln{z}$=8.9).

At the low galactic latitude of our event, and especially given the calculated distances to the lens of the large-parallax solutions, one would expect lens bodies to be members of the galactic disk. However, at a distance of $\sim6\rm{kpc}$ (as in the c -/- and c +/+ cases), it is possible that the lens is a member of the bulge population. Our galactic modeling of c +/+ showed that it is on the order of 100 times more likely to be a member of the bulge than the disk, whereas for c -/+ this was more like 1400 times more likely to be a member of the disk than the bulge. Currently, our galactic model most highly disfavours the counter-rotating BD solutions, with disk-like distances (c +/- and w +/- with $-2\Delta\ln z$, without a light-cure likelihood bias, of 24.45 and 34.12, respectively) .

It is worth noting that $\ln z$ is based on a galactic model and therefore implicitly favors solutions matching our expectation of kinematic, mass, and density dispersions. Even the kinematic dispersions displayed in Figures \ref{F:pm} and \ref{F:V} are informed by mostly bright stars and may not be truly representative of the dispersions of much dimmer objects, of which we know very little. Some healthy skepticism needs to exist around the model's completeness, especially considering the high proportion of microlensing BDs with unusual proper motions.

To determine how representative these retrograde detections are of the BD population as a whole, we must must first have a good understanding of the innate selection biases in microlensing events, for or against these extreme proper motions. However, if we were to downweight the light-curve likelihood based on the knowledge that our errors are not Gaussian, we will generally favour the low parallax solutions.


\section{Discussion} \label{S:Discussion}
In our analysis of event OGLE-2017-BLG-1038, we fit a binary lens model including higher order effects: lens orbital motion and parallax. We include space-based data from Spitzer and  model systematic errors in these data. We have a resulting eightfold solution degeneracy in this event. These solutions have total lens masses ranging from $0.027-0.27M_\odot$. We also included in our probability comparison a galactic probability for each lens configuration. After these processes we find that our most probable solutions are the c +/+ and c -/-, both with masses of $m_1\simeq170\,M_J$ and $m_2=110^{+20}_{-30}M_J$ ($0.16$ and $0.11 \, M_{\odot}$), separated by $1.7\,{\rm au}$, at a distance of $6.0\,{\rm kpc}$. The companion masses for these solutions are near the upper limit for BDs (the hydrogen burning limit). The lens systems for the c +/+ and c -/- solutions have tangential velocities of $v_{{\rm L},hel}(l,b)=(-358,-126)\,{\rm km}\,{\rm s}^{-1}$ and $v_{{\rm L},hel}(l,b)=(9,113)\,{\rm km}\,{\rm s}^{-1}$, respectively.

The c -/- solution has a minutely higher galactic probability than c +/+ with $-2\Delta\ln{\mathcal L}=1.09$. They are equally likely when considered in the context of both the light-curve fit and the galactic model.

Favouring these solutions over the large-parallax, close-family solutions ($m_1\simeq22.5$ and $m_2\simeq13.7$; $D_{\rm L}=2.33$; $v_{{\rm L},hel}(l,b)=(-11,88)\,{\rm km}\,{\rm s}^{-1}$ and $v_{{\rm L},hel}(l,b)=(-174,-21)\,{\rm km}\,{\rm s}^{-1}$ for c -/+ and c +/-, respectively) relies on our being confident in the galactic model for very-low-mass objects. Evidence from other microlensing events suggest that we do not understand the kinematic structure of BDs at distances of $D<4\,{\rm kpc}$.  To date, three BD systems have been discovered using microlensing that appear to be counterrotating with respect to the disk \citep{Chung2019,Shvartzvald2019,Shvartzvald2017}. These microlensing members lie very much in the plane of the disk and explanations for their characteristics, which we consider here, are that they are members of the disk with extreme motions; they are halo members with a coincidental disk alignment; they are members of a counterrotating population of very-low-mass objects (as suggested by \citealt{Shvartzvald2019}); or, they are evidence of an oversimplified galactic model. The physical parameters of the lens of this event raise the question as to whether or not  OGLE-2017-BLG-1038 is another member of this group. 

One explanation for extreme kinematics for a low-mass disk lens is that the disk may have a larger velocity dispersion for lower mass objects. If we assume that the lens was born in a cluster, it may have received a kick from an interaction with a star, and a binary will have a higher scattering cross section for such an interaction. %
Cluster dissolution has been extensively modeled (e.g., \citealt{Hurley2005, Wang2016}). Most stars are believed to come from open clusters, however expulsion velocities from open clusters are small compared to Galactic rotation. For example, the open cluster simulations of \cite{Jorgensen2020} show most stars escaping with velocities $<10\,{\rm km\,s^{-1}}$, relative to their parent cluster. It is therefore very unlikely that such an escapee would be travelling $\sim100\,{\rm km\,s^{-1}}$, or more, opposed to the disk. For globular clusters, higher mass objects preferentially wind up in tight binaries, whose members can be expelled at very high velocities \citep{Hut1992-2, Hut1992}, but such expulsions are likely to account for a tiny fraction of all stars. This appears an unlikely origin for these counterrotating low-mass objects.

Another aspect of the galactic model that may be misunderstood is the bulge density model. We propose that a mass dependent
spatial cutoff could explain the observed abundance of counterrotating BDs. If we consider that the bulge extends further for lower mass objects, then at $D<4\,{\rm kpc}$ the mass independent model would significantly underrepresent lower mass objects belonging to the bulge population and therefore having extreme (when compared to neighbouring disk stars) kinematics. Density models are fit to observational data and therefore are specifically fit to objects much larger than our inferred lens and those of the aforementioned retrograde BDs.

Another explanation may be that the lens is a halo star. Halo stars are known to have a much larger velocity dispersion, and their mean galactic rotation is much smaller than the disk \citep{Halo2018Du, Halo2018Posti}.  While this large velocity dispersion could explain the kinematics of the other retrograde BD stars, it is a leap to make that assumption here, when it is not unlikely that the lens belongs to the bulge.

Are these retrograde BD detections the first members of a new class of object? At this stage, the characterization of these events as an independent population is speculative. Their existence as a discrete population affects the way we view the galactic probability of this solution, because such a population is not represented in the galactic model. Even if a misunderstood selection effect or aspect of the galactic model is responsible for their overabundance in detection, such an effect is not included in our current probability calculations. More needs to be known about this retrograde group before the significance of this solution can be truly understood. 

The analysis of more low-mass lens events will provide new insights into the very-low-mass end of the mass function and its density and kinematics.  There is little observational evidence to constrain any of these distributions at present. It is always possible that low-mass BDs are far more numerous than currently known and are currently represented by our galactic model. 

Whatever the case, for low-mass lenses, we believe that selection of a solution based on typical disk kinematic arguments is unlikely to be valid. The same reasoning leads us to believe that we cannot categorically claim this lens as a member of either a bulge, halo, or retrograde BD population. A more complex consideration of selection biases and possible population dynamics (beyond the scope of this paper) would be required.

A more empirical means of confirming the small parallax configuration would be to observe the lens photometrically. The hydrogen burning host and likely hydrogen-burning companion, corresponding to the small-parallax, close-family solutions, are bright enough to be visible at their implied lens distances ($D_{\rm L}=6$ kpc). Given the relative proper motions of these solutions ($\mu_{rel,hel} = 9.0 \,{\rm mas \, yr^{-1}}$), we could expect the separation of source and lens to be sufficient for them to be resolved with the advent of infrared adaptive optics imaging from the coming generation of 40 m class telescopes. This is not true of the solutions near the planet-BD boundary, which are too dim to be resolved, no matter the angular separation between source and lens. 

We expect first light for Multi-AO Imaging Camera for Deep Observations (MICADO) on the $39\,{\rm m}$ European Extremely Large Telescope (EELT) to be 2030. \cite{Kim2021} have argued, by scaling the work of \cite{Bowler2015} with the Keck coronograph, that an EELT coronograph could achieve $\Delta K = 11$ contrast at 77 mas. By 2030 the angular separation of the lens and source will be $\sim115$ mas. Using the mass-luminosity function of \cite{Just2015} and the previously calculated source-star K magnitude, we estimate $\Delta K=9.2$ between the source star and the primary lens body for the M-dwarf solutions (c -/- and c +/+). Therefore the composition of this lens, be it BD or M-dwarf, can be verified with astrometric follow-up at the expected first light of MICADO on EELT.



\section{Summary}

In this paper we report our analysis of microlensing event OGLE-2017-BLG-1038, with data from KMTNet, OGLE, and Spitzer. Ground-based data show the event is due to a giant source passing across a fold and cusp of a resonant caustic, due to a rotating binary lens. The analysis of the combined Spitzer, KMT, and OGLE light-curve data resulted in eight degenerate satellite-parallax solutions. With a GP model fit to the Spitzer data to account for systematic effects, the best solutions are the four belonging to the close family. Of these solutions the small- parallax solutions both have masses of $M_1 \simeq 170^{+40}_{-50} M_J$ (an M-dwarf) and $m_2=110^{+20}_{-30}\,M_J$ (at the BD/M-dwarf cutoff). The large-parallax solutions are both comprised of a BD binary with $m_1=22\pm2\,M_J$ and $m_2=14\pm1\,M_J$. Inclusion of a detection probability based on a galactic model favors the small-parallax solutions. However, this approach to appraising solutions may be biased by an incomplete description of the distribution of very-low-mass objects in the galaxy and should not rule out solutions with similar light-curve-fit likelihoods. Late-time imaging could be used to reject these low-mass BD solutions, since an M dwarf should be visible given sufficient lens-source separation, but a low-mass BD binary will not.

\acknowledgments{
This research has made use of the KMTNet system operated by the Korea Astronomy and Space Science Institute (KASI) and the data were obtained at three host sites of CTIO in Chile, SAAO in South Africa, and SSO in Australia.

Work by Cheongho Han was supported by the grants of National Research
Foundation of Korea (2020R1A4A2002885 and 2019R1A2C2085965).

This work has made use of data from the European Space Agency (ESA) mission
{\it Gaia} (\url{https://www.cosmos.esa.int/gaia}), processed by the {\it Gaia}
Data Processing and Analysis Consortium (DPAC,
\url{https://www.cosmos.esa.int/web/gaia/dpac/consortium}). Funding for the DPAC
has been provided by national institutions, in particular the institutions
participating in the {\it Gaia} Multilateral Agreement.

Software: matplotlib \citep{matplotlib}, numpy \citep{numpy}, scipy \citep{scipy}}

\bibliography{OB171038}

\begin{thebibliography}{}
\expandafter\ifx\csname natexlab\endcsname\relax\def\natexlab#1{#1}\fi
\providecommand{\url}[1]{\href{#1}{#1}}

\bibitem[{{Alard} \& {Lupton}(1998)}]{Alard1998}
{Alard}, C., \& {Lupton}, R.~H. 1998, \apj, 503, 325

\bibitem[{{Albrow}(2017)}]{PyDIA}
{Albrow}, M.~D. 2017, {Michaeldalbrow/Pydia: Initial Release On Github.},
  vv1.0.0,  Zenodo, doi:10.5281/zenodo.268049

\bibitem[{{Albrow} {et~al.}(2018){Albrow}, {Yee}, {Udalski}, {Calchi Novati},
  {Carey}, {Henderson}, {Beichman}, {Bryden}, {Gaudi}, {Shvartzvald}, {Spitzer
  Team}, {Szyma{\'n}ski}, {Mr{\'o}z}, {Skowron}, {Poleski}, {Soszy{\'n}ski},
  {Koz{\l}owski}, {Pietrukowicz}, {Ulaczyk}, {Pawlak}, {OGLE Collaboration},
  {Chung}, {Gould}, {Han}, {Hwang}, {Jung}, {Ryu}, {Shin}, {Zhu}, {Cha}, {Kim},
  {Kim}, {Kim}, {Lee}, {Lee}, {Lee}, {Park}, {Pogge}, \& {KMTNet
  Collaboration}}]{albrow2018}
{Albrow}, M.~D., {Yee}, J.~C., {Udalski}, A., {et~al.} 2018, \apj, 858, 107

\bibitem[{{An} {et~al.}(2002){An}, {Albrow}, {Beaulieu}, {Caldwell}, {DePoy},
  {Dominik}, {Gaudi}, {Gould}, {Greenhill}, {Hill}, {Kane}, {Martin},
  {Menzies}, {Pogge}, {Pollard}, {Sackett}, {Sahu}, {Vermaak}, {Watson}, \&
  {Williams}}]{An2002}
{An}, J.~H., {Albrow}, M.~D., {Beaulieu}, J.~P., {et~al.} 2002, \apj, 572, 521

\bibitem[{{Bennett}(2010)}]{ICIRAS}
{Bennett}, D.~P. 2010, \apj, 716, 1408

\bibitem[{{Bensby} {et~al.}(2011){Bensby}, {Ad{\'e}n}, {Mel{\'e}ndez}, {Gould},
  {Feltzing}, {Asplund}, {Johnson}, {Lucatello}, {Yee}, {Ram{\'\i}rez},
  {Cohen}, {Thompson}, {Bond}, {Gal-Yam}, {Han}, {Sumi}, {Suzuki}, {Wada},
  {Miyake}, {Furusawa}, {Ohmori}, {Saito}, {Tristram}, \&
  {Bennett}}]{Bensby2011}
{Bensby}, T., {Ad{\'e}n}, D., {Mel{\'e}ndez}, J., {et~al.} 2011, \aap, 533,
  A134

\bibitem[{{Bessell} \& {Brett}(1988)}]{Bessel1988}
{Bessell}, M.~S., \& {Brett}, J.~M. 1988, \pasp, 100, 1134

\bibitem[{{Bowler} {et~al.}(2015){Bowler}, {Liu}, {Shkolnik}, \&
  {Tamura}}]{Bowler2015}
{Bowler}, B.~P., {Liu}, M.~C., {Shkolnik}, E.~L., \& {Tamura}, M. 2015, \apjs,
  216, 7

\bibitem[{{Bozza}(2010)}]{Bozza2010}
{Bozza}, V. 2010, \mnras, 408, 2188

\bibitem[{{Bozza} {et~al.}(2018){Bozza}, {Bachelet}, {Bartoli{\'c}}, {Heintz},
  {Hoag}, \& {Hundertmark}}]{Bozza2018}
{Bozza}, V., {Bachelet}, E., {Bartoli{\'c}}, F., {et~al.} 2018, \mnras, 479,
  5157

\bibitem[{{Calchi Novati} {et~al.}(2015{\natexlab{a}}){Calchi Novati}, {Gould},
  {Udalski}, {Menzies}, {Bond}, {Shvartzvald}, {Street}, {Hundertmark},
  {Beichman}, {Yee}, {Carey}, {Poleski}, {Skowron}, {Koz{\l}owski}, {Mr{\'o}z},
  {Pietrukowicz}, {Pietrzy{\'n}ski}, {Szyma{\'n}ski}, {Soszy{\'n}ski},
  {Ulaczyk}, {Wyrzykowski}, {OGLE Collaboration}, {Albrow}, {Beaulieu},
  {Caldwell}, {Cassan}, {Coutures}, {Danielski}, {Dominis Prester},
  {Donatowicz}, {Lon{\v{c}}ari{\'c}}, {McDougall}, {Morales}, {Ranc}, {Zhu},
  {PLANET Collaboration}, {Abe}, {Barry}, {Bennett}, {Bhattacharya},
  {Fukunaga}, {Inayama}, {Koshimoto}, {Namba}, {Sumi}, {Suzuki}, {Tristram},
  {Wakiyama}, {Yonehara}, {MOA Collaboration}, {Maoz}, {Kaspi}, {Friedmann},
  {Wise Group}, {Bachelet}, {Figuera Jaimes}, {Bramich}, {Tsapras}, {Horne},
  {Snodgrass}, {Wambsganss}, {Steele}, {Kains}, {RoboNet Collaboration},
  {Bozza}, {Dominik}, {J{\o}rgensen}, {Alsubai}, {Ciceri}, {D'Ago},
  {Haugb{\o}lle}, {Hessman}, {Hinse}, {Juncher}, {Korhonen}, {Mancini},
  {Popovas}, {Rabus}, {Rahvar}, {Scarpetta}, {Schmidt}, {Skottfelt},
  {Southworth}, {Starkey}, {Surdej}, {Wertz}, {Zarucki}, {MiNDSTEp Consortium},
  {Gaudi}, {Pogge}, {DePoy}, \& {{\ensuremath{\mu}}FUN
  Collaboration}}]{Novati2015-survey}
{Calchi Novati}, S., {Gould}, A., {Udalski}, A., {et~al.} 2015{\natexlab{a}},
  \apj, 804, 20

\bibitem[{{Calchi Novati} {et~al.}(2015{\natexlab{b}}){Calchi Novati}, {Gould},
  {Yee}, {Beichman}, {Bryden}, {Carey}, {Fausnaugh}, {Gaudi}, {Henderson},
  {Pogge}, {Shvartzvald}, {Wibking}, {Zhu}, {Spitzer Team}, {Udalski},
  {Poleski}, {Pawlak}, {Szyma{\'n}ski}, {Skowron}, {Mr{\'o}z}, {Koz{\l}owski},
  {Wyrzykowski}, {Pietrukowicz}, {Pietrzy{\'n}ski}, {Soszy{\'n}ski}, {Ulaczyk},
  \& {OGLE Group}}]{Novati2015-photometry}
{Calchi Novati}, S., {Gould}, A., {Yee}, J.~C., {et~al.} 2015{\natexlab{b}},
  \apj, 814, 92

\bibitem[{Chabrier(2003)}]{Chabrier2003}
Chabrier, G. 2003, Publications of the Astronomical Society of the Pacific,
  115, 763

\bibitem[{{Chabrier}(2005)}]{chabrier2005}
{Chabrier}, G. 2005, in Astrophysics and Space Science Library, Vol. 327, The
  Initial Mass Function 50 Years Later, ed. E.~{Corbelli}, F.~{Palla}, \&
  H.~{Zinnecker}, 41

\bibitem[{{Chabrier} \& {Baraffe}(1997)}]{HydrogenBurningVSMetalicity}
{Chabrier}, G., \& {Baraffe}, I. 1997, \aap, 327, 1039

\bibitem[{{Chabrier} {et~al.}(2014){Chabrier}, {Johansen}, {Janson}, \&
  {Rafikov}}]{chabrier2014}
{Chabrier}, G., {Johansen}, A., {Janson}, M., \& {Rafikov}, R. 2014, in
  Protostars and Planets VI, ed. H.~{Beuther}, R.~S. {Klessen}, C.~P.
  {Dullemond}, \& T.~{Henning} (Tucson, AZ: Univ. of Arizona Press), 619

\bibitem[{{Choi} {et~al.}(2013){Choi}, {Han}, {Udalski}, {Sumi}, {Gaudi},
  {Gould}, {Bennett}, {Dominik}, {Beaulieu}, {Tsapras}, {Bozza}, {Abe}, {Bond},
  {Botzler}, {Chote}, {Freeman}, {Fukui}, {Furusawa}, {Itow}, {Ling}, {Masuda},
  {Matsubara}, {Miyake}, {Muraki}, {Ohnishi}, {Rattenbury}, {Saito},
  {Sullivan}, {Suzuki}, {Sweatman}, {Suzuki}, {Takino}, {Tristram}, {Wada},
  {Yock}, {MOA Collaboration}, {Szyma{\'n}ski}, {Kubiak}, {Pietrzy{\'n}ski},
  {Soszy{\'n}ski}, {Skowron}, {Koz{\l}owski}, {Poleski}, {Ulaczyk},
  {Wyrzykowski}, {Pietrukowicz}, {OGLE Collaboration}, {Almeida}, {DePoy},
  {Dong}, {Gorbikov}, {Jablonski}, {Henderson}, {Hwang}, {Janczak}, {Jung},
  {Kaspi}, {Lee}, {Malamud}, {Maoz}, {McGregor}, {Mu{\~n}oz}, {Park}, {Park},
  {Pogge}, {Shvartzvald}, {Shin}, {Yee}, {{\ensuremath{\mu}}FUN Collaboration},
  {Alsubai}, {Browne}, {Burgdorf}, {Calchi Novati}, {Dodds}, {Fang}, {Finet},
  {Glitrup}, {Grundahl}, {Gu}, {Hardis}, {Harps{\o}e}, {Hinse}, {Hornstrup},
  {Hundertmark}, {Jessen-Hansen}, {Jrgensen}, {Kains}, {Kerins}, {Liebig},
  {Lund}, {Lundkvist}, {Maier}, {Mancini}, {Mathiasen}, {Penny}, {Rahvar},
  {Ricci}, {Scarpetta}, {Skottfelt}, {Snodgrass}, {Southworth}, {Surdej},
  {Tregloan-Reed}, {Wambsganss}, {Wertz}, {Zimmer}, {MiNDSTEp Consortium},
  {Albrow}, {Bachelet}, {Batista}, {Brillant}, {Cassan}, {Cole}, {Coutures},
  {Dieters}, {Dominis Prester}, {Donatowicz}, {Fouqu{\'e}}, {Greenhill},
  {Kubas}, {Marquette}, {Menzies}, {Sahu}, {Zub}, {PLANET Collaboration},
  {Bramich}, {Horne}, {Steele}, {Street}, \& {RoboNet
  Collaboration}}]{choi2013}
{Choi}, J.~Y., {Han}, C., {Udalski}, A., {et~al.} 2013, \apj, 768, 129

\bibitem[{{Chung} {et~al.}(2019){Chung}, {Gould}, {Skowron}, {Bond}, {Zhu},
  {Albrow}, {Jung}, {Han}, {Hwang}, {Ryu}, {Shin}, {Shvartzvald}, {Yee},
  {Zang}, {Cha}, {Kim}, {Kim}, {Kim}, {Kim}, {Lee}, {Lee}, {Lee}, {Park},
  {Pogge}, {KMTNet Collaboration}, {Udalski}, {Poleski}, {Mr{\'o}z},
  {Pietrukowicz}, {Szyma{\'n}ski}, {Soszy{\'n}ski}, {Koz{\l}owski}, {Ulaczyk},
  {Pawlak}, {The OGLE Collaboration}, {Beichman}, {Bryden}, {Calchi Novati},
  {Carey}, {Gaudi}, {Henderson}, {The Spitzer Team}, {Abe}, {Barry}, {Bennett},
  {Bhattacharya}, {Donachie}, {Fukui}, {Hirao}, {Itow}, {Kawasaki}, {Kondo},
  {Koshimoto}, {Li}, {Matsubara}, {Muraki}, {Miyazaki}, {Nagakane}, {Ranc},
  {Rattenbury}, {Suematsu}, {Sullivan}, {Sumi}, {Suzuki}, {Tristram},
  {Yonehara}, \& {MOA colllaboration}}]{Chung2019}
{Chung}, S.-J., {Gould}, A., {Skowron}, J., {et~al.} 2019, \apj, 871, 179

\bibitem[{{Dieterich} {et~al.}(2018){Dieterich}, {Weinberger}, {Boss}, {Henry},
  {Jao}, {Gagn{\'e}}, {Astraatmadja}, {Thompson}, \&
  {Anglada-Escud{\'e}}}]{HydrogenBurningLimit}
{Dieterich}, S.~B., {Weinberger}, A.~J., {Boss}, A.~P., {et~al.} 2018, \apj,
  865, 28

\bibitem[{Dominik(2006)}]{Dominik2006}
Dominik, M. 2006, Monthly Notices of the Royal Astronomical Society, 367, 669

\bibitem[{{Dong} {et~al.}(2009){Dong}, {Gould}, {Udalski}, {Anderson},
  {Christie}, {Gaudi}, {OGLE Collaboration}, {Jaroszy{\'n}ski}, {Kubiak},
  {Szyma{\'n}ski}, {Pietrzy{\'n}ski}, {Soszy{\'n}ski}, {Szewczyk}, {Ulaczyk},
  {Wyrzykowski}, {{\ensuremath{\mu}}FUN Collaboration}, {DePoy}, {Fox},
  {Gal-Yam}, {Han}, {L{\'e}pine}, {McCormick}, {Ofek}, {Park}, {Pogge}, {MOA
  Collaboration}, {Abe}, {Bennett}, {Bond}, {Britton}, {Gilmore}, {Hearnshaw},
  {Itow}, {Kamiya}, {Kilmartin}, {Korpela}, {Masuda}, {Matsubara}, {Motomura},
  {Muraki}, {Nakamura}, {Ohnishi}, {Okada}, {Rattenbury}, {Saito}, {Sako},
  {Sasaki}, {Sullivan}, {Sumi}, {Tristram}, {Yanagisawa}, {Yock}, {Yoshoika},
  {PLANET/RoboNet Collaborations}, {Albrow}, {Beaulieu}, {Brillant}, {Calitz},
  {Cassan}, {Cook}, {Coutures}, {Dieters}, {Dominis Prester}, {Donatowicz},
  {Fouqu{\'e}}, {Greenhill}, {Hill}, {Hoffman}, {Horne}, {J{\o}rgensen},
  {Kane}, {Kubas}, {Marquette}, {Martin}, {Meintjes}, {Menzies}, {Pollard},
  {Sahu}, {Vinter}, {Wambsganss}, {Williams}, {Bode}, {Bramich}, {Burgdorf},
  {Snodgrass}, {Steele}, {Doublier}, \& {Foellmi}}]{Dong2009}
{Dong}, S., {Gould}, A., {Udalski}, A., {et~al.} 2009, \apj, 695, 970

\bibitem[{{Du} {et~al.}(2018){Du}, {Li}, {Liu}, {Donlon}, \&
  {Newberg}}]{Halo2018Du}
{Du}, C., {Li}, H., {Liu}, S., {Donlon}, T., \& {Newberg}, H.~J. 2018, \apj,
  863, 87

\bibitem[{{Elmegreen}(2009)}]{elmegreen2009}
{Elmegreen}, B.~G. 2009, in The Evolving ISM in the Milky Way and Nearby
  Galaxies, 14

\bibitem[{{Fazio} {et~al.}(2004){Fazio}, {Hora}, {Allen}, {Ashby}, {Barmby},
  {Deutsch}, {Huang}, {Kleiner}, {Marengo}, {Megeath}, {Melnick}, {Pahre},
  {Patten}, {Polizotti}, {Smith}, {Taylor}, {Wang}, {Willner}, {Hoffmann},
  {Pipher}, {Forrest}, {McMurty}, {McCreight}, {McKelvey}, {McMurray}, {Koch},
  {Moseley}, {Arendt}, {Mentzell}, {Marx}, {Losch}, {Mayman}, {Eichhorn},
  {Krebs}, {Jhabvala}, {Gezari}, {Fixsen}, {Flores}, {Shakoorzadeh}, {Jungo},
  {Hakun}, {Workman}, {Karpati}, {Kichak}, {Whitley}, {Mann}, {Tollestrup},
  {Eisenhardt}, {Stern}, {Gorjian}, {Bhattacharya}, {Carey}, {Nelson},
  {Glaccum}, {Lacy}, {Lowrance}, {Laine}, {Reach}, {Stauffer}, {Surace},
  {Wilson}, {Wright}, {Hoffman}, {Domingo}, \& {Cohen}}]{Spitzer-IRAC}
{Fazio}, G.~G., {Hora}, J.~L., {Allen}, L.~E., {et~al.} 2004, \apjs, 154, 10

\bibitem[{{Forbes} \& {Loeb}(2019)}]{Forbes2019}
{Forbes}, J.~C., \& {Loeb}, A. 2019, \apj, 871, 227

\bibitem[{{Foreman-Mackey} {et~al.}(2017){Foreman-Mackey}, {Agol}, {Angus}, \&
  {Ambikasaran}}]{celerite}
{Foreman-Mackey}, D., {Agol}, E., {Angus}, R., \& {Ambikasaran}, S. 2017, arXiv
  e-prints, 1703.09710

\bibitem[{{Foreman-Mackey} {et~al.}(2013){Foreman-Mackey}, {Hogg}, {Lang}, \&
  {Goodman}}]{emcee}
{Foreman-Mackey}, D., {Hogg}, D.~W., {Lang}, D., \& {Goodman}, J. 2013, \pasp,
  125, 306

\bibitem[{{Gaia Collaboration} {et~al.}(2016){Gaia Collaboration}, {Prusti},
  {de Bruijne}, {Brown}, {Vallenari}, {Babusiaux}, {Bailer-Jones}, {Bastian},
  {Biermann}, {Evans}, {Eyer}, {Jansen}, {Jordi}, {Klioner}, {Lammers},
  {Lindegren}, {Luri}, {Mignard}, {Milligan}, {Panem}, {Poinsignon},
  {Pourbaix}, {Randich}, {Sarri}, {Sartoretti}, {Siddiqui}, {Soubiran},
  {Valette}, {van Leeuwen}, {Walton}, {Aerts}, {Arenou}, {Cropper}, {Drimmel},
  {H{\o}g}, {Katz}, {Lattanzi}, {O'Mullane}, {Grebel}, {Holland}, {Huc},
  {Passot}, {Bramante}, {Cacciari}, {Casta{\~n}eda}, {Chaoul}, {Cheek}, {De
  Angeli}, {Fabricius}, {Guerra}, {Hern{\'a}ndez}, {Jean-Antoine-Piccolo},
  {Masana}, {Messineo}, {Mowlavi}, {Nienartowicz}, {Ord{\'o}{\~n}ez-Blanco},
  {Panuzzo}, {Portell}, {Richards}, {Riello}, {Seabroke}, {Tanga},
  {Th{\'e}venin}, {Torra}, {Els}, {Gracia-Abril}, {Comoretto},
  {Garcia-Reinaldos}, {Lock}, {Mercier}, {Altmann}, {Andrae}, {Astraatmadja},
  {Bellas-Velidis}, {Benson}, {Berthier}, {Blomme}, {Busso}, {Carry},
  {Cellino}, {Clementini}, {Cowell}, {Creevey}, {Cuypers}, {Davidson}, {De
  Ridder}, {de Torres}, {Delchambre}, {Dell'Oro}, {Ducourant}, {Fr{\'e}mat},
  {Garc{\'\i}a-Torres}, {Gosset}, {Halbwachs}, {Hambly}, {Harrison}, {Hauser},
  {Hestroffer}, {Hodgkin}, {Huckle}, {Hutton}, {Jasniewicz}, {Jordan},
  {Kontizas}, {Korn}, {Lanzafame}, {Manteiga}, {Moitinho}, {Muinonen},
  {Osinde}, {Pancino}, {Pauwels}, {Petit}, {Recio-Blanco}, {Robin}, {Sarro},
  {Siopis}, {Smith}, {Smith}, {Sozzetti}, {Thuillot}, {van Reeven}, {Viala},
  {Abbas}, {Abreu Aramburu}, {Accart}, {Aguado}, {Allan}, {Allasia},
  {Altavilla}, {{\'A}lvarez}, {Alves}, {Anderson}, {Andrei}, {Anglada Varela},
  {Antiche}, {Antoja}, {Ant{\'o}n}, {Arcay}, {Atzei}, {Ayache}, {Bach},
  {Baker}, {Balaguer-N{\'u}{\~n}ez}, {Barache}, {Barata}, {Barbier}, {Barblan},
  {Baroni}, {Barrado y Navascu{\'e}s}, {Barros}, {Barstow}, {Becciani},
  {Bellazzini}, {Bellei}, {Bello Garc{\'\i}a}, {Belokurov}, {Bendjoya},
  {Berihuete}, {Bianchi}, {Bienaym{\'e}}, {Billebaud}, {Blagorodnova},
  {Blanco-Cuaresma}, {Boch}, {Bombrun}, {Borrachero}, {Bouquillon}, {Bourda},
  {Bouy}, {Bragaglia}, {Breddels}, {Brouillet}, {Br{\"u}semeister},
  {Bucciarelli}, {Budnik}, {Burgess}, {Burgon}, {Burlacu}, {Busonero}, {Buzzi},
  {Caffau}, {Cambras}, {Campbell}, {Cancelliere}, {Cantat-Gaudin}, {Carlucci},
  {Carrasco}, {Castellani}, {Charlot}, {Charnas}, {Charvet}, {Chassat},
  {Chiavassa}, {Clotet}, {Cocozza}, {Collins}, {Collins}, {Costigan}, {Crifo},
  {Cross}, {Crosta}, {Crowley}, {Dafonte}, {Damerdji}, {Dapergolas}, {David},
  {David}, {De Cat}, {de Felice}, {de Laverny}, {De Luise}, {De March}, {de
  Martino}, {de Souza}, {Debosscher}, {del Pozo}, {Delbo}, {Delgado},
  {Delgado}, {di Marco}, {Di Matteo}, {Diakite}, {Distefano}, {Dolding}, {Dos
  Anjos}, {Drazinos}, {Dur{\'a}n}, {Dzigan}, {Ecale}, {Edvardsson}, {Enke},
  {Erdmann}, {Escolar}, {Espina}, {Evans}, {Eynard Bontemps}, {Fabre},
  {Fabrizio}, {Faigler}, {Falc{\~a}o}, {Farr{\`a}s Casas}, {Faye}, {Federici},
  {Fedorets}, {Fern{\'a}ndez-Hern{\'a}ndez}, {Fernique}, {Fienga}, {Figueras},
  {Filippi}, {Findeisen}, {Fonti}, {Fouesneau}, {Fraile}, {Fraser}, {Fuchs},
  {Furnell}, {Gai}, {Galleti}, {Galluccio}, {Garabato}, {Garc{\'\i}a-Sedano},
  {Gar{\'e}}, {Garofalo}, {Garralda}, {Gavras}, {Gerssen}, {Geyer}, {Gilmore},
  {Girona}, {Giuffrida}, {Gomes}, {Gonz{\'a}lez-Marcos},
  {Gonz{\'a}lez-N{\'u}{\~n}ez}, {Gonz{\'a}lez-Vidal}, {Granvik}, {Guerrier},
  {Guillout}, {Guiraud}, {G{\'u}rpide}, {Guti{\'e}rrez-S{\'a}nchez}, {Guy},
  {Haigron}, {Hatzidimitriou}, {Haywood}, {Heiter}, {Helmi}, {Hobbs},
  {Hofmann}, {Holl}, {Holland}, {Hunt}, {Hypki}, {Icardi}, {Irwin}, {Jevardat
  de Fombelle}, {Jofr{\'e}}, {Jonker}, {Jorissen}, {Julbe}, {Karampelas},
  {Kochoska}, {Kohley}, {Kolenberg}, {Kontizas}, {Koposov}, {Kordopatis},
  {Koubsky}, {Kowalczyk}, {Krone-Martins}, {Kudryashova}, {Kull}, {Bachchan},
  {Lacoste-Seris}, {Lanza}, {Lavigne}, {Le Poncin-Lafitte}, {Lebreton},
  {Lebzelter}, {Leccia}, {Leclerc}, {Lecoeur-Taibi}, {Lemaitre}, {Lenhardt},
  {Leroux}, {Liao}, {Licata}, {Lindstr{\o}m}, {Lister}, {Livanou}, {Lobel},
  {L{\"o}ffler}, {L{\'o}pez}, {Lopez-Lozano}, {Lorenz}, {Loureiro},
  {MacDonald}, {Magalh{\~a}es Fernandes}, {Managau}, {Mann}, {Mantelet},
  {Marchal}, {Marchant}, {Marconi}, {Marie}, {Marinoni}, {Marrese},
  {Marschalk{\'o}}, {Marshall}, {Mart{\'\i}n-Fleitas}, {Martino}, {Mary},
  {Matijevi{\v{c}}}, {Mazeh}, {McMillan}, {Messina}, {Mestre}, {Michalik},
  {Millar}, {Miranda}, {Molina}, {Molinaro}, {Molinaro}, {Moln{\'a}r},
  {Moniez}, {Montegriffo}, {Monteiro}, {Mor}, {Mora}, {Morbidelli}, {Morel},
  {Morgenthaler}, {Morley}, {Morris}, {Mulone}, {Muraveva}, {Musella},
  {Narbonne}, {Nelemans}, {Nicastro}, {Noval}, {Ord{\'e}novic},
  {Ordieres-Mer{\'e}}, {Osborne}, {Pagani}, {Pagano}, {Pailler}, {Palacin},
  {Palaversa}, {Parsons}, {Paulsen}, {Pecoraro}, {Pedrosa}, {Pentik{\"a}inen},
  {Pereira}, {Pichon}, {Piersimoni}, {Pineau}, {Plachy}, {Plum}, {Poujoulet},
  {Pr{\v{s}}a}, {Pulone}, {Ragaini}, {Rago}, {Rambaux}, {Ramos-Lerate},
  {Ranalli}, {Rauw}, {Read}, {Regibo}, {Renk}, {Reyl{\'e}}, {Ribeiro},
  {Rimoldini}, {Ripepi}, {Riva}, {Rixon}, {Roelens}, {Romero-G{\'o}mez},
  {Rowell}, {Royer}, {Rudolph}, {Ruiz-Dern}, {Sadowski}, {Sagrist{\`a}
  Sell{\'e}s}, {Sahlmann}, {Salgado}, {Salguero}, {Sarasso}, {Savietto},
  {Schnorhk}, {Schultheis}, {Sciacca}, {Segol}, {Segovia}, {Segransan},
  {Serpell}, {Shih}, {Smareglia}, {Smart}, {Smith}, {Solano}, {Solitro},
  {Sordo}, {Soria Nieto}, {Souchay}, {Spagna}, {Spoto}, {Stampa}, {Steele},
  {Steidelm{\"u}ller}, {Stephenson}, {Stoev}, {Suess}, {S{\"u}veges}, {Surdej},
  {Szabados}, {Szegedi-Elek}, {Tapiador}, {Taris}, {Tauran}, {Taylor},
  {Teixeira}, {Terrett}, {Tingley}, {Trager}, {Turon}, {Ulla}, {Utrilla},
  {Valentini}, {van Elteren}, {Van Hemelryck}, {van Leeuwen}, {Varadi},
  {Vecchiato}, {Veljanoski}, {Via}, {Vicente}, {Vogt}, {Voss}, {Votruba},
  {Voutsinas}, {Walmsley}, {Weiler}, {Weingrill}, {Werner}, {Wevers},
  {Whitehead}, {Wyrzykowski}, {Yoldas}, {{\v{Z}}erjal}, {Zucker}, {Zurbach},
  {Zwitter}, {Alecu}, {Allen}, {Allende Prieto}, {Amorim},
  {Anglada-Escud{\'e}}, {Arsenijevic}, {Azaz}, {Balm}, {Beck}, {Bernstein},
  {Bigot}, {Bijaoui}, {Blasco}, {Bonfigli}, {Bono}, {Boudreault}, {Bressan},
  {Brown}, {Brunet}, {Bunclark}, {Buonanno}, {Butkevich}, {Carret}, {Carrion},
  {Chemin}, {Ch{\'e}reau}, {Corcione}, {Darmigny}, {de Boer}, {de Teodoro}, {de
  Zeeuw}, {Delle Luche}, {Domingues}, {Dubath}, {Fodor}, {Fr{\'e}zouls},
  {Fries}, {Fustes}, {Fyfe}, {Gallardo}, {Gallegos}, {Gardiol}, {Gebran},
  {Gomboc}, {G{\'o}mez}, {Grux}, {Gueguen}, {Heyrovsky}, {Hoar}, {Iannicola},
  {Isasi Parache}, {Janotto}, {Joliet}, {Jonckheere}, {Keil}, {Kim},
  {Klagyivik}, {Klar}, {Knude}, {Kochukhov}, {Kolka}, {Kos}, {Kutka}, {Lainey},
  {LeBouquin}, {Liu}, {Loreggia}, {Makarov}, {Marseille}, {Martayan},
  {Martinez-Rubi}, {Massart}, {Meynadier}, {Mignot}, {Munari}, {Nguyen},
  {Nordlander}, {Ocvirk}, {O'Flaherty}, {Olias Sanz}, {Ortiz}, {Osorio},
  {Oszkiewicz}, {Ouzounis}, {Palmer}, {Park}, {Pasquato}, {Peltzer}, {Peralta},
  {P{\'e}turaud}, {Pieniluoma}, {Pigozzi}, {Poels}, {Prat}, {Prod'homme},
  {Raison}, {Rebordao}, {Risquez}, {Rocca-Volmerange}, {Rosen}, {Ruiz-Fuertes},
  {Russo}, {Sembay}, {Serraller Vizcaino}, {Short}, {Siebert}, {Silva},
  {Sinachopoulos}, {Slezak}, {Soffel}, {Sosnowska}, {Strai{\v{z}}ys}, {ter
  Linden}, {Terrell}, {Theil}, {Tiede}, {Troisi}, {Tsalmantza}, {Tur},
  {Vaccari}, {Vachier}, {Valles}, {Van Hamme}, {Veltz}, {Virtanen}, {Wallut},
  {Wichmann}, {Wilkinson}, {Ziaeepour}, \& {Zschocke}}]{GaiaMission}
{Gaia Collaboration}, {Prusti}, T., {de Bruijne}, J.~H.~J., {et~al.} 2016,
  \aap, 595, A1

\bibitem[{{Gaia Collaboration} {et~al.}(2020){Gaia Collaboration}, {Klioner},
  {Mignard}, {Lindegren}, {Bastian}, {McMillan}, {Hern{\'a}ndez}, {Hobbs},
  {Ramos-Lerate}, {Biermann}, {Bombrun}, {de Torres}, {Gerlach}, {Geyer},
  {Hilger}, {Lammers}, {Steidelm{\"u}ller}, {Stephenson}, {Brown}, {Vallenari},
  {Prusti}, {de Bruijne}, {Babusiaux}, {Creevey}, {Evans}, {Eyer}, {Hutton},
  {Jansen}, {Jordi}, {Luri}, {Panem}, {Pourbaix}, {Randich}, {Sartoretti},
  {Soubiran}, {Walton}, {Arenou}, {Bailer-Jones}, {Cropper}, {Drimmel}, {Katz},
  {Lattanzi}, {van Leeuwen}, {Bakker}, {Casta{\~n}eda}, {De Angeli},
  {Ducourant}, {Fabricius}, {Fouesneau}, {Fr{\'e}mat}, {Guerra}, {Guerrier},
  {Guiraud}, {Jean-Antoine Piccolo}, {Masana}, {Messineo}, {Mowlavi},
  {Nicolas}, {Nienartowicz}, {Pailler}, {Panuzzo}, {Riclet}, {Roux},
  {Seabroke}, {Sordo}, {Tanga}, {Th{\'e}venin}, {Gracia-Abril}, {Portell},
  {Teyssier}, {Altmann}, {Andrae}, {Bellas-Velidis}, {Benson}, {Berthier},
  {Blomme}, {Brugaletta}, {Burgess}, {Busso}, {Carry}, {Cellino}, {Cheek},
  {Clementini}, {Damerdji}, {Davidson}, {Delchambre}, {Dell'Oro},
  {Fern{\'a}ndez-Hern{\'a}ndez}, {Galluccio}, {Garc{\'\i}a-Lario},
  {Garcia-Reinaldos}, {Gonz{\'a}lez-N{\'u}{\~n}ez}, {Gosset}, {Haigron},
  {Halbwachs}, {Hambly}, {Harrison}, {Hatzidimitriou}, {Heiter}, {Hestroffer},
  {Hodgkin}, {Holl}, {Jan{\ss}en}, {Jevardat de Fombelle}, {Jordan},
  {Krone-Martins}, {Lanzafame}, {L{\"o}ffler}, {Lorca}, {Manteiga}, {Marchal},
  {Marrese}, {Moitinho}, {Mora}, {Muinonen}, {Osborne}, {Pancino}, {Pauwels},
  {Recio-Blanco}, {Richards}, {Riello}, {Rimoldini}, {Robin}, {Roegiers},
  {Rybizki}, {Sarro}, {Siopis}, {Smith}, {Sozzetti}, {Ulla}, {Utrilla}, {van
  Leeuwen}, {van Reeven}, {Abbas}, {Abreu Aramburu}, {Accart}, {Aerts},
  {Aguado}, {Ajaj}, {Altavilla}, {{\'A}lvarez}, {{\'A}lvarez Cid-Fuentes},
  {Alves}, {Anderson}, {Anglada Varela}, {Antoja}, {Audard}, {Baines}, {Baker},
  {Balaguer-N{\'u}\textbackslash-nez}, {Balbinot}, {Balog}, {Barache},
  {Barbato}, {Barros}, {Barstow}, {Bartolom{\'e}}, {Bassilana}, {Bauchet},
  {Baudesson-Stella}, {Becciani}, {Bellazzini}, {Bernet}, {Bertone}, {Bianchi},
  {Blanco-Cuaresma}, {Boch}, {Bossini}, {Bouquillon}, {Bramante}, {Breedt},
  {Bressan}, {Brouillet}, {Bucciarelli}, {Burlacu}, {Busonero}, {Butkevich},
  {Buzzi}, {Caffau}, {Cancelliere}, {C{\'a}novas}, {Cantat-Gaudin}, {Carballo},
  {Carlucci}, {Carnerero}, {Carrasco}, {Casamiquela}, {Castellani},
  {Castro-Ginard}, {Castro Sampol}, {Chaoul}, {Charlot}, {Chemin}, {Chiavassa},
  {Comoretto}, {Cooper}, {Cornez}, {Cowell}, {Crifo}, {Crosta}, {Crowley},
  {Dafonte}, {Dapergolas}, {David}, {David}, {de Laverny}, {De Luise}, {De
  March}, {De Ridder}, {de Souza}, {de Teodoro}, {del Peloso}, {del Pozo},
  {Delgado}, {Delgado}, {Delisle}, {Di Matteo}, {Diakite}, {Diener},
  {Distefano}, {Dolding}, {Eappachen}, {Enke}, {Esquej}, {Fabre}, {Fabrizio},
  {Faigler}, {Fedorets}, {Fernique}, {Fienga}, {Figueras}, {Fouron},
  {Fragkoudi}, {Fraile}, {Franke}, {Gai}, {Garabato}, {Garcia-Gutierrez},
  {Garc{\'\i}a-Torres}, {Garofalo}, {Gavras}, {Giacobbe}, {Gilmore}, {Girona},
  {Giuffrida}, {Gomez}, {Gonzalez-Santamaria}, {Gonz{\'a}lez-Vidal}, {Granvik},
  {Guti{\'e}rrez-S{\'a}nchez}, {Guy}, {Hauser}, {Haywood}, {Helmi}, {Hidalgo},
  {H{\l}adczuk}, {Holland}, {Huckle}, {Jasniewicz}, {Jonker}, {Juaristi
  Campillo}, {Julbe}, {Karbevska}, {Kervella}, {Khanna}, {Kochoska},
  {Kordopatis}, {Korn}, {Kostrzewa-Rutkowska}, {Kruszy{\'n}ska}, {Lambert},
  {Lanza}, {Lasne}, {Le Campion}, {Le Fustec}, {Lebreton}, {Lebzelter},
  {Leccia}, {Leclerc}, {Lecoeur-Taibi}, {Liao}, {Licata}, {Lindstr{\o}m},
  {Lister}, {Livanou}, {Lobel}, {Madrero Pardo}, {Managau}, {Mann}, {Marchant},
  {Marconi}, {Marcos Santos}, {Marinoni}, {Marocco}, {Marshall}, {Polo},
  {Mart{\'\i}n-Fleitas}, {Masip}, {Massari}, {Mastrobuono-Battisti}, {Mazeh},
  {Messina}, {Michalik}, {Millar}, {Mints}, {Molina}, {Molinaro}, {Moln{\'a}r},
  {Montegriffo}, {Mor}, {Morbidelli}, {Morel}, {Morris}, {Mulone}, {Munoz},
  {Muraveva}, {Murphy}, {Musella}, {Noval}, {Ord{\'e}novic}, {Orr{\`u}},
  {Osinde}, {Pagani}, {Pagano}, {Palaversa}, {Palicio}, {Panahi}, {Pawlak},
  {Pe\textbackslash-nalosa Esteller}, {Penttil{\"a}}, {Piersimoni}, {Pineau},
  {Plachy}, {Plum}, {Poggio}, {Poretti}, {Poujoulet}, {Pr{\v{s}}a}, {Pulone},
  {Racero}, {Ragaini}, {Rainer}, {Raiteri}, {Rambaux}, {Ramos}, {Re Fiorentin},
  {Regibo}, {Reyl{\'e}}, {Ripepi}, {Riva}, {Rixon}, {Robichon}, {Robin},
  {Roelens}, {Rohrbasser}, {Romero-G{\'o}mez}, {Rowell}, {Royer}, {Rybicki},
  {Sadowski}, {Sagrist{\`a} Sell{\'e}s}, {Sahlmann}, {Salgado}, {Salguero},
  {Samaras}, {Sanchez Gimenez}, {Sanna}, {Santove{\~n}a}, {Sarasso},
  {Schultheis}, {Sciacca}, {Segol}, {Segovia}, {S{\'e}gransan}, {Semeux},
  {Siddiqui}, {Siebert}, {Siltala}, {Slezak}, {Smart}, {Solano}, {Solitro},
  {Souami}, {Souchay}, {Spagna}, {Spoto}, {Steele}, {S{\"u}veges}, {Szabados},
  {Szegedi-Elek}, {Taris}, {Tauran}, {Taylor}, {Teixeira}, {Thuillot},
  {Tonello}, {Torra}, {Torray}, {Turon}, {Unger}, {Vaillant}, {van Dillen},
  {Vanel}, {Vecchiato}, {Viala}, {Vicente}, {Voutsinas}, {Weiler}, {Wevers},
  {Wyrzykowski}, {Yoldas}, {Yvard}, {Zhao}, {Zorec}, {Zucker}, {Zurbach}, \&
  {Zwitter}}]{GaiaEDR3}
{Gaia Collaboration}, {Klioner}, S.~A., {Mignard}, F., {et~al.} 2020, arXiv
  e-prints, arXiv:2012.02036

\bibitem[{{Gould}(1994)}]{Gould1994}
{Gould}, A. 1994, \apjl, 421, L75

\bibitem[{Gould(2008)}]{hexadecapole}
Gould, A. 2008, \apj, 681, 1593

\bibitem[{{Gould}(2020)}]{AndyRich2020}
{Gould}, A. 2020, Journal of Korean Astronomical Society, 53, 99

\bibitem[{{Gould} {et~al.}(2009){Gould}, {Udalski}, {Monard}, {Horne}, {Dong},
  {Miyake}, {Sahu}, {Bennett}, {Wyrzykowski}, {Soszy{\'n}ski}, {Szyma{\'n}ski},
  {Kubiak}, {Pietrzy{\'n}ski}, {Szewczyk}, {Ulaczyk}, {OGLE Collaboration},
  {Allen}, {Christie}, {DePoy}, {Gaudi}, {Han}, {Lee}, {McCormick}, {Natusch},
  {Park}, {Pogge}, {{\ensuremath{\mu}}FUN Collaboration}, {Allan}, {Bode},
  {Bramich}, {Burgdorf}, {Dominik}, {Fraser}, {Kerins}, {Mottram}, {Snodgrass},
  {Steele}, {Street}, {Tsapras}, {RoboNet Collaboration}, {Abe}, {Bond},
  {Botzler}, {Fukui}, {Furusawa}, {Hearnshaw}, {Itow}, {Kamiya}, {Kilmartin},
  {Korpela}, {Lin}, {Ling}, {Masuda}, {Matsubara}, {Muraki}, {Nagaya},
  {Ohnishi}, {Okumura}, {Perrott}, {Rattenbury}, {Saito}, {Sako}, {Skuljan},
  {Sullivan}, {Sumi}, {Sweatman}, {Tristram}, {Yock}, {MOA Collaboration},
  {Albrow}, {Beaulieu}, {Coutures}, {Calitz}, {Caldwell}, {Fouque}, {Martin},
  {Williams}, \& {PLANET Collaboration}}]{No1BD}
{Gould}, A., {Udalski}, A., {Monard}, B., {et~al.} 2009, \apjl, 698, L147

\bibitem[{{Gould} {et~al.}(2020){Gould}, {Ryu}, {Calchi Novati}, {Zang},
  {Albrow}, {Chung}, {Han}, {Hwang}, {Jung}, {Shin}, {Shvartzvald}, {Yee},
  {Cha}, {Kim}, {Kim}, {Kim}, {Lee}, {Lee}, {Lee}, {Park}, {Pogge}, {Beichman},
  {Bryden}, {Carey}, {Gaudi}, {Henderson}, {Zhu}, {Fouque}, {Penny}, {Petric},
  {Burdullis}, \& {Mao}}]{kb180029Gould}
{Gould}, A., {Ryu}, Y.-H., {Calchi Novati}, S., {et~al.} 2020, Journal of
  Korean Astronomical Society, 53, 9

\bibitem[{{Grether} \& {Lineweaver}(2006)}]{Grether2006}
{Grether}, D., \& {Lineweaver}, C.~H. 2006, \apj, 640, 1051

\bibitem[{Han \& Gould(2003)}]{Han2003}
Han, C., \& Gould, A. 2003, The Astrophysical Journal, 592, 172

\bibitem[{{Han} {et~al.}(2017){Han}, {Udalski}, {Bozza}, {Szyma{\'n}ski},
  {Soszy{\'n}ski}, {Skowron}, {Mr{\'o}z}, {Poleski}, {Pietrukowicz},
  {Koz{\l}owski}, {Ulaczyk}, {Wyrzykowski}, {OGLE Collaboration}, {Calchi
  Novati}, {D'Ago}, {Dominik}, {Hundertmark}, {Jorgensen}, {Scarpetta}, \&
  {MiNDSTEp Consortium}}]{han2017a}
{Han}, C., {Udalski}, A., {Bozza}, V., {et~al.} 2017, \apj, 843, 87

\bibitem[{{Han} {et~al.}(2020){Han}, {Lee}, {Udalski}, {Gould}, {Bond},
  {Bozza}, {Albrow}, {Chung}, {Hwang}, {Jung}, {Ryu}, {Shin}, {Shvartzvald},
  {Yee}, {Zang}, {Cha}, {Kim}, {Kim}, {Kim}, {Lee}, {Lee}, {Park}, {Pogge},
  {Jee}, {Kim}, {KMTNet Collaboration}, {Mr{\'o}z}, {Szyma{\'n}ski}, {Skowron},
  {Poleski}, {Soszy{\'n}ski}, {Pietrukowicz}, {Koz{\l}owski}, {Ulaczyk},
  {Rybicki}, {Iwanek}, {Wrona}, {OGLE Collaboration}, {Abe}, {Barry},
  {Bennett}, {Bhattacharya}, {Donachie}, {Fujii}, {Fukui}, {Itow}, {Hirao},
  {Kamei}, {Kondo}, {Koshimoto}, {Li}, {Matsubara}, {Muraki}, {Miyazaki},
  {Nagakane}, {Ranc}, {Rattenbury}, {Satoh}, {Shoji}, {Suematsu}, {Sullivan},
  {Sumi}, {Suzuki}, {Tristram}, {Yamakawa}, {Yamawaki}, {Yonehara}, \& {MOA
  Collaboration}}]{2020BDs}
{Han}, C., {Lee}, C.-U., {Udalski}, A., {et~al.} 2020, \aj, 159, 134

\bibitem[{{Harris} {et~al.}(2020){Harris}, {Millman}, {van der Walt},
  {Gommers}, {Virtanen}, {Cournapeau}, {Wieser}, {Taylor}, {Berg}, {Smith},
  {Kern}, {Picus}, {Hoyer}, {van Kerkwijk}, {Brett}, {Haldane}, {del R{\'\i}o},
  {Wiebe}, {Peterson}, {G{\'e}rard-Marchant}, {Sheppard}, {Reddy}, {Weckesser},
  {Abbasi}, {Gohlke}, \& {Oliphant}}]{numpy}
{Harris}, C.~R., {Millman}, K.~J., {van der Walt}, S.~J., {et~al.} 2020, \nat,
  585, 357

\bibitem[{{Henebelle} \& Chabrier(2009)}]{hennebelle2009}
{Henebelle}, P., \& Chabrier, G. 2009, \apj, 702, 1428

\bibitem[{{Hennebelle} \& {Chabrier}(2008)}]{hennebelle2008}
{Hennebelle}, P., \& {Chabrier}, G. 2008, \apj, 684, 395

\bibitem[{{Herrera-Mart{\'\i}n} {et~al.}(2020){Herrera-Mart{\'\i}n}, {Albrow},
  {Udalski}, {Gould}, {Ryu}, {Yee}, {Chung}, {Han}, {Hwang}, {Jung}, {Lee},
  {Shin}, {Shvartzvald}, {Zang}, {Cha}, {Kim}, {Kim}, {Kim}, {Lee}, {Lee},
  {Park}, {Pogge}, {KMTNet Collaboration}, {Szyma{\'n}ski}, {Mr{\'o}z},
  {Skowron}, {Poleski}, {Soszy{\'n}ski}, {Koz{\l}owski}, {Pietrukowicz},
  {Ulaczyk}, {Rybicki}, {Iwanek}, {Wrona}, \& {OGLE
  Collaboration}}]{Herrera2020}
{Herrera-Mart{\'\i}n}, A., {Albrow}, M.~D., {Udalski}, A., {et~al.} 2020, \aj,
  159, 256

\bibitem[{{Hirao} {et~al.}(2020){Hirao}, {Bennett}, {Ryu}, {Koshimoto},
  {Udalski}, {Yee}, {Sumi}, {Bond}, {Shvartzvald}, {Abe}, {Barry},
  {Bhattacharya}, {Donachie}, {Fukui}, {Itow}, {Kondo}, {Li}, {Matsubara},
  {Matsuo}, {Miyazaki}, {Muraki}, {Nagakane}, {Ranc}, {Rattenbury}, {Suematsu},
  {Shibai}, {Suzuki}, {Tristram}, {Yonehara}, {MOA Collaboration}, {Skowron},
  {Poleski}, {Mr{\'o}z}, {Szyma{\'n}ski}, {Soszy{\'n}ski}, {Koz{\l}owski},
  {Pietrukowicz}, {Ulaczyk}, {Rybicki}, {Iwanek}, {OGLE Collaboration},
  {Albrow}, {Chung}, {Gould}, {Han}, {Hwang}, {Jung}, {Shin}, {Zang}, {Cha},
  {Kim}, {Kim}, {Kim}, {Lee}, {Lee}, {Lee}, {Park}, {Pogge}, {KMTNet
  Collaboration}, {Beichman}, {Bryden}, {Novati}, {Carey}, {Gaudi},
  {Henderson}, {Zhu}, {Spitzer Team}, {Bachelet}, {Bolt}, {Christie},
  {Hundertmark}, {Natusch}, {Maoz}, {McCormick}, {Street}, {Tan}, {Tsapras},
  {LCO and {\ensuremath{\mu}}FUN Follow-up Teams}, {J{\o}rgensen}, {Dominik},
  {Bozza}, {Skottfelt}, {Snodgrass}, {Ciceri}, {Jaimes}, {Evans}, {Peixinho},
  {Hinse}, {Burgdorf}, {Southworth}, {Rahvar}, {Sajadian}, {Rabus}, {von
  Essen}, {Fujii}, {Campbell-White}, {Lowry}, {Helling}, {Mancini}, {Haikala},
  {MindSTEp Collaboration}, {Kandori}, \& {IRSF Team}}]{ob170406Hirao}
{Hirao}, Y., {Bennett}, D.~P., {Ryu}, Y.-H., {et~al.} 2020, \aj, 160, 74

\bibitem[{{Hunter}(2007)}]{matplotlib}
{Hunter}, J.~D. 2007, Computing in Science and Engineering, 9, 90

\bibitem[{{Hurley} {et~al.}(2005){Hurley}, {Pols}, {Aarseth}, \&
  {Tout}}]{Hurley2005}
{Hurley}, J.~R., {Pols}, O.~R., {Aarseth}, S.~J., \& {Tout}, C.~A. 2005,
  \mnras, 363, 293

\bibitem[{{Hut} {et~al.}(1992{\natexlab{a}}){Hut}, {McMillan}, \&
  {Romani}}]{Hut1992-2}
{Hut}, P., {McMillan}, S., \& {Romani}, R.~W. 1992{\natexlab{a}}, \apj, 389,
  527

\bibitem[{{Hut} {et~al.}(1992{\natexlab{b}}){Hut}, {McMillan}, {Goodman},
  {Mateo}, {Phinney}, {Pryor}, {Richer}, {Verbunt}, \& {Weinberg}}]{Hut1992}
{Hut}, P., {McMillan}, S., {Goodman}, J., {et~al.} 1992{\natexlab{b}}, \pasp,
  104, 981

\bibitem[{{J{\o}rgensen} \& {Church}(2020)}]{Jorgensen2020}
{J{\o}rgensen}, T.~G., \& {Church}, R.~P. 2020, \mnras, 492, 4959

\bibitem[{{Just} {et~al.}(2015){Just}, {Fuchs}, {Jahrei{\ss}}, {Flynn},
  {Dettbarn}, \& {Rybizki}}]{Just2015}
{Just}, A., {Fuchs}, B., {Jahrei{\ss}}, H., {et~al.} 2015, \mnras, 451, 149

\bibitem[{Kass \& Raftery(1995)}]{BayesFactors}
Kass, R.~E., \& Raftery, A.~E. 1995, Journal of the American Statistical
  Association, 90, 773

\bibitem[{{Kervella} \& {Fouqu{\'e}}(2008)}]{Kervella2008}
{Kervella}, P., \& {Fouqu{\'e}}, P. 2008, \aap, 491, 855

\bibitem[{{Kim} {et~al.}(2021){Kim}, {Hwang}, {Gould}, {Yee}, {Ryu}, {Albrow},
  {Chung}, {Han}, {Kil Jung}, {Lee}, {Shin}, {Shvartzvald}, {Zang}, {Cha},
  {Kim}, {Kim}, {Lee}, {Lee}, {Park}, \& {Pogge}}]{Kim2021}
{Kim}, H.-W., {Hwang}, K.-H., {Gould}, A., {et~al.} 2021, \aj, 162, 15

\bibitem[{{Kim} {et~al.}(2016){Kim}, {Lee}, {Park}, {Kim}, {Cha}, {Lee}, {Han},
  {Chun}, \& {Yuk}}]{KMTNet}
{Kim}, S.-L., {Lee}, C.-U., {Park}, B.-G., {et~al.} 2016, JKAS, 49, 37

\bibitem[{{Koshimoto} \& {Bennett}(2019)}]{Koshimoto2019}
{Koshimoto}, N., \& {Bennett}, D.~P. 2019, arXiv e-prints, arXiv:1905.05794

\bibitem[{{Kroupa}(2001)}]{kroupa2001}
{Kroupa}, P. 2001, \mnras, 322, 231

\bibitem[{{Kroupa} {et~al.}(2013){Kroupa}, {Weidner}, {Pflamm-Altenburg},
  {Thies}, {Dabringhausen}, {Marks}, \& {Maschberger}}]{kroupa2013}
{Kroupa}, P., {Weidner}, C., {Pflamm-Altenburg}, J., {et~al.} 2013, Galactic
  Structure and Stellar Populations (Planets, Stars and Stellar Systems), Vol.
  5 (Dordrecht: Springer), 115

\bibitem[{{Li} {et~al.}(2019){Li}, {Zang}, {Udalski}, {Shvartzvald}, {Huber},
  {Lee}, {Sumi}, {Gould}, {Mao}, {Fouqu{\'e}}, {Wang}, {Dong}, {J{\o}rgensen},
  {Cole}, {Mr{\'o}z}, {Szyma{\'n}ski}, {Skowron}, {Poleski}, {Soszy{\'n}ski},
  {Pietrukowicz}, {Koz{\l}owski}, {Ulaczyk}, {Rybicki}, {Iwanek}, {Yee},
  {Calchi Novati}, {Beichman}, {Bryden}, {Carey}, {Gaudi}, {Henderson}, {Zhu},
  {Albrow}, {Chung}, {Han}, {Hwang}, {Jung}, {Ryu}, {Shin}, {Cha}, {Kim},
  {Kim}, {Kim}, {Lee}, {Lee}, {Park}, {Pogge}, {Bond}, {Abe}, {Barry},
  {Bennett}, {Bhattacharya}, {Donachie}, {Fukui}, {Hirao}, {Itow}, {Kondo},
  {Koshimoto}, {Li}, {Matsubara}, {Muraki}, {Miyazaki}, {Nagakane}, {Ranc},
  {Rattenbury}, {Suematsu}, {Sullivan}, {Suzuki}, {Tristram}, {Yonehara},
  {Christie}, {Drummond}, {Green}, {Hennerley}, {Natusch}, {Porritt},
  {Bachelet}, {Maoz}, {Street}, {Tsapras}, {Bozza}, {Dominik}, {Hundertmark},
  {Peixinho}, {Sajadian}, {Burgdorf}, {Evans}, {Figuera Jaimes}, {Fujii},
  {Haikala}, {Helling}, {Henning}, {Hinse}, {Mancini}, {Longa-Pe{\~n}a},
  {Rahvar}, {Rabus}, {Skottfelt}, {Snodgrass}, {Southworth}, {Unda-Sanzana},
  {von Essen}, {Beaulieu}, {Blackman}, \& {Hill}}]{Li2019}
{Li}, S.~S., {Zang}, W., {Udalski}, A., {et~al.} 2019, \mnras, 488, 3308

\bibitem[{{McDougall} \& {Albrow}(2016)}]{MORSE}
{McDougall}, A., \& {Albrow}, M.~D. 2016, \mnras, 456, 565

\bibitem[{{Mr{\'o}z} {et~al.}(2019){Mr{\'o}z}, {Udalski}, {Bennett}, {Ryu},
  {Sumi}, {Shvartzvald}, {Skowron}, {Poleski}, {Pietrukowicz}, {Koz{\l}owski},
  {Szyma{\'n}ski}, {Wyrzykowski}, {Soszy{\'n}ski}, {Ulaczyk}, {Rybicki},
  {Iwanek}, {Albrow}, {Chung}, {Gould}, {Han}, {Hwang}, {Jung}, {Shin}, {Yee},
  {Zang}, {Cha}, {Kim}, {Kim}, {Kim}, {Lee}, {Lee}, {Lee}, {Park}, {Pogge},
  {Abe}, {Barry}, {Bhattacharya}, {Bond}, {Donachie}, {Fukui}, {Hirao}, {Itow},
  {Kawasaki}, {Kondo}, {Koshimoto}, {Li}, {Matsubara}, {Muraki}, {Miyazaki},
  {Nagakane}, {Ranc}, {Rattenbury}, {Suematsu}, {Sullivan}, {Suzuki},
  {Tristram}, {Yonehara}, {Maoz}, {Kaspi}, \& {Friedmann}}]{2FFPs}
{Mr{\'o}z}, P., {Udalski}, A., {Bennett}, D.~P., {et~al.} 2019, \aap, 622, A201

\bibitem[{{Mr{\'o}z} {et~al.}(2020){Mr{\'o}z}, {Poleski}, {Han}, {Udalski},
  {Gould}, {Szyma{\'n}ski}, {Soszy{\'n}ski}, {Pietrukowicz}, {Koz{\l}owski},
  {Skowron}, {Ulaczyk}, {Gromadzki}, {Rybicki}, {Iwanek}, {Wrona}, {OGLE
  Collaboration}, {Albrow}, {Chung}, {Hwang}, {Ryu}, {Jung}, {Shin},
  {Shvartzvald}, {Yee}, {Zang}, {Cha}, {Kim}, {Kim}, {Kim}, {Lee}, {Lee},
  {Lee}, {Park}, {Pogge}, \& {KMT Collaboration}}]{2020FFPMroz}
{Mr{\'o}z}, P., {Poleski}, R., {Han}, C., {et~al.} 2020, \aj, 159, 262

\bibitem[{{Nataf} {et~al.}(2013){Nataf}, {Gould}, {Fouqu{\'e}}, {Gonzalez},
  {Johnson}, {Skowron}, {Udalski}, {Szyma{\'n}ski}, {Kubiak},
  {Pietrzy{\'n}ski}, {Soszy{\'n}ski}, {Ulaczyk}, {Wyrzykowski}, \&
  {Poleski}}]{Nataf2013a}
{Nataf}, D.~M., {Gould}, A., {Fouqu{\'e}}, P., {et~al.} 2013, \apj, 769, 88

\bibitem[{{Nordgren} {et~al.}(2002){Nordgren}, {Lane}, {Hindsley}, \&
  {Kervella}}]{Nordgren2002}
{Nordgren}, T.~E., {Lane}, B.~F., {Hindsley}, R.~B., \& {Kervella}, P. 2002,
  \aj, 123, 3380

\bibitem[{{Paczy\'nski}(1986)}]{FFPProposed}
{Paczy\'nski}, B. 1986, \apj, 304, 1

\bibitem[{{Padoan} \& {Nordlund}(2002)}]{padoan2002}
{Padoan}, P., \& {Nordlund}, {\r{A}}. 2002, \apj, 576, 870

\bibitem[{{Park} {et~al.}(2004){Park}, {DePoy}, {Gaudi}, {Gould}, {Han},
  {Pogge}, {muFun Collaboration}, {Abe}, {Bennett}, {Bond}, {Eguchi}, {Furuta},
  {Hearnshaw}, {Kamiya}, {Kilmartin}, {Kurata}, {Masuda}, {Matsubara},
  {Muraki}, {Noda}, {Okajima}, {Rattenbury}, {Sako}, {Sekiguchi}, {Sullivan},
  {Sumi}, {Tristram}, {Yanagisawa}, {Yock}, \& {MOA Collaboration}}]{Park2004}
{Park}, B.~G., {DePoy}, D.~L., {Gaudi}, B.~S., {et~al.} 2004, \apj, 609, 166

\bibitem[{{Pejcha} \& {Heyrovsk{\'y}}(2009)}]{pejcha2009}
{Pejcha}, O., \& {Heyrovsk{\'y}}, D. 2009, \apj, 690, 1772

\bibitem[{{Posti} {et~al.}(2018){Posti}, {Helmi}, {Veljanoski}, \&
  {Breddels}}]{Halo2018Posti}
{Posti}, L., {Helmi}, A., {Veljanoski}, J., \& {Breddels}, M.~A. 2018, \aap,
  615, A70

\bibitem[{{Refsdal}(1966)}]{Refsdal1966}
{Refsdal}, S. 1966, \mnras, 134, 315

\bibitem[{Rosell {et~al.}(2019)Rosell, Santiago, Ponte, Burningham, da~Costa,
  James, Marshall, McMahon, Bechtol, De~Paris, {et~al.}}]{rosell2019}
Rosell, A.~C., Santiago, B., Ponte, M.~d., {et~al.} 2019, \mnras, 489, 5301

\bibitem[{{Rybizki} {et~al.}(2021){Rybizki}, {Green}, {Rix}, {El-Badry},
  {Demleitner}, {Zari}, {Udalski}, {Smart}, \& {Gould}}]{Ryb2021}
{Rybizki}, J., {Green}, G., {Rix}, H.-W., {et~al.} 2021, arXiv e-prints,
  arXiv:2101.11641

\bibitem[{{Shin} {et~al.}(2017){Shin}, {Udalski}, {Yee}, {Calchi Novati},
  {Han}, {Skowron}, {Mr{\'o}z}, {Soszy{\'n}ski}, {Poleski}, {Szyma{\'n}ski},
  {Koz{\l}owski}, {Pietrukowicz}, {Ulaczyk}, {Pawlak}, {OGLE Collaboration},
  {Albrow}, {Gould}, {Chung}, {Hwang}, {Jung}, {Ryu}, {Zhu}, {Cha}, {Kim},
  {Kim}, {Kim}, {Lee}, {Lee}, {Park}, {Pogge}, {KMTNet Group}, {Beichman},
  {Bryden}, {Carey}, {Gaudi}, {Henderson}, {Shvartzvald}, \& {Spitzer
  Team}}]{shin2017ogle}
{Shin}, I.~G., {Udalski}, A., {Yee}, J.~C., {et~al.} 2017, \aj, 154, 176

\bibitem[{{Shvartzvald} {et~al.}(2016){Shvartzvald}, {Li}, {Udalski}, {Gould},
  {Sumi}, {Street}, {Calchi Novati}, {Hundertmark}, {Bozza}, {Beichman},
  {Bryden}, {Carey}, {Drummond}, {Fausnaugh}, {Gaudi}, {Henderson}, {Tan},
  {Wibking}, {Pogge}, {Yee}, {Zhu}, {(Spitzer Team}, {Tsapras}, {Bachelet},
  {Dominik}, {Bramich}, {Cassan}, {Figuera Jaimes}, {Horne}, {Ranc}, {Schmidt},
  {Snodgrass}, {Wambsganss}, {Steele}, {Menzies}, {Mao}, {(RoboNet}, {Poleski},
  {Pawlak}, {Szyma{\'n}ski}, {Skowron}, {Mr{\'o}z}, {Koz{\l}owski},
  {Wyrzykowski}, {Pietrukowicz}, {Soszy{\'n}ski}, {Ulaczyk}, {(OGLE Group},
  {Abe}, {Asakura}, {Barry}, {Bennett}, {Bhattacharya}, {Bond}, {Freeman},
  {Hirao}, {Itow}, {Koshimoto}, {Li}, {Ling}, {Masuda}, {Fukui}, {Matsubara},
  {Muraki}, {Nagakane}, {Nishioka}, {Ohnishi}, {Oyokawa}, {Rattenbury},
  {Saito}, {Sharan}, {Sullivan}, {Suzuki}, {Tristram}, {Yonehara}, {(MOA
  Group}, {J{\o}rgensen}, {Burgdorf}, {Ciceri}, {D'Ago}, {Evans}, {Hinse},
  {Kains}, {Kerins}, {Korhonen}, {Mancini}, {Popovas}, {Rabus}, {Rahvar},
  {Scarpetta}, {Skottfelt}, {Southworth}, {Peixinho}, {Verma}, {(MiNDSTEp},
  {Sbarufatti}, {Kennea}, {Gehrels}, \& {(Swift}}]{SpitzerSwift1BD}
{Shvartzvald}, Y., {Li}, Z., {Udalski}, A., {et~al.} 2016, \apj, 831, 183

\bibitem[{{Shvartzvald} {et~al.}(2017){Shvartzvald}, {Yee}, {Calchi Novati},
  {Gould}, {Lee}, {Beichman}, {Bryden}, {Carey}, {Gaudi}, {Henderson}, {Zhu},
  {Spitzer Team}, {Albrow}, {Cha}, {Chung}, {Han}, {Hwang}, {Jung}, {Kim},
  {Kim}, {Kim}, {Lee}, {Park}, {Pogge}, {Ryu}, {Shin}, \& {KMTNet
  Group}}]{Shvartzvald2017}
{Shvartzvald}, Y., {Yee}, J.~C., {Calchi Novati}, S., {et~al.} 2017, \apjl,
  840, L3

\bibitem[{{Shvartzvald} {et~al.}(2019){Shvartzvald}, {Yee}, {Skowron}, {Lee},
  {Udalski}, {Calchi Novati}, {Bozza}, {Beichman}, {Bryden}, {Carey}, {Gaudi},
  {Henderson}, {Zhu}, {Spitzer Team}, {Bachelet}, {Bolt}, {Christie}, {Maoz},
  {Natusch}, {Pogge}, {Street}, {Tan}, {Tsapras}, {LCO}, {{\ensuremath{\mu}}FUN
  Follow-up Teams}, {Pietrukowicz}, {Soszy{\'n}ski}, {Szyma{\'n}ski},
  {Mr{\'o}z}, {Poleski}, {Koz{\l}owski}, {Ulaczyk}, {Pawlak}, {Rybicki},
  {Iwanek}, {OGLE Collaboration}, {Albrow}, {Cha}, {Chung}, {Gould}, {Han},
  {Hwang}, {Jung}, {Kim}, {Kim}, {Kim}, {Lee}, {Lee}, {Park}, {Ryu}, {Shin},
  {Zang}, {KMTNet Collaboration}, {Dominik}, {Helling}, {Hundertmark},
  {J{\o}rgensen}, {Longa-Pe{\~n}a}, {Lowry}, {Sajadian}, {Burgdorf},
  {Campbell-White}, {Ciceri}, {Evans}, {Fujii}, {Hinse}, {Rahvar}, {Rabus},
  {Skottfelt}, {Snodgrass}, {Southworth}, \& {MiNDSTEp
  Collaboration}}]{Shvartzvald2019}
{Shvartzvald}, Y., {Yee}, J.~C., {Skowron}, J., {et~al.} 2019, \aj, 157, 106

\bibitem[{{Thies} \& {Kroupa}(2007)}]{thies2007}
{Thies}, I., \& {Kroupa}, P. 2007, \apj, 671, 767

\bibitem[{{Thies} \& Kroupa(2008)}]{thies2008}
{Thies}, I., \& Kroupa, P. 2008, \mnras, 390, 1200

\bibitem[{{Tomaney} \& {Crotts}(1996)}]{tomaney1996}
{Tomaney}, A.~B., \& {Crotts}, A. P.~S. 1996, \aj, 112, 2872

\bibitem[{{Udalski} {et~al.}(1994){Udalski}, {Szymanski}, {Kaluzny}, {Kubiak},
  {Mateo}, {Krzeminski}, \& {Paczy\'nski}}]{OGLE-EWS}
{Udalski}, A., {Szymanski}, M., {Kaluzny}, J., {et~al.} 1994, \actaa, 44, 227

\bibitem[{{van Belle}(1999)}]{vanBelle1999}
{van Belle}, G.~T. 1999, in ASP Conf., Vol. 194, Working on the Fringe: Optical
  and IR Interferometry from Ground and Space, ed. S.~{Unwin} \& R.~{Stachnik},
  64

\bibitem[{Virtanen {et~al.}(2020)Virtanen, Gommers, Oliphant, Haberland, Reddy,
  Cournapeau, Burovski, Peterson, Weckesser, Bright, {van der Walt}, Brett,
  Wilson, Millman, Mayorov, Nelson, Jones, Kern, Larson, Carey, Polat, Feng,
  Moore, {VanderPlas}, Laxalde, Perktold, Cimrman, Henriksen, Quintero, Harris,
  Archibald, Ribeiro, Pedregosa, {van Mulbregt}, \& {SciPy 1.0
  Contributors}}]{scipy}
Virtanen, P., Gommers, R., Oliphant, T.~E., {et~al.} 2020, Nature Methods, 17,
  261

\bibitem[{{Wang} {et~al.}(2016){Wang}, {Spurzem}, {Aarseth}, {Giersz}, {Askar},
  {Berczik}, {Naab}, {Schadow}, \& {Kouwenhoven}}]{Wang2016}
{Wang}, L., {Spurzem}, R., {Aarseth}, S., {et~al.} 2016, \mnras, 458, 1450

\bibitem[{{Wegg} {et~al.}(2017){Wegg}, {Gerhard}, \& {Portail}}]{wegg2017}
{Wegg}, C., {Gerhard}, O., \& {Portail}, M. 2017, \apjl, 843, L5

\bibitem[{{Wozniak}(2000)}]{2000AcA....50..421W}
{Wozniak}, P.~R. 2000, \actaa, 50, 421

\bibitem[{{Yee} {et~al.}(2012){Yee}, {Shvartzvald}, {Gal-Yam}, {Bond},
  {Udalski}, {Koz{\l}owski}, {Han}, {Gould}, {Skowron}, {Suzuki}, {Abe},
  {Bennett}, {Botzler}, {Chote}, {Freeman}, {Fukui}, {Furusawa}, {Itow},
  {Kobara}, {Ling}, {Masuda}, {Matsubara}, {Miyake}, {Muraki}, {Ohmori},
  {Ohnishi}, {Rattenbury}, {Saito}, {Sullivan}, {Sumi}, {Suzuki}, {Sweatman},
  {Takino}, {Tristram}, {Wada}, {MOA Collaboration}, {Szyma{\'n}ski}, {Kubiak},
  {Pietrzy{\'n}ski}, {Soszy{\'n}ski}, {Poleski}, {Ulaczyk}, {Wyrzykowski},
  {Pietrukowicz}, {OGLE Collaboration}, {Allen}, {Almeida}, {Batista}, {Bos},
  {Christie}, {DePoy}, {Dong}, {Drummond}, {Finkelman}, {Gaudi}, {Gorbikov},
  {Henderson}, {Higgins}, {Jablonski}, {Kaspi}, {Manulis}, {Maoz}, {McCormick},
  {McGregor}, {Monard}, {Moorhouse}, {Mu{\~n}oz}, {Natusch}, {Ngan}, {Ofek},
  {Pogge}, {Santallo}, {Tan}, {Thornley}, {Shin}, {Choi}, {Park}, {Lee}, {Koo},
  \& {{\ensuremath{\mu}}FUN Collaboration}}]{renormerrors}
{Yee}, J.~C., {Shvartzvald}, Y., {Gal-Yam}, A., {et~al.} 2012, \apj, 755, 102

\bibitem[{{Yee} {et~al.}(2015){Yee}, {Gould}, {Beichman}, {Calchi Novati},
  {Carey}, {Gaudi}, {Henderson}, {Nataf}, {Penny}, {Shvartzvald}, \&
  {Zhu}}]{Spitzer}
{Yee}, J.~C., {Gould}, A., {Beichman}, C., {et~al.} 2015, \apj, 810, 155

\bibitem[{{Yee} {et~al.}(2021){Yee}, {Zang}, {Udalski}, {Ryu}, {Green},
  {Hennerley}, {Marmont}, {Sumi}, {Mao}, {Gromadzki}, {Mr{\'o}z}, {Skowron},
  {Poleski}, {Szyma{\'n}ski}, {Soszy{\'n}ski}, {Pietrukowicz}, {Koz{\l}owski},
  {Ulaczyk}, {Rybicki}, {Iwanek}, {Wrona}, {Albrow}, {Chung}, {Gould}, {Han},
  {Hwang}, {Jung}, {Kim}, {Shin}, {Shvartzvald}, {Cha}, {Kim}, {Kim}, {Lee},
  {Lee}, {Lee}, {Park}, {Pogge}, {Bachelet}, {Christie}, {Hundertmark}, {Maoz},
  {McCormick}, {Natusch}, {Penny}, {Street}, {Tsapras}, {Beichman}, {Bryden},
  {Calchi Novati}, {Carey}, {Gaudi}, {Henderson}, {Johnson}, {Zhu}, {Bond},
  {Abe}, {Barry}, {Bennett}, {Bhattacharya}, {Donachie}, {Fujii}, {Fukui},
  {Hirao}, {Ishitani Silva}, {Itow}, {Kirikawa}, {Kondo}, {Koshimoto}, {Li},
  {Matsubara}, {Muraki}, {Miyazaki}, {Olmschenk}, {Ranc}, {Rattenbury},
  {Satoh}, {Shoji}, {Suzuki}, {Tanaka}, {Tristram}, {Yamawaki}, \&
  {Yonehara}}]{ob190960Yee}
{Yee}, J.~C., {Zang}, W., {Udalski}, A., {et~al.} 2021, arXiv e-prints,
  arXiv:2101.04696

\bibitem[{{Zang} {et~al.}(2020){Zang}, {Shvartzvald}, {Udalski}, {Yee}, {Lee},
  {Sumi}, {Zhang}, {Yang}, {Mao}, {Calchi Novati}, {Gould}, {Zhu}, {Beichman},
  {Bryden}, {Carey}, {Gaudi}, {Henderson}, {Mr{\'o}z}, {Skowron}, {Poleski},
  {Szyma{\'n}ski}, {Soszy{\'n}ski}, {Pietrukowicz}, {Koz{\l}owski}, {Ulaczyk},
  {Rybicki}, {Iwanek}, {Wrona}, {Albrow}, {Chung}, {Han}, {Hwang}, {Jung},
  {Ryu}, {Shin}, {Cha}, {Kim}, {Kim}, {Kim}, {Lee}, {Lee}, {Park}, {Pogge},
  {Bond}, {Abe}, {Barry}, {Bennett}, {Bhattacharya}, {Donachie}, {Fujii},
  {Fukui}, {Hirao}, {Itow}, {Kirikawa}, {Kondo}, {Koshimoto}, {Li},
  {Matsubara}, {Muraki}, {Miyazaki}, {Ranc}, {Rattenbury}, {Satoh}, {Shoji},
  {Suzuki}, {Tanaka}, {Tristram}, {Yamawaki}, {Yonehara}, {Bachelet},
  {Hundertmark}, {Figuera Jaimes}, {Maoz}, {Penny}, {Street}, \&
  {Tsapras}}]{ob180799Zang}
{Zang}, W., {Shvartzvald}, Y., {Udalski}, A., {et~al.} 2020, arXiv e-prints,
  arXiv:2010.08732

\bibitem[{{Zhu} {et~al.}(2015){Zhu}, {Udalski}, {Gould}, {Dominik}, {Bozza},
  {Han}, {Yee}, {Calchi Novati}, {Beichman}, {Carey}, {Poleski}, {Skowron},
  {Koz{\l}owski}, {Mr{\'o}z}, {Pietrukowicz}, {Pietrzy{\'n}ski},
  {Szyma{\'n}ski}, {Soszy{\'n}ski}, {Ulaczyk}, {Wyrzykowski}, {OGLE
  Collaboration}, {Gaudi}, {Pogge}, {DePoy}, {Jung}, {Choi}, {Hwang}, {Shin},
  {Park}, {Jeong}, \& {{\ensuremath{\mu}}FUN
  Collaboration}}]{2015ApJ...805....8Z}
{Zhu}, W., {Udalski}, A., {Gould}, A., {et~al.} 2015, \apj, 805, 8

\bibitem[{{Zhu} {et~al.}(2017){Zhu}, {Udalski}, {Novati}, {Chung}, {Jung},
  {Ryu}, {Shin}, {Gould}, {Lee}, {Albrow}, {Yee}, {Han}, {Hwang}, {Cha}, {Kim},
  {Kim}, {Kim}, {Kim}, {Lee}, {Park}, {Pogge}, {KMTNet Collaboration},
  {Poleski}, {Mr{\'o}z}, {Pietrukowicz}, {Skowron}, {Szyma{\'n}ski},
  {KozLowski}, {Ulaczyk}, {Pawlak}, {OGLE Collaboration}, {Beichman}, {Bryden},
  {Carey}, {Fausnaugh}, {Gaudi}, {Henderson}, {Shvartzvald}, {Wibking}, \&
  {Spitzer Team}}]{Zhu2017}
{Zhu}, W., {Udalski}, A., {Novati}, S.~C., {et~al.} 2017, \aj, 154, 210

\end{thebibliography}

\end{document}